\documentstyle[color,preprint,aps,harvard,floats,epsfig]{revtex}
\citationstyle{dcu}
\newcommand{\red}[1]{#1}
\newcommand{\figref}[1]{Fig.~\protect\ref{#1}}
\newcommand{\secref}[1]{Sec.~\protect\ref{#1}}
\newcommand{\eqref}[1]{Eq.~(\protect\ref{#1})}
\newcommand{\xref}[1]{\protect\ref{#1}}

\newcommand{\fmref}[1]{(\protect\ref{#1})}
\newcommand{\eg}{e.g.\ }    
\newcommand{\ie}{i.e.\ }    
\newcommand{\tdvp}{time-dependent variational principle}
\newcommand{\se}{Schr\"odinger equation}
\newlength{\CaptionWidth}
\newcommand{\mycaption}[2]{
\begin{center}\begin{minipage}{\CaptionWidth}
\caption[]{#1}\label{#2}
\end{minipage}\end{center}
}
\newcommand{\element}[2]{$^{#1}$#2}
\def\geap{\raisebox{-.6ex}{$\stackrel {>}{\sim}$}} 
\def\leap{\raisebox{-.6ex}{$\stackrel {<}{\sim}$}} 
\def\V0{\stackrel{\circ}{V}}
\def\v0{\stackrel{\circ}{v}}
\def\half{{\frac{1}{2}}\;}

\newcommand{\MeV}{\mbox{\,MeV}}
\newcommand{\dint}{\mbox{d}}

\renewcommand{\Re}{\mbox{Re}}
\renewcommand{\Im}{\mbox{Im}}

\newcommand{\sign}{\mbox{sgn}}
\newcommand{\Tr}{\mbox{Tr}}

\newcommand{\op}[1]{%
    \fontdimen12\textfont3=2pt\fontdimen12\scriptfont3=1.4pt%
    \!\null\mathop{\vphantom{#1}\smash{#1}}\limits_{\sim}\null\!}
\newcommand{\Operator}[1]{\smash{\raisebox{-1.1ex}{
$\!\!\stackrel{\displaystyle #1}{\sim}$}}}
\newcommand{\LittleOperator}[1]{\smash{\raisebox{-1.1ex}{
$\!\!\stackrel{\scriptstyle #1}{\sim}$}}}
\newcommand{\EinsOp}
           {\;\smash{\raisebox{-1.1ex}{$\!\!\stackrel{\!\mbox{1}
            \hspace{-0.4ex}\rule[0.0ex]{0.06ex}{1.60ex}}{\sim}$}}}
\newcommand{\OpHHO}
           {\Operator{H}_{{\vphantom{A}}^{\mbox{\scriptsize HO}}}}
\newcommand{\OphHO}
           {\Operator{h}_{{\vphantom{A}}^{\mbox{\scriptsize HO}}}}

\newcommand{\EnsembleMean}[1]{\langle\langle \, {#1}\, 
            \rangle\rangle_{\big|\,\scriptstyle{T}} \,}
\newcommand{\ErgodicMean}[1]{\overline{\langle \, {#1}^{\vphantom{A}}
            \,\rangle}_{\big|
            \scriptstyle{\erw{\LittleOperator{H}}}}\,}

\def\bra#1{\langle \, {#1} \, | \,}
\def\ket#1{\, | \, {#1} \, \rangle}
\newcommand{\braket}[2]{\langle \, {#1} \, | \, {#2} \, \rangle}

\def\erw#1{\,\langle \, {#1} \, \rangle\,}

\def\OO{\Big( {\cal O}_{mk}{\cal O}_{nl} - {\cal O}_{ml}{\cal O}_{nk} \Big)}
\def\brak{\langle q_k | \;}

\def\ketl{\; | q_l \rangle}
\def\bram{\langle q_m | \;}

\def\ketn{\; | q_n \rangle}
\def\brakl{\langle q_k q_l | \;}

\def\ketmn{\; | q_m q_n \rangle}

\newcommand{\pp}[2]{\frac{\partial \, {#1}}{\partial \, {#2}}}
\newcommand{\dd}[2]{\frac{{d}\, {#1}}{{d} {#2}}}

\newcommand{\ddt}{\frac{d}{dt}}

\newcommand{\vek}[1]{{\!\vec{\,#1}}}

\newcommand{\veck}{\vek{k}\,}

\newcommand{\vecp}{\vek{p}\,}
\newcommand{\vecpk}{\vek{p}_{k}\,}
\newcommand{\vecr}{\vek{r}\,}
\newcommand{\vecrk}{\vek{r}_{k}\,}

\newcommand{\vecx}{\vek{x}\,}

\newcommand{\Hcal}{{\mathcal H}}

\def\prodkl{\langle q_k|q_l\rangle}
\newcommand{\Rerg}
           {\Operator{R}_{{\vphantom{A}}^{\mbox{\scriptsize erg}}}}
\newcommand{\HHO}
           {\Operator{H}_{{\vphantom{A}}^{\mbox{\scriptsize HO}}}}

\newcommand{\bvec}{\vek{b}\,}

\newcommand{\xvec}{\vek{x}\,}
\newcommand{\Xvec}{\vek{X}\,}
\newcommand{\dotvecr}{\dot{\vek{r}}}

\newcommand{\dotvecp}{\dot{\vek{p}}}

\newcommand{\kvecsim}{\raisebox{-1.1ex}{
$\!\!\vec{\stackrel{\displaystyle k}{\sim}}$}}

\newcommand{\xvecsim}{\raisebox{-1.1ex}{
$\!\!\vec{\stackrel{\displaystyle x}{\sim}}$}}

\begin{document}

\title{Molecular Dynamics for Fermions}
\author{Hans Feldmeier}
\address{Gesellschaft f\"ur Schwerionenforschung mbH\\ 
D-64220 Darmstadt}
\author{J\"urgen Schnack}
\address{Universit\"at Osnabr\"uck, Fachbereich Physik \\
Barbarastr. 7, D-49069 Osnabr\"uck}
\maketitle

\begin{abstract}
The time-dependent variational principle for many-body trial
states is used to discuss the relation between the 
approaches of different molecular dynamics models to describe
indistinguishable fermions. Early attempts to include 
effects of the Pauli principle by means of nonlocal potentials
as well as more recent models 
which work with antisymmetrized many-body states are reviewed 
under these premises.

\vspace{2ex}

\noindent{\it PACS:}
02.70.Ns; 
05.30.-d; 
05.30.Ch; 
05.30.Fk; 
05.60.+w; 
24.10.Cn

\vspace{1ex}

\noindent{\it Keywords:}
Many-body theory; Fermion system; Molecular dynamics;
Wave-packet dynamics; 
Time-dependent variational principle; Statistical properties;
Canonical ensemble; Ergodicity; Time averaging
\end{abstract}

\vspace*{2cm}
To be published in July 2000 issue of
{\it Reviews of Modern Physics}\\
{\copyright} 2000 The American Physical Society

\newpage
\tableofcontents
\newpage

%
\section{Introduction and Summary}
\label{I}

When correlations and fluctuations become important in the
dynamical evolution of a many-body system and mean field
approximations are not sufficient, molecular dynamics methods are
invoked frequently. 
Molecular dynamics means that the constituents (molecules) of
the many-body system are represented by few classical degrees of 
freedom (center of mass position and momentum, angles of
rotation etc.) and interact through potentials. The
equations of motion (Newton's or Hamilton's type)
are solved numerically. The interactions which are used range
from purely phenomenological to sophisticated ab initio quantal
potentials. The major advantage of molecular dynamics
simulations is that they do not rely on quasi-particle
approximations but include 
both, mean-field effects and many-body correlations. Therefore
they can provide insight into complex systems with correlations
on different 
scales.   

During the past decade one can observe an increasing interest in
the dynamics of many-fermion systems in which correlations are
important. In nuclear as well as in 
atomic physics collisions of composite fermion systems like
nuclei or atomic clusters are demanding many-body models which can
account for a large variety of phenomena. 
Depending on the energy one
observes fusion, dissipative reactions, fragmentation and even
multi-fragmentation, vaporisation or evaporation and
ionization. Also phase transitions in small systems are of
current interest.

Classical molecular dynamics is applicable if the de Broglie
wave-length of the molecules is small compared to the length scale of
typical variations of the interaction, otherwise the quantal
uncertainty relation becomes important. If the molecules are
identical fermions or bosons the de Broglie wave-length should
also be small compared to the mean inter-particle distance in
phase space, otherwise the Pauli or Bose principle is violated
and the model will have wrong statistical properties.

For nucleons in a nucleus, for example, both conditions
neccessary for classical mechanics are not fulfilled. The same
holds for electrons in bound states or at high densities and low
temperatures. Nevertheless one would like to utilize the merits
of a molecular dynamics model for indistingushable particles in
the quantum regime.

This article reviews attempts that are made to combine
Fermi-Dirac statistics with a semi-quantal trajectory picture
from the viewpoint of the quantal time-dependent variational
principle.  The closest quantum analogue to a point in
single-particle phase space representing a classical particle is
a wave packet well localized in phase space. The analogue to a
point in many-body phase space representing several classical
particles is a many-body state which is a product of localized
single-particle packets. If the particles are identical
fermions this product state has to be antisymmetrized, in the
case of bosons it has to be symmetrized.

For models which are formulated in terms of trial states and a
Hamilton operator, both, static and dynamical properties can be
obtained from appropriate quantum variational principles. Ground
states can be determined with the help of the Ritz variational
principle and the equations of motion can be accessed through
the time-dependent variational principle
which allows to derive approximations to the time-dependent \se\ on
different levels of accuracy. 

In chapter \xref{TDVP} we first discuss the
time-dependent variational principle in general and then show
how it works for various trial 
states and how classical mechanics can be obtained from quantum
mechanics by an appropriate choice of the dynamical variables.

Using wave packets automatically guarantees
that the Heisenberg uncertainty principle is not violated by the
model. That is actually a great problem in
classical simulations of Coulomb systems, where for instance the
hydrogen atom for a point like proton and electron is infinitely bound.
 
Using antisymmetrized many-body trial states automatically
guarantees that the Pauli exclusion principle is respected by the model.
Phenomena like shell structure or Fermi-Dirac statistics emerge in a
natural way and are discussed in chapter \xref{Antisymm} and
\xref{sec-4-0}.

Although antisymmetrized product states of single-particle
gaussian wave packets possess already major relevant degrees of
freedom, some processes like disintegration of wave packets, which
can occur in coordinate space (evaporation, capture or tunneling) 
and in momentum space (large momentum transfer due to collisions),
are poorly described by the equations of motion. 
One therefore often represents the system with a
mixture of trial states between which random transitions may
occur. This branching procedure, which is employed in atomic as
well as in nuclear physics, will be explained in
\secref{quantbranch}.

In chapter \xref{models} models used in nuclear and atomic
physics are reviewed from the general point of view of chapter \xref{TDVP} 
and \xref{Antisymm}. When energetic collisions between heavy
atomic nuclei became available classical molecular dynamics models 
were developed to describe the various phenomena observed.
To simulate the effect of antisymmetrization on the classical
trajectories many authors add to the hamiltonian a two-body
``Pauli potential" which is supposed to keep fermions apart from
each other in phase space.

Quantum Molecular Dynamics (QMD)
attributes to each fermion a gaussian wave packet with fixed
width instead of a point, but still uses a simple product state
for the many-body wave function and therefore obtains classical
equations of motion with two-body forces acting on the centroids
of the wave packets. These Newtonian forces are 
supplemented by random forces which simulate hard collisions. In
these collision terms Pauli blocking is included. The
statistical properties are however mainly those of
distinguishable particles. 

Antisymmetrized Molecular Dynamics (AMD) also uses gaussian wave
packets with fixed width, but antisymmetrizes the many-body state.
Like in QMD a
collision term is added to account in a phenomenological way for
branching into other Slater determinants.
In Fermionic Molecular Dynamics (FMD) 
in addition the width degree of freedom is considered. This
non-classical degree of freedom is important for phenomena like
evaporation and it also plays an important r\^{o}le in
statistical properties of the model.

Section \xref{secap} explains the different use of the term Quantum
Molecular Dynamics in atomic and nuclear physics. It then refers
to applications of trajectory calculations for individual
electrons and ions where the density of the electrons is too
large to neglect their fermion character. Quantum branching in
the atomic context is also briefly discussed.  

Although designed for non-equilibrium simulations like
collisions, molecular dynamics models are also used to
simulate systems in thermal equilibrium. In chapter
\xref{sec-4-0} their statistical properties are investigated by
means of time averaging. As applications the nuclear liquid-gas
phase transition or the hydrogen plasma 
under extreme conditions are discussed.

In conclusion, the review shows that it is possible to extend
the classical trajectory picture to identical fermions by means
of localized wave packets. When the phase space density
increases the classical notion of positions $\vec{r}_k$ and
momenta $\vec{p}_k$ as mean values of narrow wave packets has to
be reinterpreted as parameters which identify an antisymmetrized
many-body state
$\ket{Q}=\ket{\vec{r}_1,\vec{p}_1,\vec{r}_2,\vec{p}_2, \cdots}$.
When the individual packets overlap in phase space $\vec{r}_k$
and $\vec{p}_k$ can no longer be identified with the classical
variables.  Calculating all observables with the many-body state
$\ket{Q}$ as quantum expectation values, \eg ${\mathcal
H}(\vec{r}_1,\vec{p}_1,\vec{r}_2,\vec{p}_2, \cdots)=
\bra{Q}\op{H}\ket{Q}/\braket{Q}{Q}$, and not misinterpreting them  
as classical expressions, includes the Pauli exclusion principle
and Heisenberg's uncertainty principle in a natural and correct
way. In the equations of motion the exclusion principle causes a
complicated metric in the $N$-particle parameter space
\red{in the sense that canonical pairs of variables can only be
defined locally.}

\red{The antisymmetrization of localized wave packets, which brings
together the Pauli principle, Heisenberg's uncertainty relation and
the classical trajectory picture, leads to many, some times unexpected,
quantal features. Nevertheless one has to be aware that one is still
dealing with a very simplified trial state and other degrees of freedom
may be important. Especially when the interaction is not smooth across
a wave packet it may want to change its shape to one which is not in
the allowed set, for example split with certain
amplitudes into different parts which after some time evolve
independently. Or more generally, branching into other trial states
away from the one which follows the
approximate time evolution of a pure state leads to a mixture of
antisymmetrized wave packets.
The consistent treatment of this aspect of
quantum branching needs further attention in the literature.}


Common to all models discussed is the anticorrelation between
its degree of consistent derivation and its computational effort.

\newpage
\section{Time-dependent variational principle}
\label{TDVP}

\subsection{General remarks}
\label{gen.remarks}

The time evolution of a state in quantum mechanics is given by
the time-dependent Schr\"odinger equation 
\footnote{
Throughout the article all operators are underlined with a tilde,
\eg $\op{H}$, and expectation values
are denoted by caligraphic letters, \eg ${\mathcal H}$. If not
needed explicitly $\hbar$ is taken to be one.}
\begin{equation} \label{Schroedinger}
i\frac{d}{dt}\ket{\Psi(t)}=\op{H}\ket{\Psi(t)}\ ,
\end{equation}
where $\op{H}$ is the hamiltonian and $\ket{\Psi(t)}$ the
many-body state which describes the physical system.  This
equation of motion can be obtained from the variation of the
action \cite{KeK76,KrS81,DPC86,BLK88}
\begin{equation} \label{actionprime}
{\mathcal A}^\prime=\int^{t_2}_{t_1} dt \ 
       {\mathcal L}^\prime(\Psi(t)^*,\Psi(t),\dot{\Psi}(t))\ ,
\end{equation}
keeping the variations fixed at the end points: 
$\delta \Psi(t_1)=\delta \Psi(t_2)=
\delta \Psi^*(t_1)=\delta \Psi^*(t_2)=0$.
The Lagrange function  
\begin{eqnarray} \label{Lagrange}
{\mathcal L}^\prime(\Psi(t)^*,\Psi(t),\dot{\Psi}(t))&=&
\bra{\Psi(t)}i\frac{d}{dt}\ket{\Psi(t)}
-\bra{\Psi(t)}\op{H}\ket{\Psi(t)}  
\end{eqnarray}
is a function of the dynamical variables, denoted by the set
$\Psi(t)$. Usually they are chosen to be complex so that
$\Psi(t)$ and $\Psi^*(t)$  may be regarded as independent
variables. If the mapping of $\Psi(t)$ on the many-body state
$\ket{\Psi(t)}$ is analytic, $\ket{\Psi(t)}$ depends only on 
$\Psi(t)$ and
the hermitian adjoint state $\bra{\Psi(t)}$ only on the 
complex conjugate set $\Psi^*(t)$.
The Lagrange function depends on the first time 
derivatives $\dot{\Psi}(t)$ through $\frac{d}{dt}\ket{\Psi(t)}$. 

In general, the set $\Psi(t)$ contains infinitely many dynamical
degrees of freedom, for example  
the complex coefficients of an orthonormal basis or the values
of a $N$-body wave function on a $3N$-dimensional grid in coordinate space.

The Lagrange function (\ref{Lagrange}) is appropriate if the set of
variables $\Psi(t), \Psi^*(t)$ contains a complex overall factor
to $\ket{\Psi(t)}$ which takes care of norm and phase.
If $\ket{\Psi(t)}$ cannot be normalized by means of the set
$\Psi(t)$ the Lagrange function defined in \eqref{LagrNorm}
should be used. We shall employ both forms.

For linearly independent variables $\Psi(t)$ and $\Psi^*(t)$
the variation of $\Psi^*(t)$ yields for the extremal action   
\begin{equation} \label{extremal}
0=\delta{\mathcal A}^\prime=\int_{t_1}^{t_2} dt \quad 
\bra{\delta\Psi(t)}i\ddt-\op{H}\ket{\Psi(t)}\ .
\end{equation}
If $\ket{\Psi(t)}$ represents the most general state in Hilbert space,
the variation $\bra{\delta\Psi(t)}$ is unrestricted and
\eqref{extremal} can only be fulfilled if 
\begin{equation} 
\Big(i\ddt-\op{H}\Big)\ket{\Psi(t)} = 0 \ ,
\end{equation}
which is just the Schr\"odinger equation (\ref{Schroedinger}).

The variation of $\Psi(t)$ yields 
\begin{equation} 
0=\delta{\mathcal A}^\prime=\int_{t_1}^{t_2} dt \quad 
\bra{\Psi(t)}i\ddt-\op{H}\ket{\delta\Psi(t)} \ ,
\end{equation}
which after partial integration over $t$ with fixed endpoints
$\ket{\delta\Psi(t_1)}=\ket{\delta\Psi(t_2)}=0$ results in the
hermitian adjoint of the Schr\"odiger equation
\begin{equation}
\bra{\Psi(t)}{\left(-i\stackrel{\longleftarrow}{\ddt}-
\stackrel{\longleftarrow}{\op{H}}\right)}=0\ .
\end{equation}

As $\Psi(t)$ is linearly independent of $\Psi^*(t)$, both have
to be varied. Therefore the time integral which was missing in
early attempts, used by \cite{Fre34} and others, is necessary.

The reason for the reformulation of a differential equation as a
variational principle is of course the anticipation that a
suitably chosen restriction in the dynamical variables will lead
to a useful approximation of the full problem.
\red{For more general considerations on the construction of variational
principles see \cite{GRS83,BaV88} and references therein.}

\subsubsection{Restricted set of variables}

Let a restricted choice of variables be denoted by the complex
set $Q^\prime(t)=\{q_0(t),q_1(t),q_2(t),\cdots\}$ which
specifies the many-body state $\ket{Q^\prime(t)}$.  It is
presumed that $q_0(t)$ is always a complex overall factor in the
sense
\begin{equation} \label{QQprime} 
\ket{Q^\prime(t)}=q_0(t)\ket{Q(t)}=
q_0(t)\ket{q_1(t),q_2(t),\cdots} \ .
\end{equation}
Furthermore, the time-dependence of $\ket{Q^\prime(t)}$ is
supposed to be only implicit through the variables $Q^\prime(t)$.

It should be noted that the manifold $\ket{Q^\prime(t)}$ is in general
only a subset of the Hilbert space and needs not to form a subspace.  
 
Variation of the action (\ref{actionprime}) with the Lagrange
function (\ref{Lagrange}) with respect to $Q^{\prime *}(t)$ leads to
\begin{equation} \label{varpinciple}
0=\delta{\mathcal A}^\prime=\int_{t_1}^{t_2} dt 
\sum_\nu \delta q_\nu^*(t) \left( \pp{}{q_\nu^*}\bra{Q^\prime(t)}\right) 
\left( i\ddt-\op{H}\right) \ket{Q^\prime(t)}\ .
\end{equation}
As $\delta q_\nu^*(t)$ are arbitrary functions the action is
extremal if the following equations of motion are fulfilled 
\begin{equation} \label{eomQ} 
\left( \pp{}{q_\nu^*}\bra{Q^\prime(t)}\right) 
\left( i\ddt-\op{H} \right) \ket{Q^\prime(t)}=0
\end{equation}
or
\begin{equation}
\label{E-1-1-2}
i\,\sum_{\mu}\;
{\mathcal C}_{\nu\mu}^\prime\,
\dot{q}_{\mu}\,
=
\pp{}{q_{\nu}^*}\;
\bra{Q^\prime(t)} \op{H} \ket{Q^\prime(t)}
\ ,
\end{equation}
where
\begin{equation}
\label{E-1-1-3}
{\mathcal C}_{\nu\mu}^\prime
=
\frac{\partial^2}{\partial\,q_{\nu}^*\;\partial\,q_{\mu}}
\braket{Q^\prime(t)}{Q^\prime(t)}
\ .
\end{equation}
Different from the unrestricted variation one obtains 
equations of motion for the complex parameters which in turn
define the time evolution of the trial state $\ket{Q^\prime(t)}$
in Hilbert space.

After partial integration over time variation with respect to
$Q^\prime(t)$, with fixed end points, 
$\delta q_\nu(t_1)=\delta q_\nu(t_2)=0$, results in equations of motion
which are the complex conjugate of Eqs. \fmref{eomQ} - \fmref{E-1-1-3}. 

In the following, for sake of simplicity, the explicit indication of
the time dependence is sometimes omitted.

The time evolution of $q_0$ can be expressed in terms of the other
variables. For that the Lagrangian ${\mathcal L}^\prime$ may be written as
\begin{eqnarray}
{\mathcal L}^\prime & = & \bra{Q^\prime}i\ddt-\op{H}\ket{Q^\prime} 
\nonumber \\
             & = & i \, q_0^* \dot{q_0} \, \braket{Q}{Q}  
+ q_0^* q_0 \, \braket{Q}{Q}\, {\mathcal L}(Q,Q^*,\dot{Q},\dot{Q}^*)   
- \frac{i}{2} \ddt\Big(q_0^* q_0 \, \braket{Q}{Q}\Big) \ ,
\end{eqnarray}
where the set $Q$ does not contain $q_0$ anymore (see
\eqref{QQprime}) and a new Lagrangian ${\mathcal L}(Q^*,Q,\dot{Q}^*,\dot{Q})$ is
defined by
\begin{eqnarray} \label{LagrNorm}
{\mathcal L}(Q^*,Q,\dot{Q}^*,\dot{Q})&=&
\frac{i}{2} \left(
\frac{\braket{Q}{\dot{Q}}-\braket{\dot{Q}}{Q}}{\braket{Q}{Q}}\right)
-\frac{\bra{Q}\op{H}\ket{Q}}{\braket{Q}{Q}} \nonumber
\\
&\equiv& {\mathcal L}_0(Q^*,Q,\dot{Q}^*,\dot{Q}) - 
{\mathcal H}(Q^*,Q) 
\end{eqnarray}
with
\begin{eqnarray} 
\ket{\dot{Q}}\equiv\ddt\ket{Q}=\sum_\nu\dot{q}_\nu
\pp{}{q_\nu}\ket{Q} \qquad\mbox{and}\qquad
\bra{\dot{Q}}\equiv\ddt\bra{Q}=\sum_\nu\dot{q}_\nu^*
\pp{}{q_\nu^*}\bra{Q} \ .
\end{eqnarray}
The new Lagrange function ${\mathcal L}$ contains the norm explicitly
and is made real by subtracting the total time
derivative. ${\mathcal H}(Q^*,Q)$ is the expectation value of the
hamiltonian $\op{H}$ and will be called Hamilton function.

It is easy to verify that the solution of the equation of motion
(\ref{eomQ}) for $q_0$ is
\begin{eqnarray} \label{q0}
q_0(t)=\frac{1}{\braket{Q(t)}{Q(t)}^\half}
\exp\left\{i\int^tdt'{\mathcal L}(t') \right\}
\end{eqnarray}
and analogue for the complex conjugate $q_0^*(t)$. Thus, the
variational freedom of an overall factor $q_0(t)$ is used by the
time-dependent variational principle to provide a state
$\ket{Q^\prime(t)}$ with a time-independent norm and an the
additional phase $\int^tdt'{\mathcal L}(t')$ (${\mathcal L}$ is real by
construction).  Furthermore, insertion of the solution
(\ref{q0}) into ${\mathcal L}^\prime$ shows that along the trajectory
\begin{eqnarray}
{\mathcal L}^\prime(Q^{\prime*}(t),Q^\prime(t),\dot{Q}^\prime(t))=0
\ ,
\end{eqnarray}
irrespective of the choice of the remaining degrees of freedom
in $\ket{Q}$. 

The equations of motion (\ref{eomQ}) and their complex conjugate
ones can of course also be expressed as Euler Lagrange equations
\begin{eqnarray} \label{EulLagrprime}
\dd{}{t}\pp{{\mathcal L}^\prime}{\dot{q}_\nu^*}-
\pp{{\mathcal L}^\prime}{q_\nu^*}=0 
\qquad  \mbox{and} \qquad
\dd{}{t}\pp{{\mathcal L}^\prime}{\dot{q}_\nu}-
\pp{{\mathcal L}^\prime}{q_\nu}=0 \ . 
\end{eqnarray}
For the remaining variables $\{q_\nu^*,q_\nu;\nu\neq0\}$
they can be written in terms of the new Lagrange function
${\mathcal L}$ as 
\begin{eqnarray} \label{LLprime}
\ddt\pp{{\mathcal L}^\prime}{\dot{q}_\nu^*}-\pp{{\mathcal L}^\prime}{q_\nu^*}
&=&
\left(\ddt\pp{{\mathcal L}}{\dot{q}_\nu^*}-\pp{{\mathcal L}}{q_\nu^*}\right)
q_0^*q_0\braket{Q}{Q} + \nonumber \\  
& & \ddt(q_0^*q_0\braket{Q}{Q})\pp{{\mathcal L}}{\dot{q}_\nu^*}-
(iq^*_0\dot{q}q_0+q_0^*q_0{\mathcal L})\pp{\braket{Q}{Q}}{q_\nu^*}=0
\end{eqnarray}
and the analogue complex conjugate ones. The last two terms
vanish when the general solution (\ref{q0}) for $q_0$ is
inserted. Hence for $\nu\neq0$ the Euler Lagrange equations
with ${\mathcal L}$ as given in (\ref{LagrNorm}) are equivalent to the
ones with ${\mathcal L}^\prime$.   

Therefore, the action 
\begin{eqnarray} \label{action}
{\mathcal A}=\int_{t_1}^{t_2}dt \ {\mathcal L}(Q^*,Q,\dot{Q}^*,\dot{Q}) \ ,
\end{eqnarray}
which one also finds often as a starting point
\cite{KeK76,KrS81,DPC86,BLK88}, is the appropriate one if the
trial state $\ket{Q}$ is not normalized and the phase is
disregarded\footnote{${\mathcal L}$ may also be written with normalized trial
states $\ket{Q}/\sqrt{\braket{Q}{Q}}$. The only difference is a total
time derivative emerging from ${\mathcal L}_0$ so that the Lagrange
functions are equivalent.}.
This form will turn out to be more convenient when
dealing with antisymmetrized states in section \ref{Antisymm}.

One should however keep in mind that the simple inclusion of
$q_0$ provides automatically norm and phase.

The Euler Lagrange equations, which result from variation of the
action (\ref{action})
\begin{eqnarray} \label{EulLagr}
\dd{}{t} \pp{{\mathcal L}}{\dot{q}_\nu^*} =\pp{{\mathcal L}}{q_\nu^*} 
\qquad \mbox{and} \qquad 
\dd{}{t} \pp{{\mathcal L}}{\dot{q}_\nu} =\pp{{\mathcal L}}{q_\nu} 
\end{eqnarray}
can be written in terms of
the Hamilton function 
${\mathcal H}=\bra{Q}\op{H}\ket{Q}/\braket{Q}{Q}$ as generalized
Hamilton's equations
\begin{eqnarray} \label{eom}
  i \sum_{\nu} {\mathcal C}_{\mu\nu} \, \dot{q}_{\nu} =
  \pp{{\mathcal H}}{q^*_{\mu}} \qquad \mbox{and} \qquad 
  - i \sum_{\nu} {\mathcal C}^*_{\mu\nu} \, {\dot{q}}^*_{\nu} = 
  \pp{{\mathcal H}}{q_{\mu}} \ . 
\end{eqnarray}
The nonnegative hermitian matrix ${\mathcal C}$ depends in
general on $Q$ and is given by   
\begin{eqnarray}  \label{cmatrix}
  {\mathcal C}_{\mu\nu}  =
  \frac{\pp{}{q^*_{\mu}}\pp{}{q_{\nu}}\braket{Q}{Q}}{\braket{Q}{Q}} 
  -  \frac{\pp{}{q^*_{\mu}}\braket{Q}{Q}}{\braket{Q}{Q}} 
  \frac{\pp{}{q_{\nu}}\braket{Q}{Q}}{\braket{Q}{Q} }
 = \pp{}{q^*_{\mu}}\pp{}{q_{\nu}} \ln \braket{Q}{Q} 
  \ .
\end{eqnarray}
It plays the r\^{o}le of a metric on the manifold of dynamical
variables $Q$.

There is no need to assume that the set $Q$ contains only
complex variables. But as will be seen later in the context of
classical mechanics, the real and imaginary parts of $q_\nu$
play the r\^{o}le of canonical pairs of variables. On the other
hand for quantum mechanics it is only natural to work with
complex variables as the quantum state $\ket{Q}$ is necessarily
complex.
The assumption of an analytic mapping of $Q$ onto the trial state
$\ket{Q}$ is also not compulsory but makes the equations usually
more transparent.

\subsubsection{Deviation from the exact solution}
\label{S-1-1-2}

The static analogue to the time-dependent variational principle
is the Rayleigh-Ritz variational principle for the energy. Here one
can show that the overlap between the true ground state of the
hamiltonian and the trial state $\ket{Q^\prime}$ is increasing when the
energy $\bra{Q^\prime}\op{H}\ket{Q^\prime}$ is
decreasing. Since any additional degree of freedom lowers the
energy, more degrees of freedom imply always an
improved description. General statements like that cannot be
made for the time-dependent variational principle, although one
would believe that more degrees of freedom will result in a time
evolution which is closer to that of the Schr\"odinger equation.

In order to understand better in which sense the \tdvp\
optimizes the evolution of the trial state, we consider the
deviation of the approximate solution from the exact one, which
develops during a short time $\delta t$
\begin{eqnarray}  \nonumber 
\ket{\Delta(t, \delta t)}
&=&
\ket{\Psi_{exact}(t+\delta t)} - \ket{Q^\prime(t+\delta t)}
=\exp(-i\op{H}\delta t)\ket{Q^\prime(t)}-
   \ket{Q^\prime(t+\delta t)}\\
\label{error}
&=&
-i\left(\op{H} - i \sum_\nu\, \dot{q}_\nu \pp{}{q_\nu}\right)
       \ket{Q^\prime(t)} \delta t + \mbox{order}(\delta t^2)\ .  
\end{eqnarray}
The equations of motion (\ref{eomQ}) which result from the
time-dependent variational principle demand that 
\begin{eqnarray} \label{errortang}
\left(\pp{}{q^*_\nu}\bra{Q^\prime(t)}\right)
\ket{\Delta(t,\delta t)} = 0 \ ,
\end{eqnarray}
which means that the deviation is orthogonal to all tangent states
$\pp{}{q_\nu}\!\!\ket{Q^\prime}$, see \figref{F-1-3-1}. In
other words the approximate equations of motions evolve
$Q^\prime(t)$ to that point $Q^\prime(t+\delta t)$ in the
manifold where any small change in all possible directions
increases the distance $\braket{\Delta(t, \delta t)}{\Delta(t,
\delta t)}$ between true $\ket{\Psi_{exact}(t+\delta t)}$ and
approximate solution $\ket{Q^\prime(t+\delta t)}$. 
\begin{figure}[t]
\unitlength1mm
\begin{picture}(120,80)
\put(0,0){\epsfig{file=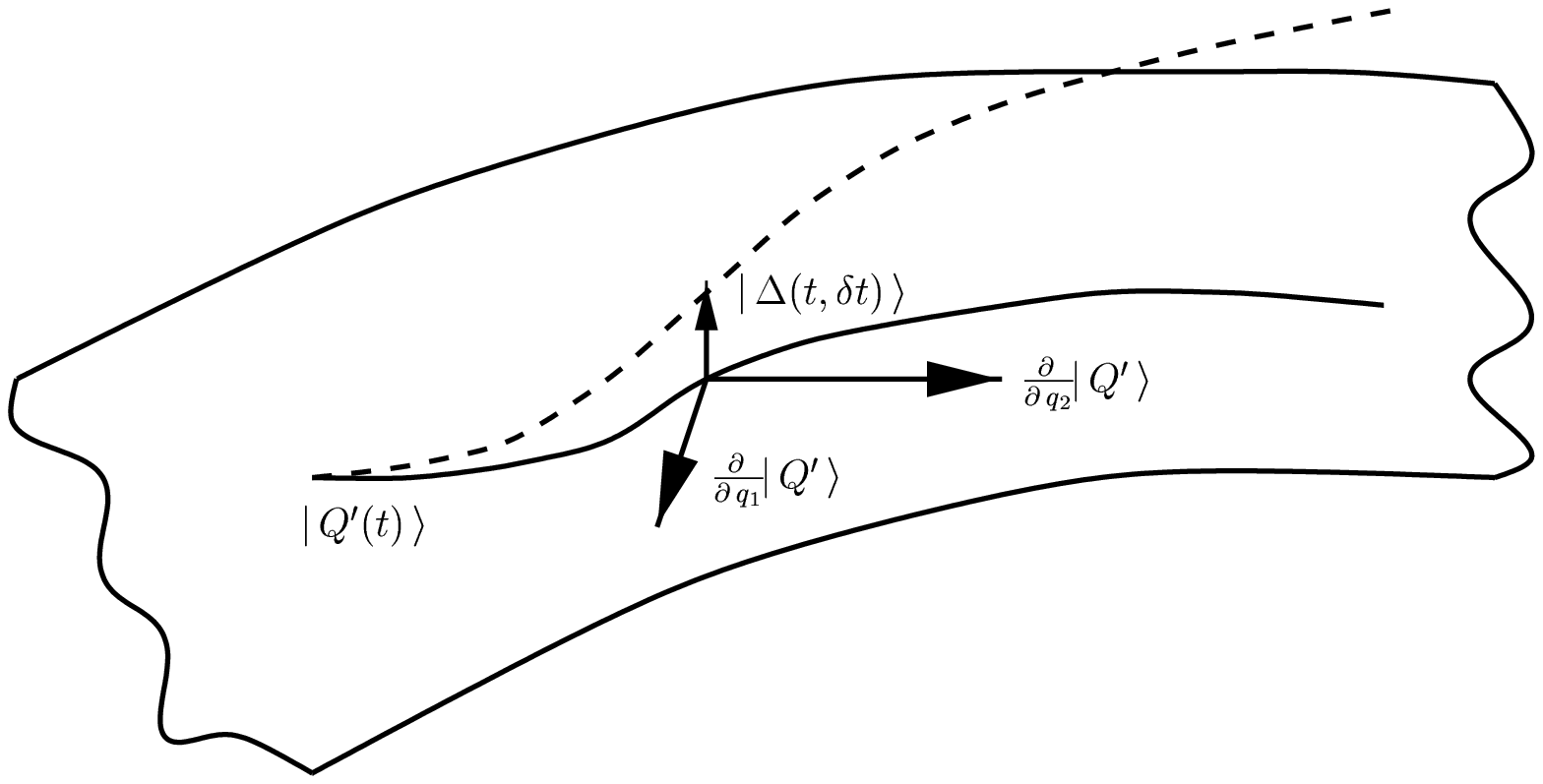,height=80mm}}
\end{picture}
\mycaption{Sketch of the manifold of trial states with the
approximate time evolution (solid line) and the solution of the
Schr\"odinger equation (dashed line). The error
$\ket{\Delta(t,\delta t)}$ is orthogonal to all tangent states
$\pp{}{q_\nu}\!\!\ket{Q^\prime}$.}{F-1-3-1}
\end{figure} 

The minimum condition 
\begin{eqnarray} 
0
&=&
\pp{}{q^*_\nu(t+\delta t)}\braket{\Delta(t, \delta t)}{\Delta(t, \delta t)}
\nonumber
\\
&=&
-\left(\pp{}{q^*_\nu(t+\delta t)}\bra{Q^\prime(t+\delta t)}\right)
\ket{\Delta(t, \delta t)}
\nonumber
\\
\label{dist} 
&=&
-\left(\pp{}{q^*_\nu(t)}\bra{Q^\prime(t)}\right)
\ket{\Delta(t, \delta t)} 
+ \mbox{order}(\delta t^3) \ .
\end{eqnarray}
is fulfilled because of \eqref{errortang}.
For variations with respect to $q_\nu$ the complex conjugate of
Eqs. (\ref{errortang}) and (\ref{dist}) have to be used.
Of course after many time
steps $\delta t$ the deviation may become large.

\subsubsection{Poisson brackets and canonical variables}

With the help of ${\mathcal C}$ it is possible to
introduce Poisson brackets. Using the equations of motion
(\ref{eom}) the time derivative of the expectation value of a
time-independent observable $\op{B}$ can be written as
\begin{eqnarray}
    \dot{{\mathcal B}} 
&=& \ddt \frac{\bra{Q}\op{B}\ket{Q}}{\braket{Q}{Q}} 
= \sum_{\nu} \biggl( 
\dot{q}^*_\nu \pp{{\mathcal B}}{q^*_{\nu}} + \,
\dot{q}_{\nu} \pp{{\mathcal B}}{q_\nu} \biggr) 
\nonumber \\
&=& i\, \sum_{\mu,\nu} 
\biggl(
      \pp{{\mathcal H}}{q_{\mu}} {\mathcal C}_{\mu\nu}^{-1}
      \pp{{\mathcal B}}{q_\nu^*} 
- \, \pp{{\mathcal B}}{q_\mu}
      {\mathcal C}_{\mu\nu}^{-1} \pp{{\mathcal H}}{q_\nu^*}
\biggr) 
\nonumber \\
\label{Poisson}
&=&: \{ {\mathcal H}, {\mathcal B} \} \; .
\end{eqnarray}
The real and imaginary parts of $q_\nu$ form pairs of canonical
variables if ${\mathcal C}$ is the unit matrix.
In the following section \ref{qutoclass} two examples are given
for this case, the Schr\"odinger equation and Hamilton's
equation of motion.  

In the general case, in which ${\mathcal C}_{\mu\nu}$ is not diagonal
and depends on $Q$, pairs of canonical variables exist locally
according to Darboux's theorem \cite{Arn89}.
One possible transformation is given by
\begin{eqnarray}
\label{E-1-1}
\dint r_{\mu} 
= 
\sum_{\nu}\;{\mathcal C}_{\mu\nu}^{1/2}(Q^*,Q)\,\dint q_{\nu} 
\quad\mbox{and}\quad
\dint r_{\mu}^*
= 
\sum_{\nu}\;\dint q_{\nu}^*\,{\mathcal C}_{\mu\nu}^{1/2}(Q^*,Q)
\ .
\end{eqnarray}
Written with the new variables $r_{\mu}$ and $r_{\mu}^*$ the
Poisson bracket \fmref{Poisson} takes the form
\begin{eqnarray}
\{ {\mathcal H}, {\mathcal B} \}
&=&
i\, \sum_{\mu} 
\biggl(
      \pp{\widehat{\mathcal H}}{r_{\mu}}\, \pp{\widehat{\mathcal B}}{r_\mu^*} 
-     \pp{\widehat{\mathcal B}}{r_\mu}\, \pp{\widehat{\mathcal H}}{r_\mu^*}
\biggr) 
\ ,
\end{eqnarray}
so that $(r_{\mu},r_{\mu}^*)$ form pairs of canonical
variables. The problem is, however, that in nontrivial cases the
transformation \fmref{E-1-1} cannot be written in a global way
as $R(Q)=\{r_0(q_0, q_1,\cdots), r_1(q_0, q_1,\cdots),\cdots\}$
and the Hamilton
function ${\mathcal H}$ or the observable ${\mathcal B}$ cannot
be expressed in the new variables
\begin{eqnarray}
\widehat{\mathcal H}(R^*,R)
=
{\mathcal H}(Q^*,Q)
\quad ,\qquad
\widehat{\mathcal B}(R^*,R)
=
{\mathcal B}(Q^*,Q)
\ .
\end{eqnarray}
A set of canonical pairs which are real is given by
\begin{eqnarray}
\rho_\mu
=
\frac{1}{\sqrt{2}}\left(r_\mu^* + r_\mu \right)
\quad\mbox{and}\quad
\pi_\mu
=
\frac{i}{\sqrt{2}}\left(r_\mu^* - r_\mu \right)
\ ,
\end{eqnarray}
which yields the standard Poisson brackets
\begin{eqnarray}
\{ {\mathcal H}, {\mathcal B} \}
&=&
\sum_{\mu} 
\biggl(
      \pp{\widehat{\widehat{\mathcal H}}}{\pi_{\mu}}\, 
\pp{\widehat{\widehat{\mathcal B}}}{\rho_\mu} 
-     \pp{\widehat{\widehat{\mathcal B}}}{\pi_\mu}\, 
\pp{\widehat{\widehat{\mathcal H}}}{\rho_\mu}
\biggr) 
\ ,
\end{eqnarray}
where $\widehat{\widehat{\mathcal H}}$ and 
$\widehat{\widehat{\mathcal B}}$ are now functions of $\rho_\mu$
and $\pi_\mu$.
Besides the trivial examples discussed in \secref{qutoclass} we
give one nontrivial example in \secref{sec-2-1-3}. 

The mere fact that according to Darboux's theorem canonical
pairs $(\rho_\mu,\pi_\mu)$ exist, admits to guess or make an
ansatz for the Hamilton function 
$\widehat{\widehat{\mathcal H}}(\rho_1,\rho_2,\dots,\pi_1,\pi_2,\dots)$
and use Hamilton's equations of motion. But with such a guess the
connection to the trial state $\ket{Q}$ is lost and the physical
meaning of $\rho_\mu$ and $\pi_\mu$ is obscured. This will
become obvious when we discuss trial states for
indistinguishable particles in \secref{Antisymm}.

\subsubsection{Conservation laws}

An expectation value is conserved if \cite{BLL89}
\begin{eqnarray}
\dot{{\mathcal B}}
=
\{ {\mathcal H},{\mathcal B} \}
&=& i\, \sum_{\mu,\nu} 
\biggl(
      \pp{{\mathcal H}}{q_{\mu}} {\mathcal C}_{\mu\nu}^{-1}
      \pp{{\mathcal B}}{q_\nu^*} 
- \, \pp{{\mathcal B}}{q_\mu}
      {\mathcal C}_{\mu\nu}^{-1} \pp{{\mathcal H}}{q_\nu^*}
\biggr) 
=
0
\ .
\end{eqnarray}
Hence the energy
$\mathcal H$ itself is always conserved by the equations of
motion, provided they are derived from the variational
principle. This is completely independent on the choice of the
trial state.
 
In the following it is shown how to identify other constants of
motion and how the trial state has to be chosen in order to
ensure desired conservation laws.  For that we consider a
unitary transformation with the hermitian generator $\op{G}$
\begin{equation}
\op{U} = \exp\left(\,i\,\varepsilon \; \op{G} \, \right) \quad , \quad
\varepsilon \quad \mbox{real} \ .
\end{equation}
If $\op{U}$ maps the set of trial states onto itself
\begin{equation}
\op{U} \ket{Q^\prime}\quad \in \quad \left\{ \ket{Q^\prime} \right\}\ ,
\end{equation}
then the special infinitesimal variation 
$\ket{Q^\prime(t)+\delta Q^\prime(t)}=\exp\{i\,\delta\varepsilon(t)\,\op{G}\}
\ket{Q^\prime(t)}$
of the action \fmref{actionprime} yields
\begin{eqnarray} 
0
&=&
\int_{t_1}^{t_2}\dint t\;
\bra{Q^\prime(t)}\exp\{-i\,\delta\varepsilon(t)\,\op{G}\}
\left(
i\,\ddt - \op{H}
\right)
\exp\{i\,\delta\varepsilon(t)\,\op{G}\}\ket{Q^\prime(t)}
\\
&=&
\int_{t_1}^{t_2}\dint t\;
\delta\varepsilon(t)\,
\left\{
\ddt \bra{Q^\prime(t)}\op{G}\ket{Q^\prime(t)}
-
\bra{Q^\prime(t)}i\left[\op{H},\op{G} \right]\ket{Q^\prime(t)}
\right\}
\nonumber \\
&&
+ \mbox{total time derivative} + \mbox{order}(\delta\varepsilon^2)
\nonumber \ .
\end{eqnarray}
As $\delta\varepsilon(t)$ is arbitrary and vanishes at the end
points one obtaines
\begin{eqnarray} 
\ddt {\mathcal G} 
&=& 
\ddt \bra{Q^\prime(t)} \op{G} \ket{Q^\prime(t)} 
\nonumber
\\
&=&
\{ {\mathcal H},{\mathcal G} \}
=  
\bra{Q^\prime(t)} i\left[\op{H},\op{G} \right]\ket{Q^\prime(t)} \ .
\label{gener}
\end{eqnarray}
That means that for this class of generators the generalized
Poisson bracket is just the expectation value of the
commutator with $i\op{H}$.

Relation (\ref{gener}) is very useful for two reasons.  First,
if $\op{G}$ commutes with the hamiltonian $\op{H}$ and
$\exp(i\varepsilon \op{G}) \ket{Q_1^\prime}=\ket{Q_2^\prime}$
then $\bra{Q^\prime(t)}\op{G} \ket{Q^\prime(t)}$ is
automatically a constant of motion.  Second, this relation is an
important guidance for the choice of the trial state $\ket{Q^\prime}$.
If one wants the model to obey certain conservation laws then
the set of trial states should be invariant under unitary
transformations generated by the constants of motion. For
example total momentum conservation implies that a translated
trial state is again a valid trial state. Conservation of total
spin $\vek{\op{J}}=\vek{\op{L}}+\vek{\op{S}}$ is guaranteed when
a rotation of the trial state in coordinate and spin space results
again in a trial state.
 
Relation (\ref{gener}) also sheds some light on the quality of
the variational principle.  It says that under the premises that
$\exp(i\varepsilon \op{G})$ does not map out of the set of trial
states the expectation value ${\mathcal G}(t)$ of $\op{G}$
develops for short times like the exact solution. From
\eqref{gener} follows that along the trajectory $\ket{Q^\prime(t)}$
the time derivative of ${\mathcal G}(t)$ equals the exact one.
\newpage
\subsection{From quantum to classical mechanics}
\label{qutoclass}

This section demonstrates that the time-dependent variational
principle, discussed in general in the previous section,
represents a method to go in a smooth way from quantum physics
to classical physics by appropriately choosing the dynamical
degrees of freedom in the trial state. In contrast to
Ehrenfest's theorem this method also works for identical
particles and in finite dimensional spin-spaces.

\subsubsection{Quantum mechanics}

As a first illustration let us represent the trial state in
terms of an orthonormal basis $\ket{n}$ in many-body space. 
We may write a general state $\ket{Q^\prime}$ as  
\begin{equation} \label{trial-1}
\ket{Q^\prime}=\ket{\rho_1, \rho_2, \cdots, \pi_1, \pi_2,\cdots}=
\sum_n \ \frac{1}{\sqrt{2}}(\rho_n+i\pi_n)\ket{n} 
\equiv\sum_n \ c_n \ket{n} \ ,
\end{equation}
where the complex amplitudes are written in terms of their real and
imaginary parts $\rho_n$ and $\pi_n$. It is easy to verify that the
Lagrange function for normalizable states defined in
\eqref{Lagrange} is given by 
\begin{eqnarray} \label{Lagr-1}
{\mathcal L}^\prime= 
\sum_n \  \frac{1}{2}\,(\pi_n \dot{\rho}_n- \rho_n \dot{\pi}_n) 
+\dd{}{t}\sum_n \frac{i}{4}\,(\rho_n^2 + \pi_n^2) 
-\Hcal(\rho_1,\rho_2,\cdots,\pi_1,\pi_2,\cdots) \ . 
\end{eqnarray}
The real Hamilton function, expressed with the real and
imaginary parts of the matrix elements
$H_{nk} \equiv \bra{n}\op{H}\ket{k}$, is bilinear in 
$\rho$ and $\pi$
\begin{eqnarray} \label{Ham-1}
\Hcal(\rho_1,\rho_2,\cdots,\pi_1,\pi_2,\cdots)=\frac{1}{2}
\sum_{k,n} \,\left[(\rho_k \rho_n +\pi_k \pi_n) \, \Re{H_{kn}} +
(\pi_k \rho_n - \rho_k \pi_n)\, \Im{H_{kn}} \right] \ .
\end{eqnarray}
The Euler Lagrange equations 
\begin{eqnarray} \label{EulLagr-1}
\dd{}{t} \pp{{\mathcal L}^\prime}{\dot{\rho}_n} =
\pp{{\mathcal L}^\prime}{\rho_n} 
\ \ \ \mbox{and} \ \ \ 
\dd{}{t} \pp{{\mathcal L}^\prime}{\dot{\pi}_n} =
\pp{{\mathcal L}^\prime}{\pi_n} 
\end{eqnarray}
yield
\begin{eqnarray} \label{Bewgl-1}
\dd{}{t} \pi_n = - \pp{\Hcal}{\rho_n} 
\ \ \ \mbox{and} \ \ \ 
\dd{}{t} \rho_n =  \pp{\Hcal}{\pi_n} \ .
\end{eqnarray}
There is a very important message to be learned from this little
exercise: Eqs. (\ref{Bewgl-1}) look exactly like Hamilton's
equations of motion in which $(\rho_n,\pi_n)$ are pairs of
canonical variables and 
$\Hcal(\rho_1,\rho_2,\cdots,\pi_1 \pi_2,\cdots)$ 
is the Hamilton function, bilinear in the coordinates and
momenta of the system. But these seemingly classical equations are just 
a representation of the Schr\"{o}dinger equation as can easily
been seen by rewriting the two real Eqs. (\ref{Bewgl-1}) in
terms of the complex coefficients $c_n=\frac{1}{\sqrt{2}}(\rho_n+i\pi_n)$ 
as one complex equation, namely 
\begin{eqnarray} 
\label{Bewgl-2}
i \dd{}{t} c_n = \pp{\Hcal}{c_n^*}=
\sum_k \ H_{nk} \, c_k
\quad\mbox{or}\quad
i \ddt \ket{Q^\prime} =  \op{H} \ket{Q^\prime}\ .
\end{eqnarray}
The mere fact that the equations of motion (\ref{Bewgl-1})
appear in a classical form does not necessarily imply that the
system is classical and for example violates the uncertainty
relation or, in case of indistinguishable fermions, Fermi-Dirac
statistics. This will be discussed in detail in
\secref{sec-4-0}. 

The symplectic structure of Eqs. (\ref{Bewgl-1}) is fundamental
to all energy conserving dynamical theories, classical, quantum
or quantum field theories, see e.g. \cite{Kat65}.  Only the
physical meaning of the dynamical variables $\rho_n$ and $\pi_n$
and of the Hamilton function $\Hcal$ determine which kind of
physical system one is dealing with.

\red{
In the example (\ref{trial-1}) $\rho_n$ and $\pi_n$ are not
the $6N$ positions and momenta of the $N$ particles but infinitely many
variables which specify the many-body state $\ket{Q^\prime}$.
A truncation to a finite number results in an approximation of
the exact time evolution. The quality depends on how well the selected set
of basis states, $\{\ket{n}\}$, can represent the part of the Hilbert space
which is occupied by the physical system under consideration.
}

\subsubsection{Classical mechanics}
\label{classicalmech}

A second example shows \cite{Hel75} that one gets the
classical Hamilton equations of motion
\red{for the $6N$ positions and momenta}
{from} the time-dependent
variational principle by choosing the following dynamical
variables.  The normalized trial state $\ket{Q}$ is set up to
describe $N$ distinguishable particles which are localized in
phase space.
\begin{eqnarray} \label{prodstate}
\ket{Q} = \ket{\vek{r}_1,\vek{p}_1}\otimes
\ket{\vec{r}_2,\vek{p}_2}\otimes\cdots\otimes
\ket{\vek{r}_N,\vek{p}_N} 
\end{eqnarray}
The single-particle states $\ket{\vecrk,\vecpk}$ in the direct product are
taken to be the closest quantum analogue to classical particles,
namely gaussian wave packets of minimum uncertainty,
i.e. coherent states \cite{KlS85}:
\begin{eqnarray} \label{gaussian}
\braket{\vecx}{\vecrk,\vecpk}&=&
\left( \frac{1}{\pi a_0}\right)^{3/4}
\exp\left\{ \, -\; \frac{(\,\vecx-\vecrk)^2}{2\,a_0}
 + i\, \vecpk \vecx \right\}
\\
\braket{\veck}{\vecrk,\vecpk}&=&
\left( \frac{a_0}{\pi}\right)^{3/4}
\exp\left\{ \, -\; \frac{a_0\,(\,\veck-\vecpk)^2}{2}
 - i\, \vecrk  (\veck-\vecpk) \right\} \ .
\end{eqnarray}
The dynamical variables $\vecrk$ and $\vecpk$ are just the mean
values of the position and momentum operator, respectively 
\begin{eqnarray} 
\label{rp}
\vecrk=\bra{\vecrk,\vecpk}\op{\vek{x}}\ket{\vecrk,\vecpk} \ 
\ , \ \
\vecpk=\bra{\vecrk,\vecpk}\op{\vek{k}}\ket{\vecrk,\vecpk} \ . 
\end{eqnarray}
The width parameter $a_0$ is a real fixed number here. In later
applications it will also be taken as a complex dynamical variable.

The evaluation of the Lagrange function (\ref{LagrNorm}) is
again simple and yields  
\begin{eqnarray}
{\mathcal L}
= 
-\sum_{k=1}^N
\vecrk \dot{\!\vec{\,p}}_k 
- \Hcal \ ,
\end{eqnarray}
where the Hamilton function $\Hcal=\bra{Q}\op{H}\ket{Q}$  
\begin{eqnarray}
\label{E-1-2-3}
\Hcal = \sum_{k=1}^N
\left( \frac{\vek{p}^2_{k}}{2m_k} + \frac{3}{4m_k a} \right) +
\sum_{k<l=1}^N
\bra{\vec{r}_k,\vec{p}_k}\otimes\bra{\vec{r}_l,\vec{p}_l}\op{V}(1,2)
\ket{\vec{r}_k,\vec{p}_k}\otimes\ket{\vec{r}_l,\vec{p}_l}
\end{eqnarray}
is the expectation value of the hamiltonian
\begin{eqnarray} 
\op{H} &=& \sum_{l=1}^N \frac{\op{\vek{k}}^2\!(l)}{2m_l} + 
\sum_{k<l=1}^N \op{V}(k,l)  \ . 
\end{eqnarray}
The Euler Lagrange equations
\begin{eqnarray}
\frac{d}{dt}\pp{{\mathcal L}}{\dot{\vek{r}}_k}=
\pp{{\mathcal L}}{\vec{r}_k} \quad &\rightarrow& \quad 
0=-\dot{\vek{p}}_k-\pp{\Hcal}{\vek{r}_k} \\
\frac{d}{dt}\pp{{\mathcal L}}{\dot{\!\vec{\,p}}_k}=
\pp{{\mathcal L}}{\vek{p}_k} \quad &\rightarrow& \quad 
-\dot{\vek{r}}_k=-\pp{\Hcal}{\vek{p}_k}\ ,
\end{eqnarray}
result in 
\begin{eqnarray} \label{Hamiltoneof}
\ddt \vek{p}_k=-\pp{\Hcal}{\vek{r}_k} \quad \mbox{and} \quad
\ddt \vek{r}_k=\pp{\Hcal}{\vek{p}_k} \ . 
\end{eqnarray}
In a situation where classical mechanics holds, i.e. the wave
packets are narrow enough so that one can approximate the
expectation value
\begin{eqnarray}
\bra{Q}\op{H}\ket{Q} 
=\Hcal(\vecr_1,\vecr_2,\cdots,\vecp_1,\vecp_2,\cdots)
\approx
H(\vecr_1,\vecr_2,\cdots,\vecp_1,\vecp_2,\cdots)
\end{eqnarray}
by replacing the momentum and position operators in the
hamiltonian by their respective mean values of the wave packets,
Eqs. (\ref{Hamiltoneof}) become the classical
Hamilton's equations of motion. 

Up to this point the time-dependent variational principle leads
to the same results as Ehrenfest's theorem which usually
establishes the connection between quantum and classical
systems. As will be demonstrated in section \ref{Antisymm} the
prescription to replace the operators in the
Heisenberg equation by their mean values does not work for
indistinguishable particles. 
But the  time-dependent variational principle for an
antisymmetrized trial state $\ket{Q}$ provides the molecular
dynamics equations for identical fermions. For bosons one would
of course use a symmetrized state.

The canonical pair $(\vecr_k,\vecp_k)$ of real position and momentum
can be combined to a complex variable 
$z_k=\sqrt{\frac{1}{2a_0}}\,\vecr_k+i \sqrt{\frac{a_0}{2}}\,\vecp_k$.
The Hamilton equations (\ref{Hamiltoneof}) written in terms of 
$z_k$ and $z^*_k$ take the form
\begin{eqnarray}
\label{E-1-2-2}
i \ddt z_k = \pp{}{z^*_k}
\Hcal(z^*_1,z^*_2,\cdots,z_1,z_2,\cdots)\ ,
\end{eqnarray}
which formally looks like the Schr\"odinger equation
(\ref{Bewgl-2}). 

This and the previous example show that one cannot decide from
the form of the equations of motion alone whether the system, they are
describing, is classical or quantal. Furthermore, the
time-dependent variational principle can provide both, classical
and quantal many-body equations of motion, depending how the
trial state $\ket{Q}$ is chosen. In the following third example
an intermediate situation is sketched, where some quantum
effects are included.

\subsubsection{Semi-classical, semi-quantal}
\label{sec-1-2-3}

A third example which has part of quantum effects included is a
trial state in which the width parameter $a$ is a dynamical
variable and complex, $a=a_R + i\,a_I$ \cite{TsF91}.
Take the $N$-body trial state to be a product state
\begin{eqnarray} \label{product-state}
\ket{Q}=\ket{q_1}\otimes\ket{q_2}\otimes\cdots\otimes\ket{q_N}
\end{eqnarray}
of single-particle states, which are gaussians
\begin{eqnarray}
\label{E-1-2-1}
\braket{\vecx}{q_l}
=
\braket{\vecx}{\vek{r}_l,\vek{p}_l,a_l}
=
\left(2\pi\, \frac{a_l^*\,a_l}{a_l^* + a_l}\right)^{-3/4}
\exp\left\{ \, -\; \frac{(\,\vecx-\vek{r}_l)^2}{2\,a_l}
 + i\, \vek{p}_l \, \vecx + i\, \phi_l\right\}
\end{eqnarray}
characterized by their mean positions $\vek{r}_l$, mean momenta
$\vek{p}_l$, widths $a_l$ and phases $\phi_l$. 

For a single-particle hamiltonian which contains in addition to the
kinetic energy a harmonic oscillator potential
\begin{eqnarray}
\OpHHO
=
\sum_{l=1}^N \,\OphHO(l)
=
\sum_{l=1}^N \left( \frac{\op{\vek{k}}^2\!(l)}{2m_l} + 
\frac{1}{2} m_l\, \omega^2\, \op{\vek{x}}^2\!(l) \right) 
\ .
\end{eqnarray}
the equations of motion are
\begin{eqnarray}
\label{HOEq-1}
\ddt \vek{r}_l &=& \frac{\vek{p}_l}{m_l}
\, , \quad 
\ddt \vek{p}_l = - m_l\, \omega^2\, \vek{r}_l\\
\label{HOEq-2}
\ddt a_l       &=& \frac{i}{m_l}\,- i m_l\, \omega^2\, a_l^2 
\, , \quad 
\ddt \phi_l = 
-\frac{\vek{p}_l^2}{2\,m_l}
-\frac{3\,a_{R\,l}}{2\,m_l\,|a_l|^2}
-\frac{m_l}{2}\,\omega^2\,\vek{r}_l^2
\ .
\end{eqnarray}
These equations, although looking classically for
$\vek{r}_l$ and $\vek{p}_l$, represent the exact solution of the
Schr\"odinger equation, provided the wave function is a gaussian
wave packet at time zero. The centers of the wave packets as well
as the mean momenta oscillate harmonically with the frequency
$\omega$. Due to the time dependence of the widths the packets
also breath but with twice the frequency. The solution is
fully quantum mechanical, although described by only a few
(classically looking) parameters. Free motion without a
potential ($\omega=0$) is of course also exact.
\red{
For general potentials the trial state (\ref{E-1-2-1})
may serve as an approximation if locally,
in the region where $\braket{\vecx}{q_l}$ does not vanish,
the potential is well represented by a harmonic oscillator.
}

In general one can say, that a trial state provides the exact
solution of the Schr\"odinger equation if the action of the
Hamilton operator $\op{H}$ on the trial state can be expressed
in terms of parameters and first derivatives with respect to the
parameters like in the following example
\begin{eqnarray}
\op{\vek{k}} \ket{q}
=
\left(
i\,\pp{}{\vek{r}}
- i\,\vek{p}\,\pp{}{\phi}
\right)\ket{q}
\end{eqnarray}
and
\begin{eqnarray}
\op{\vek{k}}^2 \ket{q}
=
\left(
2\,i\,\pp{}{a_I}
+
2\,i\,\vek{p}\cdot\pp{}{\vek{r}}
-
i\,
\left[
\vek{p}^2
+
3\,\frac{a_R}{a_R^2+a_I^2}
\right]
\,\pp{}{\phi}
\right)\ket{q}
\ .
\end{eqnarray}
For the harmonic oscillator the Schr\"odinger equation takes the
form (index $l$ omitted)
\begin{eqnarray} \label{E-1-2-4}
i\,\ddt \ket{q}
&=&
i\,\sum_{\nu} \dot{q}_{\nu} \pp{}{{q}_{\nu}} \ket{q}
= \OphHO \ket{q}
\\
&=&
\Bigg\{
i\,\left(\frac{\vek{p}}{m}
\right)\pp{}{\vek{r}}
-
i\,\left(m\, \omega^2\, \vek{r}
\right)\pp{}{\vek{p}}
\nonumber \\
&& \quad
+
i\,\left(2\,m\,\omega^2\, a_R\,a_I
\right)\pp{}{a_R} 
+
i\,\left(\frac{1}{m}\,- m\, \omega^2\, (a_R^2-a_I^2) 
\right)\pp{}{a_I} 
\nonumber \\
&& \quad
-
i\,\left(
\frac{\vek{p}^2}{2\,m}
+
\frac{3\,a_R}{2\,m\,|a|^2}
-
\frac{m}{2}\,\omega^2\,\vek{r}^2
\right)\pp{}{\phi}
\Bigg\}
\ket{q}
\nonumber \ .
\end{eqnarray}
{From} \eqref{E-1-2-4} the equations of motion \fmref{HOEq-1} and
\fmref{HOEq-2} follow at once.
Because gaussian wave packets are an exact solution of the
Schr\"odinger equation for these one-body hamiltonians, also
the respective product states are an exact solution for the
corresponding many-body problem. Moreover, since
antisymmetrization and symmetrization commute with the exact time
evolution, the equations of motion \fmref{HOEq-1} remain the
same for antisymmetric product states (identical fermions)
and symmetric product states (identical bosons).

The time-dependent width parameter $a$, which describes the
variances in coordinate and momentum space, provides the first
nonclassical degree of freedom in the parameter manifold. It
completes the classical equations \fmref{HOEq-1} to the full
quantum solution for spherical harmonic oscillator potentials as
well as for free motion. In section \ref{sec-4-0} it is shown
that also in thermodynamic considerations the inclusion of
finite widths leads to quantum statistics.

\newpage
\subsection{Further remarks}
\label{sec-1-3}

In the previous \secref{qutoclass} several examples are
presented to get acquainted with the numerous aspects of the
\tdvp\ and its function as a bridge between quantum and
classical physics. We add now further remarks which are again of
general type and useful to understand the different models
referred to later.

\subsubsection{Selfconsistency and nonlinearity}

The equations of motion \fmref{E-1-1-2} (or \fmref{eom}) determine
the time evolution of the parameters $q_\nu(t)$ in parameter
space. But, even if these parameters have an intuitive physical
meaning, they first of all determine the trial state
$\ket{Q^\prime(t)}$ from which all physical observables have to be
calculated in a quantum fashion. Therefore, we derive here the
corresponding equation of motion for $\ket{Q^\prime(t)}$ in
Hilbert space.

The Hamilton operator $\op{H}_0$, which evolves the trial state
$\ket{Q^\prime(t)}$ in time according to the equations of motion
\fmref{E-1-1-2}, has to fulfill
\begin{eqnarray} 
\op{H}_0\ket{Q^\prime}=i \ket{\dot{Q}^\prime}
\equiv i\,\sum_{\nu}\dot{q}_{\nu}\pp{}{{q}_{\nu}} \ket{Q^\prime}
\ .
\end{eqnarray}
It is given by
\begin{eqnarray} 
\label{E-1-3-4}
\op{H}_0=
i\Big(\ket{\dot{Q}^\prime}\bra{Q^\prime}-
\ket{Q^\prime}\bra{\dot{Q}^\prime}\Big)+
\frac{i}{2}\ket{Q^\prime}
\Big(\braket{\dot{Q}^\prime}{Q^\prime}-
\braket{Q^\prime}{\dot{Q}^\prime}\Big)\bra{Q^\prime}
\ ,
\end{eqnarray}
where $\ket{\dot{Q}^\prime}$ stands for
\begin{eqnarray} 
\label{E-1-3-3}
\ket{\dot{Q}^\prime}
=
\ddt \ket{Q^\prime}
=
-i\, \sum_{\mu,\nu} 
\pp{}{q_\mu}\ket{Q^\prime}\,
{\mathcal C}_{\mu\nu}^{\prime\;-1}\, 
\pp{{\mathcal H}}{q_\nu^*}
\end{eqnarray}
and  analogue for $\bra{\dot{Q}^\prime}$.
Thus $\op{H}_0$ itself depends on $Q^{\prime\,*}$ and
$Q^\prime$. Using the fact that $\ket{Q^\prime}$ is always
normalized it is easy to show that $\op{H}_0$ defined in
\fmref{E-1-3-4} is the generator of the approximate time evolution 
\begin{eqnarray} 
\label{E-1-3-1}
i\,\ddt\ket{Q^\prime}
=
\op{H}_0\left(Q^{\prime\,*},Q^\prime\right)\ket{Q^\prime}
\ .
\end{eqnarray}
The equation of motion \fmref{E-1-3-1} for the trial state in
Hilbert space is the counterpart for the equation of motion
\fmref{E-1-1-2} in parameter space
\begin{equation}
i\,
\dot{q}_{\mu}\,
=\sum_{\nu}\;
{\mathcal C}_{\mu\nu}^{\prime-1}\left(Q^{\prime\,*},Q^\prime\right)\,
\pp{}{q_{\nu}^*}\;
{\mathcal H}\left(Q^{\prime\,*},Q^\prime\right)
\ .
\end{equation}
\eqref{E-1-3-1} is selfconsistent in the sense that the
hamiltonian depends on the actual state
$\ket{Q^\prime(t)}$. For example, if $\ket{Q^\prime(t)}$ is a
single Slater determinant, but otherwise unrestricted,
$\op{H}_0\left(Q^{\prime\,*},Q^\prime\right)$ is the Hartree-Fock
hamiltonian \cite{KeK76}. Furthermore, the approximate
\eqref{E-1-3-1} is usually not a linear equation like the exact
\se\ and therefore violates the superposition principle of
quantum mechanics.

Both, selfconsistency and nonlinearity are also common to
classical molecular dynamics where the classical Hamilton
function $H(r_1(t),\cdots,p_1(t),\cdots)$ depends on the actual
physical state. The nonlinearity of
$H(r_1(t),\cdots,p_1(t),\cdots)$ plays an important r\^{o}le for
statistical properties like equilibration and chaotic behaviour,
see \secref{sec-4-0}. In quantum mechanics these properties can
only be investigated by breaking the superposition principle
by coarse graining or phase averaging.

\subsubsection{Quantum branching}
\label{quantbranch}

The time-dependent variational principle creates an approximate
quantum dynamics in a manifold of trial states.  Due to the
severe restrictions it is very likely that areas of the Hilbert
space are not reachable by the approximate equations of motion
which would be visited by the solution of the Schr\"odinger
equation. The idea for an improved model is to allow branching
from one trajectory
$\ket{Q_i^\prime(t)}$ to another one $\ket{Q_j^\prime(t)}$,
i.e. to jump in the parameter manifold with a certain
probability. 

If the local deviation between the exact solution and the
solution of the \tdvp, see \secref{S-1-1-2}, is not so large,
the operator
\begin{eqnarray} 
\nonumber
\Delta \op{H}&=&(\op{H}-i \sum_\nu\, \dot{q}_\nu \pp{}{q_\nu})
\\ 
\label{deltaH-1}
&\equiv&\op{H}-\op{H}_0
\ ,
\end{eqnarray}
which appears in \eqref{error} for the error, may be regarded as a
perturbation, which causes transitions between the trial states
$\ket{Q_i^\prime(t)}$, while each $\ket{Q_i^\prime(t)}$ follows the
approximate time evolution as given in
 Eqs.~\fmref{E-1-3-4} and \fmref{E-1-3-3} with its
selfconsistent hamiltonian
$\op{H}_0\left(Q_i^{\prime\,*},Q_i^\prime\right)$.
As in usual pertubation theory the transition 
$\ket{Q_i^\prime(t)}\rightarrow\ket{Q_j^\prime(t)}$
should be related to the amplitude
$\bra{Q_j^\prime(t)}\Delta\op{H}\ket{Q_i^\prime(t)}$
\cite{Tul90,THA97,LCA98}.
One would, however, like to work with probabilities instead of
amplitudes. This should be possible if the physical system is in
an energy regime with high level density where statistical
arguments may be used. 

In literature one often employs phenomenological arguments (see
\secref{models}) for random jumps in parameter space. The above
sketched procedure which refers to $\Delta \op{H}$ has several
advantages. The most important is selfconsistency, the
transition depends on the actual situation and the choice of the
trial states $\ket{Q_i^\prime(t)}$. If for example
$\ket{Q_i^\prime(t)}$ is already a solution of the \se ,
$\Delta \op{H}\ket{Q_i^\prime(t)}=0$, no branching occures.

Another important aspect is the breaking of symmetries \red{\cite{CoC98}}.
Whereas the exact quantum state can retain dynamically conserved
symmetries by linear superpositions of states, the nonlinearity
and selfconsistency of the approximate hamiltonian
\fmref{E-1-3-4} usually does not permit that. 
Take for example the mirror symmetry of a molecule or a nucleus
which breaks into pieces. Each measured final channel which
corresponds to one $\ket{Q_i^\prime}$ does not possess mirror symmetry
but the ensemble of all measured channels does. If the initial
trial state $\ket{Q^\prime(t=0)}$ has this symmetry and $\op{H}$ does
not break it explicitly the symmetry will be kept in
$\ket{Q^\prime(t)}$. Here, quantum branching into a nonsymmetric trial
state $\ket{Q_j^\prime}$ and its mirrored counterpart $\ket{-Q_j^\prime}$ with
equal probabilities would resolve the problem of spurious cross
channel coupling observed in selfconsistent
approximations \cite{GLD80}.
The transitions may also allow quantum fluctuations to
configurations which are close in Hilbert space but cannot
be reached by the 
approximate time evolution, for example tunneling.

\begin{figure}[hhht]
\begin{center}
{\epsfig{file=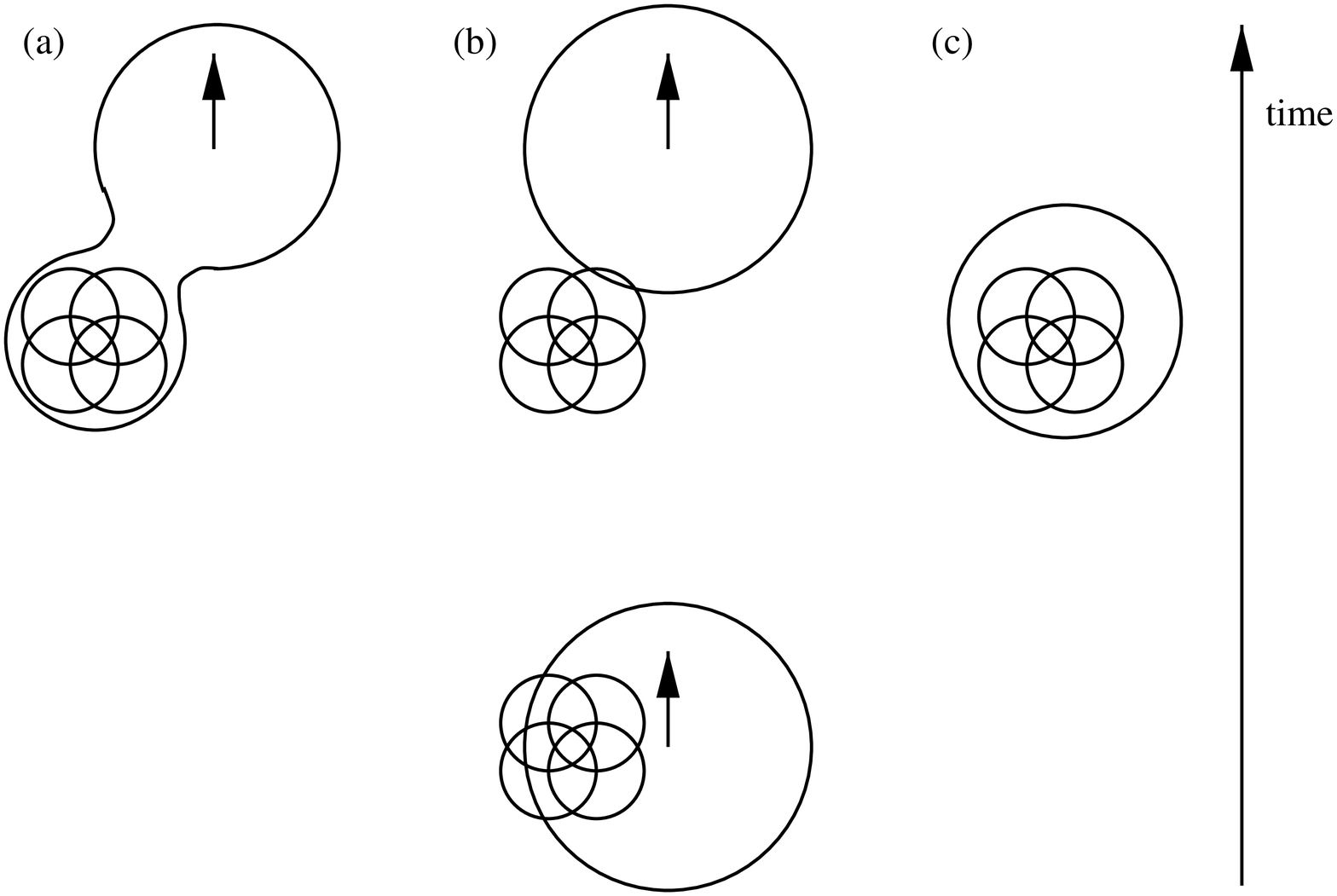,height=90mm}}
\end{center}
\mycaption{
Sketch of a four particle cluster and a
passing wave packet. Circles indicate half-density contours. The
\red{molecular dynamics} equations of motion for the centroids
and the widths of the packets can only
influence the wave packet as a whole,
either it passes by (b) or it is captured (c). The exact
solution allows to split the packet into a captured and proceeding
part (a) (see text).
}{F-1-4-2}
\end{figure} 

Another noteworthy example of quantum branching is displayed
in \figref{F-1-4-2} which in a pictorial way shows four
gaussian wave packets $\ket{q_l(t)}$ (see \eqref{E-1-2-1})
forming a bound cluster and a wave packet (large circle) of an
unbound particle which passes by.  
The time-dependent width of this wave packet has spread in
coordinate space because it was moving freely for some
time. Such situations occurr frequently in simulations. In
the exact case a piece of this wave packet would stick to the
four particle cluster indicating that,  with a
certain probability, a five particle cluster has been
created. The remaining part of the wave packet would move
on. Because the gaussian single-particle state
does not permit this freedom the wave packet can
only be bound to the cluster in total or escape. 
However, in order to become bound its width has to shrink to the
size of the cluster otherwise the matrix element of the
interaction with the other wave packets is too weak. It is
obvious that beyond a certain size of the passing packet the
overlap with the cluster will be too small to create a
sufficiently large generalized force on its width parameter to
let it shrink. Therefore, small capture probabilities cannot be
described. 

A branching procedure 
would in the above given example divide the time-evolution into
a branch of a five particle cluster and one of a four particle
cluster and a freely moving wave packet with the respective
probabilities.
\red{In general one can say whenever the interaction varies
strongly across the region of the wave packet additional
shape degrees of freedom should be considered. Otherwise
local actions are washed out and only mean field properties
survive.}

But quantum branching faces also problems. In
going from amplitudes 
to probabilities one should not violate conservation laws. For
instance, one should not alter the energy distribution of the
system. The mean value and the variance of the energy of the
ensemble of trial states, which are populated during the
branching should be the same as in the initial state.
The same is true for other conserved quantities like total
momentum or angular momentum.
\red{It is quite conceivable that variational principles with the
appropriate constraints could be helpful \cite{BaV88,LCA98}.}

\subsubsection{Approximation or new dynamical model?}

At this point some philosophical thoughts on the meaning of the
results from the time-dependent variational principle seem to be
appropriate. One could of course always claim that any
restriction of the degrees of freedom in the trial state
$\ket{Q^\prime}$ leads to approximations of the Schr\"odinger
equation, the more restraint the trial state the worse the
approximation. But the simple examples in \secref{qutoclass} and
those for antisymmetric states in the following
\secref{Antisymm} demand a more differentiated point of view. 

Let us exemplify this by restricting step by step the degrees of
freedom in the general quantum state for a number of atoms.
First we make a restricted ansatz for the trial state
which contains only the coordinates and spins of nucleons
and electrons and we disregard completely the internal quark
and gluon degrees of 
freedom for the nucleons and the quanta of the electromagnetic
field. For low excitation energies this is certainly a good
approximation.

The next step is to neglect the internal degrees of freedom of
the nuclei assuming that they are all in their ground states and
to retain only their center of mass coordinates and the electron
variables.
But this is still too complex to solve the Schr\"odinger
equation. Therefore we describe
the c.m. motion of the heavy nuclei by gaussian
packets, which leads to the classical equations of motion
\fmref{Hamiltoneof} for the nuclei with a Hamilton function
${\mathcal H}$ which 
couples to the quantal electrons. The state for the electrons may
then be constrained to a single Slater determinant with no
further restrictions on the single-particle states. Now we have
arrived at the time-dependent Hartree Fock model which is a
mean-field theory. The selfconsistent hamiltonian $\op{H}_0$ is
a one-body operator.
 
But we can of course go further and disregard the internal
degrees of freedom of the atoms or the molecules altogether and
treat only their center of mass motion by means of fixed-width
gaussians. At this point we obtain classical molecular dynamics
with a Hamilton function ${\mathcal H}$ which contains two-body
potentials between the molecules.

If the molecules are big, may be junks of crystals, may be even
stars hold together by gravity, their center of mass coordinates
may be the adequate parameters for very narrow gaussians. The
time-dependent variational principle will now provide Newton's
equations for macroscopic objects.

We don't believe that Newtons equations should be viewed as a
bad approximation (bad because of the many constraints) of
Quantum-Chromo-Dynamics where we started.\footnote{The
time-dependent variational principle can also be 
formulated in quantum field theory and it reduces to the
nonrelativistic one discussed here.} 
They should also not be regarded as an approximation to
time-dependent Hartree Fock, which formally one could plead for,
because further constraints on the Slater determinant led
there. 

The question is rather, which are the relevant degrees of
freedom for a physical system. When those are identified the
time-dependent variational principle provides a dynamical model
which is selfconsistent and has all properties of a Lagrange
formalism, like existence of canonical variables, Noether's
theorem, etc.

Since the restriction of the trial state $\ket{Q^\prime}$ can be made
in finer or coarser steps it is a matter of debate at which
point one wants to speak about a new dynamical model and when of an
approximation.
\newpage
\section{Antisymmetrization}
\label{Antisymm}

The previous chapter demonstrated how to derive classical
equations of motion from quantum mechanics by means of the
time-dependent variational principle and localized
single-particle states. It was also shown that there is no clear
boundary between quantum and classical mechanics. This will
be even more the case in this chapter where we are dealing with
indistinguishable fermions.

The natural generalization of the product state
\fmref{prodstate} (which led to classical molecular dynamics) to
a trial state for identical fermions is the antisymmetrized
product
\begin{eqnarray} 
\label{E-2-0-1}
\ket{Q} &=& \op{A}\ \ket{q_{1}}\otimes\ket{q_{2}}
  \otimes\cdots\otimes\ket{q_{N}} \\
        &=& \frac{1}{N!}\sum_{all\ P} \mbox{sgn}(P)
             \ \ket{q_{P(1)}}\otimes\ket{q_{P(2)}}
              \otimes\cdots\otimes\ket{q_{P(N)}}
\nonumber \ .
\end{eqnarray}
The sum runs over all permutations $P$ and $\mbox{sgn}(P)$ is
the sign of the permutation. The localized single-particle
states $\ket{q_{l}}$ contain for spin-$\half$ particles also
two-component spinors $\ket{\chi_l}$
\begin{eqnarray}
\label{E-2-0-2}
\braket{\vek{x}}{q_l}
&=&
\exp\left\{ \, -\; \frac{(\,\vek{x}-\vek{b}_l)^2}{2\,a_l}
\right\}
\otimes\!\ket{\chi_l}
\ ,\qquad
\vek{b}_l
=
\vek{r}_l
+ i a_l \vek{p}_l
\ .
\end{eqnarray}
The spinor may for example be represented by the two complex
parameters $\chi_l^\uparrow=\braket{\uparrow}{\chi_l}$ and
$\chi_l^\downarrow=\braket{\downarrow}{\chi_l}$, where $\ket{\uparrow}$ and
$\ket{\downarrow}$ are the eigenstates of the spin operator
$\op{s}_z$. 

The general Lagrangian $\mathcal L$, as given in
\eqref{LagrNorm}, calculated with the antisymmetrized trial
state \fmref{E-2-0-1} provides through the Euler Lagrange
equations \fmref{EulLagr} or \fmref{eom} the desired molecular
dynamics equations for fermions
\begin{eqnarray}
\label{E-2-0-3}
i \sum_{\nu} {\mathcal C}_{\mu\nu}\left(Q^*,Q \right)\, \dot{q}_{\nu} 
=
\pp{{\mathcal H}\left(Q^*,Q \right)}{q^*_{\mu}}
\ .
\end{eqnarray}
The hermitian matrix ${\mathcal C}_{\mu\nu}\left(Q^*,Q \right)$,
\eqref{cmatrix}, is not as simple as in the classical case,
where ${\mathcal C}_{\mu\nu}=\delta_{\mu\nu}$, compare
\eqref{E-1-2-2}, but depends on all parameters contained in
$Q$. 

Since the trial state is antisymmetric all consequences of the
Pauli principle are incorporated in the equations of motion
\fmref{E-2-0-3}. ${\mathcal C}_{\mu\nu}\left(Q^*,Q \right)$,
which is the second logarithmic derivative of the determinant
$\braket{Q}{Q}=\mbox{det}\{\braket{q_k}{q_l}\}$, plays the
r\^{o}le of a metric and will lead for example to large
velocities $\dot{q}_{\nu}$ when the fermions get close to Pauli
forbidden regions in phase space. In the energy 
${\mathcal H}\left(Q^*,Q \right)=\bra{Q}\op{H}\ket{Q}/\braket{Q}{Q}$ the
Pauli principle causes exchange terms which induce for example
additional momentum dependences. It should also be noted that the
determinantal structure together with the nonorthogonality of
the single-particle states result in an expression for the
kinetic energy
\begin{equation}
\frac{\bra{Q}\op{T}\ket{Q}}{\braket{Q}{Q}} 
= \sum^N_{k,l=1} \bra{q_k}\op{t}\ket{q_l}\ {\mathcal O}_{lk}
\ ,
\label{E-2-0-4}
\end{equation}
which contains a two-fold summation over all states (particles)
compared to the single summation in classical molecular dynamics,
\eqref{E-1-2-3}. ${\mathcal O}_{lk}$ is the inverse of the
overlap matrix
\begin{equation}
\left({\mathcal O}^{-1}\right)_{kl} := \braket{q_k}{q_l}\ ;\quad
k,l=1,\cdots,N \ .
\label{Ovlap}
\end{equation}
The expectation value of a two-body operator like the
interaction is a four-fold sum
\begin{eqnarray} \label{E-2-0-5}
\frac{\bra{Q}\op{V}\ket{Q}}{\braket{Q}{Q}} 
= 
\frac{1}{2} \sum^N_{k,l,m,n=1} 
\brakl \op{v} \ketmn \OO
\ .
\end{eqnarray}
The generalized forces $-\pp{}{q_{\mu}}{\mathcal
H}\left(Q^*,Q\right)$  are therefore rather involved expressions
which reflect the fact that the antisymmetrization $\op{A}$ is a
$N$-body operation which correlates all $N$ particles
simultaneously.

The apparent large numerical effort led different authors to
propose approximations which will be discussed in
\secref{models}. 

First we explain in some length the two-body case because many
of the new features concerning antisymmetrization and
indistinguishability can be understood in this simple
study. Furthermore, attempts to approximate the effects of the
Pauli principle in molecular dynamics by so called Pauli
potentials are based on considerations in two-body space. 

The consequences of antisymmetrization in many-body space, which
are discussed in \secref{sec-2-2}, are even more
intricate. Depending on how much the single-particle states
overlap the antisymmetrization can change the properties of the
trial state completely. There is for example Fermi motion even
if the gaussian single-particle states have no mean momentum. 

\newpage
\subsection{Effects of antisymmetrization in two-body space}
\label{sec-2-1}

\subsubsection{Static considerations}
\label{sec-2-1-1}

For two distinguishable particles the most simple trial state
which leads in the proper limit to classical mechanics is the
product of two gaussians $\ket{q_1}\otimes\ket{q_2}$. The
corresponding state for indistinguishable fermions is the
projection \fmref{E-2-0-1} onto the antisymmetric component
\begin{equation}
\label{E-2-1-1}
\ket{Q} = \op{A}\ket{q_1}\otimes\ket{q_2}
= \frac{1}{2!}\left\{
\ket{q_1}\otimes\ket{q_2}-\ket{q_2}\otimes\ket{q_1}\right\}
\ .
\end{equation}
In order to simplify the following discussion we do not use this
Slater determinant but a trial state which separates center of
mass and relative motion
\begin{equation}
\label{3.2.1}
\ket{Q} = \ket{q_{cm}}\otimes\ket{q}
\ .
\end{equation}
The center-of-mass wave function
is parameterized by $q_{cm}=\{A,\vek{B}\}$ as
\begin{equation} \label{cm-state}
\langle \vek{X}| q_{cm}(t) \rangle 
=
\exp \left\{-\frac{(\vek{X}-\vek{B}(t))^2}{2A(t)} \right\} 
\; ,\quad
\vek{B}(t) = \vek{R}(t)+i A(t) \vek{P}(t)
\ ,
\end{equation}
where $\Xvec= \half\left(\vek{x}_1+\vek{x}_2 \right)$ 
is the center of mass coordinate and the parameter set
$q_{cm}=\{A,\vek{B}\}$ contains the mean c.m. position $\vek{R}$ the
mean c.m. momentum $\vek{P}$ combined in the complex parameter
$\vek{B}$ and the complex width $A$. 

The wave packet $\ket{q}$ for the relative motion contains also
the spins, thus $q=\{\,a, \vek{b}, \chi_1, \chi_2\}.$ Its
parametrized form in coordinate space reads 
\begin{eqnarray} 
\label{relwp}
\langle \, \xvec \ket{q(t)} =
& \left[ \
\displaystyle{
 \exp \left\{-\frac{(\xvec-\bvec(t))^2}{2a(t)} \right\}  }
\ket{\chi_1(t)}\otimes\ket{\chi_2(t)}
\right.&  \nonumber \\
  & - \left.
\displaystyle{
\exp \left\{ -\frac{(\xvec+\bvec(t))^2}{2a(t)} \right\}
\ket{\chi_2(t)}\otimes\ket{\chi_1(t)} } \
\right]& 
\ ,
\end{eqnarray}
where small letters denote the relative coordinates and
parameters. In relative coordinate space the exchange 
of the two particles, $1\leftrightarrow 2$, is equivalent to the
parity operation $\vek{x}=\vek{x}_1-\vek{x}_2\,\leftrightarrow
-\vek{x}=\vek{x}_2-\vek{x}_1$.

The two-body wave packet \fmref{3.2.1} is in general not a
single Slater determinant because the center of mass motion for
a Slater determinant does not separate if the single-particle
packets have different width parameters. In this section we choose
the widths $A(t)$ for the center of mass motion and $a(t)$ for
the relative packet to be independent in order to decouple the
center-of-mass degree of freedom. It should also be noted that
trial state \fmref{relwp} is a linear combination of total spin
$S=0$ and $S=1$.

The relation between $\bvec$ and the more intuitive
quantities ``distance" $\vek{r}$ and ``relative momentum"
$\vek{p}$ is (time-dependence not indicated any longer)
\begin{eqnarray}
\vek{b}_R = \vek{r} - a_I \vek{p}
\quad&\mbox{and}&\quad
\vek{b}_I = a_R\vek{p} \; ,
\\
\vek{r}
=
\frac{a^* \vek{b} +a \vek{b}^*}{a + a^*}
\quad&\mbox{and}&\quad
\vek{p}
=
\frac{1}{i}
\frac{\vek{b} - \vek{b}^*}{a + a^*}
\ .
\end{eqnarray}
Here and throughout this section the
real and imaginary part of complex numbers are denoted by
the indices $R$ and $I$, respectively. 

It is important to realize that $\vek{r}$ and
$\vek{p}$ have their classical meaning only if the two
particles are far away in phase space. We illustrate this effect
of antisymmetrization in \figref{F-2-1-3} where on the
l.h.s. the relative wave function 
\begin{eqnarray}
\braket{\vek{x}}{q_d}
&=&
\left( 2\,\pi \frac{a^*\,a}{a^* + a}\right)^{-3/4}
\exp\left\{ \, -\; \frac{(\,\vek{x}-\vek{r})^2}{2\,a}
 + i\, \vek{p} \, \vek{x} \right\}
\end{eqnarray}
is plotted along the $\vek{r}$-direction. This trial state
describes two distinguishable particles at relative distance
$\vek{r}=\bra{q_d}\op{\vek{x}}\ket{q_d}$. For a single gaussian
the expectation value of
$\op{\vek{x}}=\op{\vek{x}}(1)-\op{\vek{x}}(2)$ is always equal
to the parameter $\vek{r}$ independent of $\vek{p}$ and $a_R$.

\begin{figure}[ht]
\begin{center}
\epsfig{file=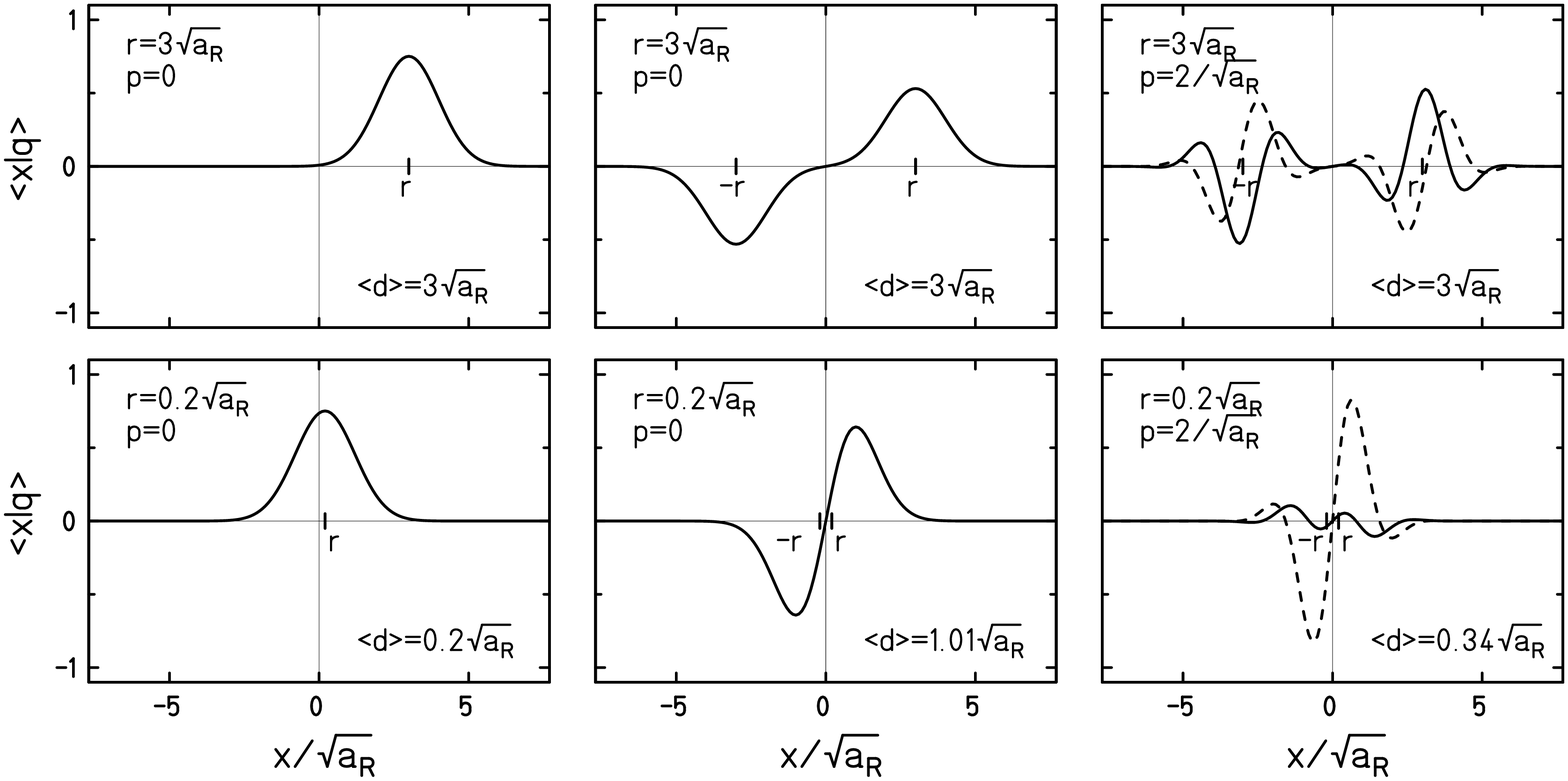,width=\textwidth}
\mycaption{Real (solid line) and imaginary part (dashed line) of
the relative wave function for distinguishable particles (left
column) and fermions (middle and right column) at different
values of the parameters $\vek{r}=(r,0,0)$, $\vek{p}=(p,0,0)$
and $a=a_R$. $\erw{d}$ measures the
distance between particles, see \eqref{E-2-1-14}.}{F-2-1-3}
\end{center}
\end{figure}

This is a typical example where Ehrenfest's theorem can be used
to derive the classical equations of motion from quantum
mechanics. The method is to replace the
expectation values of the Heisenberg equation
\begin{eqnarray}
\label{E-2-1-10}
\bra{q_d}\ddt\,\op{\vek{x}}\ket{q_d}
=
\bra{q_d}i\left[\frac{1}{2\,\mu}\op{\vek{k}}^2,\op{\vek{x}}\right]\ket{q_d}
=
\frac{1}{\mu}\bra{q_d}\op{\vek{k}}\ket{q_d}
\end{eqnarray}
for narrow wave packets by the mean values 
\begin{eqnarray}
\label{E-2-1-11}
\ddt\,\vek{r}
=
\frac{\vek{p}}{\mu}
\ .
\end{eqnarray}
Analogue for the relative momentum operator
$\op{\vek{k}}=\half\!\left(\op{\vek{k}}(1)-\op{\vek{k}}(2)\right)$
\begin{eqnarray}
\label{E-2-1-12}
\bra{q_d}\ddt\,\op{\vek{k}}\ket{q_d}
=
\bra{q_d}i\left[V(\op{\vek{x}}),\op{\vek{k}}\right]\ket{q_d}
=
-\bra{q_d}\pp{}{\vek{x}}V(\op{\vek{x}})\ket{q_d}
\end{eqnarray}
or
\begin{eqnarray}
\label{E-2-1-13}
\ddt\,\vek{p}
\approx
\pp{}{\vek{r}}
V(\vek{r})
\ .
\end{eqnarray}
While \fmref{E-2-1-11} is exact for $\ket{q_d}$,
\fmref{E-2-1-13} is only an approximate expression which can be
improved. 

However, the antisymmetrized packet
\begin{eqnarray}
\braket{\vek{x}}{q}
&=&
\left[
\exp\left\{ \, -\; \frac{(\,\vek{x}-\vek{r})^2}{2\,a}
 + i\, \vek{p} \, \vek{x} \right\}
-
\exp\left\{ \, -\; \frac{(\,\vek{x}+\vek{r})^2}{2\,a}
- i\, \vek{p} \, \vek{x} \right\}
\right]
\end{eqnarray}
displayed for two values of $\vek{r}$ and $\vek{p}$ in the
middle and right column leads to different results. First, for
indistinguishable particles the operator
$\op{\vek{x}}=\op{\vek{x}}(1)-\op{\vek{x}}(2)$ is not an
observable anymore, because it is not symmetric with respect to
particle exchange. Due to its negative parity the expectation
value $\bra{q}\op{\vek{x}}\ket{q}/\braket{q}{q}$ is always zero
even for particles, which are far away from each other. The same
holds true for the relative momentum,
$\bra{q}\op{\vek{k}}\ket{q}/\braket{q}{q}=0$. Therefore
Ehrenfest's theorem is changed into a triviality because
Eqs.~\fmref{E-2-1-10} and \fmref{E-2-1-12} become
meaningless ($0=0$).
The conclusion is that Ehrenfest's theorem cannot be used to
derive classical equations of motion for indistinguishable
particles.

An observable which coincides with $|\vek{r}|$ at large
distances is the root of the square radius minus the intrinsic
width of the packet
\begin{eqnarray}
\label{E-2-1-14}
\erw{d}=\left(\frac{\bra{q}\op{\vek{x}}^2\ket{q}}{\braket{q}{q}}
         -\frac{3\, |a|^2}{2\,a_R}\right)^{\half}\ .
\end{eqnarray}
In the middle of \figref{F-2-1-3}, which is for $\vek{p}=0$, it
becomes evident that the observable distance $\erw{d}$
between two fermions with equal spin is larger than $|\vek{r}|$
and deviates most when $\vek{r}^2 \ll a_R$. The r.h.s. with
$|\vek{p}|=2/\sqrt{a_R}$ demonstrates that the distance
$\erw{d}$ depends also on the relative momentum. For large
relative momenta, $\vek{p}^2\,a_R \gg 1$, the Pauli principle is
again less effective and $\erw{d}\approx|\vek{r}|$. The
explicit expression for the mean square radius in terms of
parameters is 
\begin{eqnarray}
\label{E-2-1-15}
\frac{\bra{q}\op{\vek{x}}^2\ket{q}}{\braket{q}{q}}
=
\vek{r}^2 + \frac{3\, |a|^2}{2\,a_R} 
+\frac{|a|^2}{a_R} f_{ex}(\xi,S_{12})
\ ,
\end{eqnarray}
where
\begin{eqnarray}
\label{E-2-1-16}
f_{ex}(\xi,S_{12})
&=&
\frac{\xi\,e^{-\xi}\,S_{12}}{1 - e^{-\xi}\,S_{12}}
\quad \mbox{and}\quad
\xi
=
\frac{|\vek{b}|^2}{a_R}
=
\frac{(\vek{r}-a_I\,\vek{p})^2}{a_R}
+ a_R\,\vek{p}^2
\ .
\end{eqnarray}
The first two terms in \fmref{E-2-1-15} are the same as for
distinguishable particles. The remaining part is the exchange
term which contains the quantity $\xi$ and the spin overlap
$S_{12}=|\braket{\chi_1}{\chi_2}|^2$. From \eqref{E-2-1-16} it
is evident that $\xi$ may be regarded as a dimensionless measure
for the distance in phase space where $\sqrt{a_R}$ is the length
scale. For large $\xi\geap 4$ the exchange vanishes and the
Pauli principle is not active. For small $\xi$ the distance
between the fermions is always larger than $|\vek{r}|$ because
the exchange term is positive definite. This effect is sometimes
called ``Pauli repulsion", but this is not a force between the
particles. 

The analogue expression for the distance in momentum space is
\begin{eqnarray}
\erw{\kappa}=\left(
\frac{\bra{q}\op{\vek{k}}^2\ket{q}}{\braket{q}{q}}
- \frac{3}{2\,a_R} \right)^{\half} 
\end{eqnarray}
with the expectation value for the square momentum
\begin{eqnarray}
\label{E-2-1-17}
\frac{\bra{q}\op{\vek{k}}^2\ket{q}}{\braket{q}{q}}
=
\vek{p}^2 + \frac{3}{2\,a_R} 
+\frac{1}{a_R}\,f_{ex}(\xi,S_{12})
\ .
\end{eqnarray}
Again a term proportional to $f_{ex}(\xi,S_{12})$ appears so
that there is complete analogy between coordinate and momentum
space. The exchange term is again positive and
causes a ``Pauli repulsion" in relative momentum.

It is interesting to note that for a mininum uncertainty state
$\ket{q}$ with $a_I=0$ and equal spins ($S_{12}=1$) the sum of
the observable distances in coordinate and momentum space fulfill
\begin{equation}
\erw{d}^2/a_R+\erw{\kappa}^2a_R
=
\xi\,
\frac{1+e^{-\xi}}{1-e^{-\xi}} 
\ge 2
\ \ \mbox{for}\ \ a_I=0 \ .
\end{equation}
Different from the measure $\xi$ the observable distance in
phase space can never get smaller than 2, for details
see subsection \secref{sec-2-1-4}. 

Expression \fmref{E-2-1-17} tells also that the kinetic energy
of relative motion
\begin{eqnarray}
\label{E-2-1-19}
{\mathcal T}
=
\frac{1}{2\mu}\frac{\bra{q}\op{\vek{k}}^2\ket{q}}{\braket{q}{q}} 
=
\frac{\vek{p}^2}{2\mu}
+
\frac{3}{4\mu a_R}
+
\frac{1}{2\mu a_R}\, f_{ex}(\xi,S_{12})
\end{eqnarray}
consists of the classical part ${\vek{p}^2}/(2\mu)$, the
contribution from the uncertainty $3/(4\mu a_R)$ and an
additional potential $f_{ex}(\xi,S_{12})/(2\mu a_R)$ which
depends on $\vek{r},\vek{p},a$ and the spin overlap $S_{12}$.

An analogue expression can be obtained for a spin-independent
potential which is smooth
\begin{eqnarray} \label{E-2-1-20}
{\mathcal V}
=
\frac{\bra{q} V(\op{\vek{x}}) \ket{q}}{\braket{q}{q}} 
\approx
V(\vek{r}) 
+ \frac{1}{6}\, \Delta V(\vek{r})\, \frac{3\, |a|^2}{2\,a_R}
+ 
\frac{1}{6}\, 
\frac{|a|^2}{a_R}\; \Delta V(\vek{r}=0)\;
f_{ex}(\xi,S_{12})
\ .
\end{eqnarray}
This expression is exact if $V(\op{\vek{x}})=V_0 + V_2\,\op{\vek{x}}^2$,
and a good approximation if the Taylor expansion up to second
order of $V(\op{\vek{x}})$ around $\vek{x}=\vek{r}$ is adequate
within the range of the wave packet. Please note that due to
rotational symmetry $V(\op{\vek{x}})$ depends only on $\op{\vek{x}}^2$. 

Combining
Eqs.~\fmref{E-2-1-19} and \fmref{E-2-1-20} yields a Hamilton
function which splits into three parts
\begin{eqnarray} \label{E-2-1-21}
{\mathcal H}&=&{\mathcal T} + {\mathcal V}
={\mathcal H}_{classical} + {\mathcal V}_{uncertainty} + {\mathcal V}_{Pauli}
\\
&=&\left[ \frac{\vek{p}^2}{2\mu}+V(\vek{r}) \right]
+\left[ \frac{3}{4\mu a_R}+\frac{|a|^2}{4\,a_R}\, \Delta V(\vek{r}) \right]
+\left[ \frac{1}{2\mu a_R}+\frac{|a|^2}{6\,a_R}\; \Delta V(\vek{r}=0)\,
\right]\,f_{ex}(\xi,S_{12})
\nonumber
\ .
\end{eqnarray}
\eqref{E-2-1-21} is a basis for the concept used by several
authors in nuclear
\cite{WHK77,WYC78,BoG88,DDR87,DoR87,PRS91,MOH92,NCM95} and
atomic physics \cite{KTR94A,KTR94B,EbM97} to incorporate the
uncertainty principle and the Pauli principle in order to extend
classical molecular dynamics.  A two-body potential
$\sum_{i<j}\,{\mathcal
V}_{uncertainty}(\vek{r}_{ij},\vek{p}_{ij},a)$ is added to
simulate the effects of the Heisenberg uncertainty principle and
$\sum_{i<j}\,{\mathcal V}_{Pauli}(\vek{r}_{ij},\vek{p}_{ij},a,S_{ij})$ 
is added to
imitate the effects of the Pauli principle. The explicit form
needs not to be the one given in \fmref{E-2-1-21}, but it is
usually adapted to the specific use. The method is quite
successful in calculating energies \cite{DDR87,DoR87}, but we
want to advise caution in using the Hamilton function
(\ref{E-2-1-21}) naively in equations of motion like
\begin{eqnarray}
\label{E-2-1-22}
\dot{\vek{r}}_i
= 
\pp{}{\vek{p}_i}{\mathcal H}
\quad\mbox{and}\quad
\dot{\vek{p}}_i
= 
- \pp{}{\vek{r}_i}{\mathcal H}
\ .
\end{eqnarray}
The reason is that $\vek{r}_i$ and $\vek{p}_i$ are no longer
canonical variables. As discussed earlier, although they still
define the trial state $\ket{Q}$ uniquely, they loose their
intuitive meaning when the particles are indistinguishable. The operator 
$\op{\vek{x}}(i)$, ``position of particle $i$", is meaningless
and $\bra{q}\op{\vek{x}}(i)\ket{q}/\braket{q}{q}$ is not
$\vek{r}_i\,$! The analogue holds for the momentum.
Instead of postulating \eqref{E-2-1-22} we have to go back to the
Lagrangian \fmref{LagrNorm} and derive the equations of motion
\fmref{E-2-0-3}. It is evident that the matrix ${\mathcal C}$
will also be changed by the antisymmetrization.

\subsubsection{Center of mass motion}

Before discussing the relative motion in the antisymmetric case
we consider first the center of mass motion.
The Lagrange function \fmref{LagrNorm} for the center of mass
wave packet (\ref{cm-state}) is given by 
\begin{eqnarray}
{\mathcal L}_{cm}
&=&
{\mathcal L}_{0\,cm}-{\mathcal T}_{cm}
\end{eqnarray}
with 
\begin{eqnarray}
\label{L0CM}
{\mathcal L}_{0\,cm}
&=&
\frac{i}{2}
\frac{\braket{q_{cm}}{\dot{q}_{cm}} -
\braket{\dot{q}_{cm}}{q_{cm}}}{\braket{{q}_{cm}}{q_{cm}}}
\\
&=&
\frac{i}{2}
\frac{(\vek{B}^*-\vek{B})(\dot{\vek{B}}^*+\dot{\vek{B}})}{A^*+A}
-\frac{i}{4}\left[
\left(\frac{\vek{B}^*-\vek{B}}{A^*+A}\right)^2
-\frac{3}{A^*+A}\right](\dot{A}^*-\dot{A})
+\frac{3\,i}{4}
\left(\frac{\dot{A}}{A}-\frac{\dot{A}^*}{A^*}\right)
\nonumber\\
&=&
\vek{P}\cdot\dot{\vek{R}}+\frac{3}{4}\frac{\dot{A}_I}{A_R}
+\mbox{total time derivative} 
\nonumber
\end{eqnarray}
and the kinetic energy
\begin{eqnarray}
\label{Tcm}
{\mathcal T}_{cm}
&=&
\frac{\bra{q_{cm}}  \frac{1}{2M}\op{\vek{K}}^2  \ket{q_{cm}} }
{\braket{q_{cm}}{q_{cm}}}
=
\frac{1}{2M} \frac{(\vek{B}^*-\vek{B})^2}{A^*+A} + \frac{3}{2\,M\,(A^*+A)}
\ =\ \frac{\vek{P}^2}{2M} + \frac{3}{4\,M\,A_R}
\ .
\end{eqnarray}
The Euler Lagrange equations yield
\begin{equation}
\dot{\vek{B}}=0,\ \  \dot{A}=\frac{i}{M} \;   ,
\end{equation}
or if one transforms $\vek{B}$ and $\vek{B}^*$ into $\vek{R}$
and $\vek{P}$
\begin{equation}
\dot{\vek{P}}=0,\ \  \dot{\vek{R}}=\frac{\vek{P}}{M},\ \  
\dot{A}=\frac{i}{M} \;   .
\end{equation}
In the center of mass wave function $\vek{R}(t)$ and $\vek{P}(t)$
have always the classical meaning of the mean center of mass
position and momentum, respectively. 
Nevertheless, when the width parameter $A$ is included as a
dynamical variable, the wave packet
$\ket{q_{cm}(t)}=\ket{A(t),\vek{B}(t)}$ is the exact solution of
the \se .

\subsubsection{Relative motion}
\label{sec-2-1-3}

The Lagrange function for the relative motion
\begin{eqnarray}
\label{Lrel}
{\mathcal L}(q,q^*,\dot{q},\dot{q}^*)
&=&
\frac{i}{2} \left(
\frac{\braket{q}{\dot{q}}-\braket{\dot{q}}{q}}{\braket{q}{q}}\right)
-\frac{\bra{q} \kvecsim^2/{2\mu} \ket{q}}{\braket{q}{q}}
-\frac{\bra{q} V(\op{\vek{x}}) \ket{q}}{\braket{q}{q}}
\\
&\equiv& 
{\mathcal L}_0-{\mathcal T}-{\mathcal V} \nonumber
\end{eqnarray}
features the antisymmetrization not only in the kinetic and
potential energy as seen in Eqs.~\fmref{E-2-1-19} and
\fmref{E-2-1-20} but also in the metric part ${\mathcal L}_0$.

For the sake of simplicity we treat in this subsection only
equal, time-independent spins. In the general
case the potential
$V(\op{\vek{x}},\op{\vek{s}}(1),\op{\vek{s}}(2))$ may
of course depend on spin and the spin degrees of freedom.
The parameters
$\braket{\uparrow,\downarrow}{\chi_{1,2}(t)}$ would then also
appear in 
${\mathcal L}_0,{\mathcal T}$ and ${\mathcal V}$ and one would get 
equations of motion for them.  For time-independent equal spins 
${\mathcal L}_0$ is given by
\begin{eqnarray}
\label{E-2-1-23}
{\mathcal L}_0
&=&
\frac{i}{2}
\frac{(\vek{b}^*-\vek{b})(\dot{\vek{b}}^*+\dot{\vek{b}})}{a^*+a}
-
\frac{i}{4}
\left[
\left(
\frac{\vek{b}^*-\vek{b}}{a^*+a}
\right)^2
-
\frac{3}{a^*+a}
\right]
(\dot{a}^*-\dot{a})
\\
&&
+
\frac{i}{2}
\left[
\frac{\dot{\vek{b}}\cdot\vek{b}^*-\vek{b}\cdot\dot{\vek{b}}^*}
{\vek{b}^*\cdot\vek{b}} 
+
\frac{\dot{a}^*-\dot{a}}{a^*+a}
\right]\,f_{ex}(\xi,S_{12})
+\mbox{total time derivative}
\nonumber\\
&=&
\vek{p}\cdot\dot{\vek{r}}
+
\frac{3}{4}\frac{\dot{a}_I}{a_R}
+
\left[
\frac{\vek{p}\cdot\dot{\vek{r}}
-
\vek{r}\cdot\dot{\vek{p}}}{\xi}
+
\frac{\vek{p}^2\,(\dot{a}_R\,a_I-\dot{a}_I\,a_R-\vek{r}
\cdot\vek{p}\,\dot{a}_R)}{a_R}
+
\frac{\dot{a}_I}{a_R}
\right]\,f_{ex}(\xi,S_{12})
\nonumber \\
&&
+\ \mbox{total time derivative}
\ .
\nonumber
\end{eqnarray}
As expected, ${\mathcal L}_0$ contains an exchange term, besides
the terms for distinguishable particles, compare with 
${\mathcal L}_{0\,cm}$ in \eqref{L0CM}. Although 
${\mathcal L}_0$ looks very complicated the Euler Lagrange equations
\fmref{E-2-0-3}
\begin{eqnarray}
\label{E-2-1-25}
i \, {\mathcal C}\;
\left(
\begin{array}{c} 
   \dot{a} \\ \dot{\vek{b}}
\end{array}
\right)
=
\left(
\begin{array}{c} 
   \pp{\mathcal T}{a^*} \\ \pp{\mathcal T}{\vek{b}^*}
\end{array}
\right)
+
\left(
\begin{array}{c} 
   \pp{\mathcal V}{a^*} \\ \pp{\mathcal V}{\vek{b}^*}
\end{array}
\right)
\ .
\end{eqnarray}
can be partially simplified because the free motion is, like for
the c.m. state or the case of distinguishable particles, again
the exact solution of the \se 
\begin{eqnarray}
\label{E-2-1-26}
{\mathcal C}^{-1}\;
\left(
\begin{array}{c} 
   \pp{\mathcal T}{a^*} \\ \pp{\mathcal T}{\vek{b}^*}
\end{array}
\right)
=
\left(
\begin{array}{c} 
   -\frac{1}{\mu} \\ 0
\end{array}
\right)
\end{eqnarray}
and herewith
\begin{eqnarray}
\label{E-2-1-27}
\left(
\begin{array}{c} 
   \dot{a} \\ \dot{\vek{b}}
\end{array}
\right)
=
\left(
\begin{array}{c} 
   \frac{i}{\mu} \\ 0
\end{array}
\right)
-
i \, {\mathcal C}^{-1}\;
\left(
\begin{array}{c} 
   \pp{\mathcal V}{a^*} \\ \pp{\mathcal V}{\vek{b}^*}
\end{array}
\right)
\ .
\end{eqnarray}
The complicated form of ${\mathcal C}^{-1}$ combines with the
complicated derivatives $\left(\pp{\mathcal T}{a^*},\pp{\mathcal
T}{\vek{b}^*}\right)$ of the kinetic energy such that the simple
result \fmref{E-2-1-26} is obtained. Although lengthy to
calculate it is easy to understand. The reason is that both
states, $\ket{a,\vek{b}}$ and $\ket{a,-\vek{b}}$, are exact
solutions of the \se , they only differ in the initial
conditions. The antisymmetric state
$\ket{q}=\ket{a,\vek{b}}-\ket{a,-\vek{b}}$, defined in
\fmref{relwp}, is therefore also an exact solution. The deeper
reason is that both, the time derivative and the hamiltonian,
commute with the antisymmetrization operator $\op{A}$ in
\eqref{E-2-0-1}
\begin{eqnarray}
\label{E-2-1-28}
0
=
i\,\ddt\,\op{A}\,\ket{\Psi(t)} -\op{H}\,\op{A}\,\ket{\Psi(t)}
=
\op{A}\,
\left(
i\,\ddt\,\ket{\Psi(t)} -\op{H}\,\ket{\Psi(t)}
\right)
\ .
\end{eqnarray}
Therefore, if $\ket{\Psi(t)}$ is the exact solution of the \se ,
so is $\op{A}\,\ket{\Psi(t)}$. This general statement is not
true if $\ket{\Psi(t)}$ is only an approximation.

In the special case of a harmonic interaction,
$V(\op{\vek{x}})=V_0+V_2\,\op{\vek{x}}^2$, the contribution from
the potential assumes a very simple form as well, because
\begin{eqnarray}
\label{E-2-1-29}
{\mathcal C}^{-1}\;
\left(
\begin{array}{c} 
   \pp{\erw{\op{\vek{x}}^2}}{a^*} \\ \pp{\erw{\op{\vek{x}}^2}}{\vek{b}^*}
\end{array}
\right)
=
\left(
\begin{array}{c} 
   2\,a^2 \\ 2\,a\,\vek{b}
\end{array}
\right)
\ .
\end{eqnarray}
The amazing result is that Eqs.~\fmref{E-2-1-26} and
\fmref{E-2-1-29} are the same for distinguishable particles and for
indistinguishable particles, where ${\mathcal C}^{-1}$ is a
complicated matrix depending on $a^*,a$ and
$\vek{b}^*,\vek{b}$. The exchange term in the expression for
$\erw{\op{\vek{k}}^2}$, \eqref{E-2-1-17}, and for
$\erw{\op{\vek{x}}^2}$, \eqref{E-2-1-15}, compensates for the
different ${\mathcal C}$. 
It is also interesting to see that for a harmonic interaction
the parameters $\vek{r}$ and $\vek{p}$ obey the classical
equations of motion
\begin{eqnarray}
\label{E-2-1-30}
\dot{\vek{r}} = \frac{\vek{p}}{\mu}\ ,\quad
\dot{\vek{p}} = - 2\, V_2\, \vek{r}
\ ,
\end{eqnarray}
although the trial state $\ket{q}$ is antisymmetric and
describes two identical fermions. This result is, however, only
obtained if the width $a$ is at the same time a dynamical
variable with the equation of motion 
\begin{eqnarray}
\label{E-2-1-31}
\dot{a} = \frac{i}{\mu} - i\,2\, V_2\, a^2
\ .
\end{eqnarray}
The case, where $a(t)=a_0$ is supposed to be a positive
time-independent number, leads to completely different results
and is discussed in the following subsection.

\subsubsection{Relative motion with a time-independent width parameter}
\label{sec-2-1-4}

This section investigates how the equations of motion
change if the shape of the relative wave packet, \eqref{relwp},
is restricted further by removing the width degree of freedom
$a(t)=a_R(t)+ia_I(t)$ as a dynamical variable.  For simplification
only parallel spins are considered, \ie $S_{12}=1$. By setting
$a_R(t)=a_0, a_I(t)=0$ and $\dot{a}(t)=0$ in Eqs. \fmref{E-2-1-19}
and \fmref{E-2-1-23} we obtain
\begin{eqnarray} \label{L0fixed}
{\mathcal L}_0 &=& \vek{p}\cdot\dot{\vek{r}}
       +\left(\vek{p}\cdot\dot{\vek{r}}-\vek{r}\cdot\dot{\vek{p}}\right)
               \frac{e^{-\xi}}{1-e^{-\xi}} 
\\ \label{Tfixed}
{\mathcal T} &=& \frac{\vek{p}^2}{2\mu}
+ \frac{3}{4\mu a_0} + \frac{1}{2\mu a_0} \ 
\frac{\xi\,e^{-\xi}}{1-e^{-\xi}} \ ,
\end{eqnarray}
where
\begin{equation}
\xi = \vek{r}^2/a_0 + \vek{p}^2a_0 \ \ \mbox{and}  \ \
{\mathcal L} = {\mathcal L}_0 -{\mathcal T} -{\mathcal V} \; .
\end{equation}
The potential energy ${\mathcal V}$ is not given here explicitly
and the spin dynamics is also not considered for simplicity. The
equations of motion for $\vek{r}$ and $\vek{p}$ are
\begin{eqnarray}
\label{rvec}
0 &=& \ddt \frac{\partial {\mathcal L}}{\partial \dotvecp}
- \frac{\partial {\mathcal L}}{\partial \vek{p}}
\;\;\;\mbox{or} \nonumber \\
-\dotvecr &-& \frac{2 e ^{-\xi}}{1- e^{-\xi}}
\left( \dotvecr -
\frac{\vek{r}(\vek{r}\dotvecr)/a_0
+ \vek{p}(\vek{p}\dotvecr)a_0}{1-e^{-\xi}} \right)
+ \frac{2e^{-\xi}a_0}{(1-e^{-\xi})^2}
\left(\vek{r}(\vek{p} \dotvecp) -
\vek{p} (\vek{r} \dotvecp)\right) \nonumber \\
&=& -\frac{\partial}{\partial \vek{p}} {\mathcal H} (\vek{r}, \vek{p})
\end{eqnarray}
and
\begin{eqnarray}
\label{pvec}
0 &=& \ddt \frac{\partial {\mathcal L}}{\partial \dotvecr}
- \frac{\partial {\mathcal L}}{\partial \vek{r}}
\;\;\; \mbox{or} \nonumber \\
\dotvecp &+& \frac{2 e ^{-\xi}}{1- e^{-\xi}}
\left( \dotvecp -
\frac{\vek{r}(\vek{r}\dotvecp)/a_0
+ \vek{p}(\vek{p}\dotvecp)a_0}{1-e^{-\xi}} \right)
+ \frac{2e^{-\xi}}{(1-e^{-\xi})^2 a_0}
\left(\vek{r}(\vek{p} \dotvecr) -
\vek{p} (\vek{r} \dotvecr)\right) \nonumber \\
&=& - \frac{\partial}{\partial \vek{r}} {\mathcal H} (\vek{r}, \vek{p}) \; .
\end{eqnarray}
For the reduced set of variables we are able to solve
Eqs. \fmref{rvec} and \fmref{pvec} for $\dotvecr$ and $\dotvecp$.
The result is
\begin{equation}
\label{3.2.23}
\dotvecr = \alpha_1 (\xi)
\frac{\partial {\mathcal H}}{\partial \vek{p}} +
\alpha_2 (\xi) \left\{ a_0 \left(\vek{p}
\frac{\partial {\mathcal H}}{\partial \vek{p}} +
\vek{r} \frac{\partial {\mathcal H}}{\partial \vek{r}} \right)
\vek{p} + \left(\frac{\vek{r}}{a_0}
\frac{\partial {\mathcal H}}{\partial \vek{p}} - a_0
\vek{p} \frac{\partial {\mathcal H}}{\partial \vek{r}} \right)
\vek{r} \right\}
\end{equation}
and
\begin{equation}
\label{3.2.24}
\dotvecp = - \alpha_1 (\xi)
\frac{\partial {\mathcal H}}{\partial \vek{r}} +
\alpha_2 (\xi) \left\{ \left(\frac{\vek{r}}{a_0}
\frac{\partial {\mathcal H}}{\partial \vek{p}} - a_0
\vek{p} \frac{\partial {\mathcal H}}{\partial \vek{r}} \right)
\vek{p} -    \left( \vek{r}
\frac{\partial {\mathcal H}}{\partial \vek{r}} +
\vek{p} \frac{\partial {\mathcal H}}{\partial \vek{p}} \right)
\frac{\vek{r}}{a_0} \right\} ,
\end{equation}
where $\alpha_1 (\xi)$ and $\alpha_2 (\xi)$ are functions of
$\xi = \vek{r}^2/a_0 + \vek{p}^2 a_0$  and given by
\begin{equation}
\label{3.2.25}
\alpha_1 (\xi) = \frac{1-e^{-\xi}}{1+e^{-\xi}}
\end{equation}
\begin{equation}
\label{3.2.26}
\alpha_2 (\xi) = \frac{2e^{-\xi} (1-e^{-\xi})}{(1+e^{-\xi})^2
(1-e^{-\xi}) - 2 \xi e^{-\xi} (1+e^{-\xi})} .
\end{equation}
Since $\alpha_1(\xi \gg 1)=1$ and $\alpha_2(\xi \gg 1)=0$ one
recognizes for $\xi \gg 1$ Hamilton's equations of motion.
$\xi\gg 1$ means that the two fermions are far from each other
in phase space.  In that limit the identical fermions behave
like classical distinguishable particles although their wave
function is of course still antisymmetrized. When they get close
in phase space, \ie $\xi < 1$, $\alpha_1 (\xi) \rightarrow
\xi/2$ and $\alpha_2 (\xi) \rightarrow 3/\xi^2$ which means that the
Hamilton-like parts in \fmref{3.2.23} and \fmref{3.2.24} vanish
like $\xi/2$ but the remaining parts increase like $3/\xi^2$.

In this example one sees that for $\xi \leap 2$ the equations of
motion, which result from the parameterization \fmref{relwp}
with $a(t) \equiv a_0$,
cannot be cast into Hamilton's form when $\vek{r}$ and $\vek{p}$ are
regarded as canonical variables. To prove this statement let us
suppose that a Hamilton function 
${\mathcal H}_{Pauli}(\vek{r},\vek{p})$ exists such that
Eqs. \fmref{3.2.23} and \fmref{3.2.24} can be written as 
\begin{equation}
\label{Hpauli}
\dot{r}_i= \frac{\partial {\mathcal{H}}_{Pauli} }{\partial p_i}
\;\;\; \mbox{and} \;\;\;
\dot{p}_i=-\frac{\partial {\mathcal{H}}_{Pauli} }{\partial r_i}
,\; i=1,2,3 \; ,
\end{equation}
where $i$ denotes the three spatial directions.  Let us now
disprove the existence of  a function 
${\mathcal{H}}_{Pauli}$ 
by calculating the mixed derivatives 
$\frac{\partial\dot{r}_i}{\partial r_k}$ and
$\frac{\partial\dot{p}_k}{\partial p_i}$
which should add to zero if  \eqref{Hpauli} is true. From the equations of
motion \fmref{3.2.23} and \fmref{3.2.24} it is easy to verify that
\begin{equation}
\label{3.2.28}
\frac{\partial\dot{r}_i}{\partial r_k} + 
\frac{\partial\dot{p}_k}{\partial p_i} \neq 0\;\; .
\end{equation}
This
disproves the existence of a hamiltonian
${\mathcal{H}}_{Pauli} (\vek{r},\vek{p})$ as
a function of $\vek{r}$ and $\vek{p}$ which would
describe the fermionic dynamics derived from the ansatz
\fmref{relwp} with $a(t) \equiv a_0$ for the wave function.  

It is, however, possible
to find a pair of canonical variables $(\vek{\rho},\vek{\pi})$
\begin{eqnarray} \label{canonicalvar}
\vek{\rho} = \sqrt{\frac{1+e^{-\xi}}{1-e^{-\xi}}} \; \vek{r}
\;\;\; \mbox{and} \;\;\;
\vek{\pi} = \sqrt{\frac{1+e^{-\xi}}{1-e^{-\xi}}} \; \vek{p}
\ ,
\end{eqnarray}
which are non-linear functions of the original variables
$(\vek{r},\vek{p})$ \cite{SKF83} such that ${\mathcal L}_0$ in
\eqref{L0fixed} assumes the canonical form
\begin{eqnarray}
{\mathcal L}_0 &=& \half \left(
      \vek{\pi}\cdot\dot{\vek{\rho}}
       -\vek{\rho}\cdot\dot{\vek{\pi}}\right)
    \   + \half \ddt(\vek{r}\vek{p})  
\ .
\end{eqnarray}
With these new variables the
equations of motion aquire the form \fmref{Hpauli} where
${\mathcal{H}}_{Pauli}(\vek{\rho},\vek{\pi})=
{\mathcal H}(\vek{r},\vek{p})$ is the total energy expressed in the
new canonical variables. However, 
${\mathcal{H}}_{Pauli}(\vek{\rho},\vek{\pi})$ 
cannot be expressed in a closed form because
Eqs. \fmref{canonicalvar} cannot be solved for 
$\vek{r}(\vek{\rho},\vek{\pi})$ and $\vek{p}(\vek{\rho},\vek{\pi})$.

Equations \fmref{Hpauli} represent an approach often used in
literature to incorporate the effects of the Pauli principle
by means of a two-body interaction. 
Different forms of momentum-dependent potentials have been added
to the classical hamiltonian or to the lagrangian
\cite{NeF94}. Here we see that this method 
differs in several aspects from fermionic dynamics. Firstly, as
discussed above the dynamical behaviour of $\vek{r}$ and $\vek{p}$
needs not be of Hamilton's type.  Secondly, one should not
replace $\langle V(\xvecsim) \rangle$ simply by $V(\vek{r})$
because even for narrow wave packets there is an exchange term
which is not small for $\xi < 1$, see \eqref{E-2-1-21}.

\begin{figure}
\begin{center}
\epsfig{file=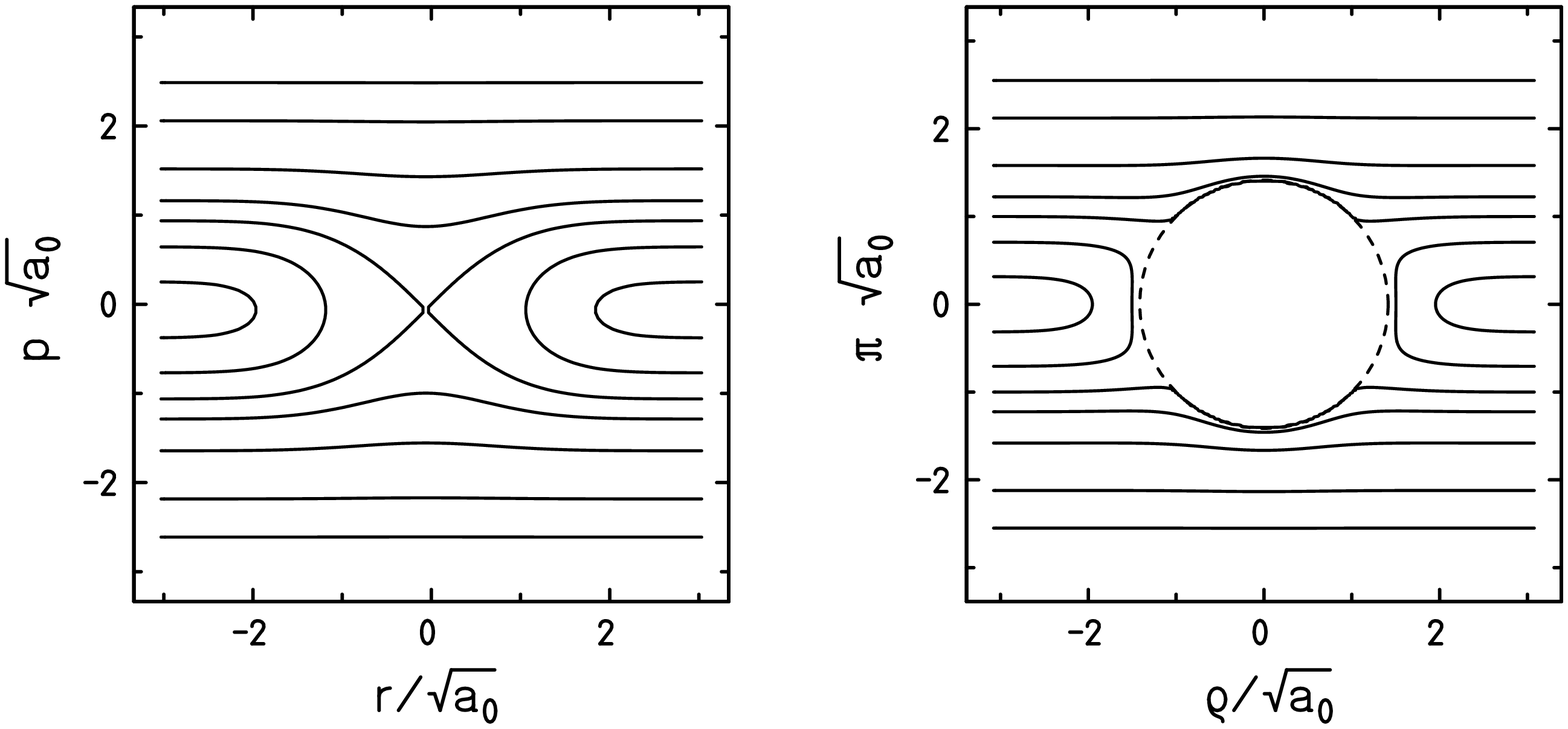,width=140mm}
\vspace*{3mm}
\mycaption{Trajectories in phase space of one-dimensional free
motion with fixed width.
L.h.s.: variables $r$ and $p$.
R.h.s.: canonical variables $\rho$ and $\pi$.}{F-2-1-1}
\end{center}
\end{figure}

The equations of motion simplify appreciably for a
time-independent width parameter $a_0$, but the price to be paid
is that free motion without interaction is not exact anymore.
For $\xi < 1$ the equations of motion differ essentially from
the expected result $\dotvecr = \vek{p} /\mu$ and $\dotvecp = 0$
(see \fmref{3.2.23} and \fmref{3.2.24}). The shape of the wave
packet is restrained too much so that the particles scatter even
if there is no interaction.

There is however the appealing feature of the canonical
variables $(\vek{\rho},\vek{\pi})$ to exhibit a geometrically
forbidden region in phase space. From \fmref{canonicalvar} it is
evident that $\vek{\rho}^2/a_0+\vek{\pi}^2 a_0 \ge 2.$
This is demonstrated in
\figref{F-2-1-1}, where in a one-dimensional example several
trajectories of the relative motion are shown for two freely
moving fermions. The l.h.s. displays the trajectories, which are
actually contours of constant kinetic energy, using $r$ and $p$
as dynamical variables, the r.h.s. shows the same trajectories
but using $\rho$ and $\pi$.  The empty area in the middle of the
right-hand figure is the Pauli forbidden region of phase space.
It corresponds to the single point at the origin ($r=0$ and
$p=0$) in the left-hand figure. The trajectories which cross
there, go around the circle in the canonical variables $\rho$ and
$\pi$.

Pauli potentials are usually chosen such that a pair
of particles acquires a high energy in the forbidden region. One should 
however be aware that the kinetic energy at the boundary is
finite, namely $5/(4\mu a_0)$ in the three-dimensional case, see
\fmref{Tfixed}. 
The deficiency, that particles scatter even if there is no
interaction, is also present in all Pauli potentials.  Inclusion
of the complex width $a(t)$ as a dynamical variable cures this
problem as has been demonstrated. 

At the end of this subsection the reader should not be left with
the impression that the Pauli principle is a two-body effect. In
fact antisymmetrization is a genuine $N$-body correlation as will be
discussed in the following.

\subsubsection*{Resum\'{e}}

\begin{enumerate}
\item The intuitive idea to include the Paul principle in a
classical description by treating the exchange terms of the
relative kinetic and potential energy of two identical fermions
as an additional ``Pauli potential" which supplements the
classical equations of motion cannot be supported. It is not
correct to regard the parameters $\vek{r}$ and $\vek{p}$ as
canonical variables if
\begin{eqnarray}
\label{E-2-1-32}
\vek{b}\cdot\vek{b}^* 
=
| \vek{r}+i\,a\,\vek{p} |^2
\quad\leap\quad
2\,|a^*+a|
\end{eqnarray}
and hence it is questionable whether the expectation value of the
hamiltonian can be used as the Hamilton 
function in Hamilton's equations of motion.
\item Heisenberg's quantum uncertainty refers to the variances
$(\erw{\op{\vek{x}}^2}-\erw{\op{\vek{x}}}^2)$ and
$(\erw{\op{\vek{k}}^2}-\erw{\op{\vek{k}}}^2)$ of the wave packet
which are given by the width parameter $a$ and are not related to
$\vek{r}^2$ or $\vek{p}^2$. It is therefore open to doubt if
inclusion of uncertainty 
into classical equations of motion can be achieved by a
potential which depends on $(\vek{r}_{ij}^2\cdot\vek{p}_{ij}^2)$.
\end{enumerate}

\newpage
\subsection{Effects of antisymmetrization in many-body space}
\label{sec-2-2}

\subsubsection{Shell structure due to antisymmetrization}
\label{sec-2-2-1}

It is not immediately obvious that an antisymmetrized product
state like \eqref{E-2-0-1} includes shell-model features,
like the nodal structure of single-particle orbits, because the
states are localized in coordinate and momentum space.  But due
to the invariance of a Slater determinant under linearly
independent transformations among the occupied single-particle
states, after antisymmetrization, any set of single-particle
states which is complete in the occupied phase space is as good
as any other. This applies also to nonorthogonal states.  To
illustrate this we take four one-dimensional real gaussians
with the same real width parameter $a_0$ and zero mean momentum
and displace them by $d=0.75\sqrt{a_0}$ (see l.h.s. of
\figref{P-3.2-1}).
\begin{figure}[bbb]
\unitlength1mm
\begin{picture}(120,50)
\put(5,-10){\epsfig{file=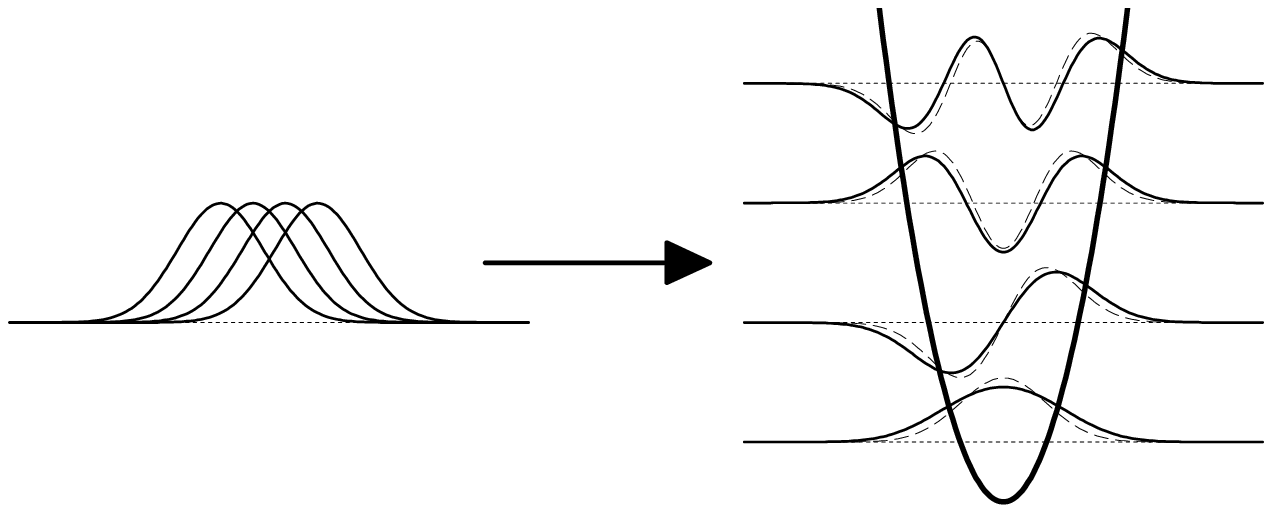,height=60mm}}
\end{picture}
\mycaption{Natural orbits.
Antisymmetrization of four displaced gaussians (l.h.s.) leads to
four occupied natural orbits (r.h.s.) which are almost
harmonic oscillator states (dashed lines).
}{P-3.2-1}
\end{figure}
The one-body density can be written in terms of orthonormal
states $\ket{\psi_m}$ as
\begin{equation}
\label{Onebds}
\op{\rho}^{(1)}
 = \sum^N_{k,l=1} \ket{q_k}\ {\mathcal O}_{kl}\  \bra{q_l} 
= \sum^N_{m=1} \ket{\psi_m} \bra{\psi_m}  \ ,
\end{equation}
where the orthonormal eigenstates of $\op{\rho}^{(1)}$,
called natural orbits, are given
by the following superposition of gaussians
\begin{equation}
\ket{\psi_m}
 = \sum^N_{k=1} \ket{q_k}\ ({\mathcal O}^{\frac{1}{2}})_{km} 
\ .
\end{equation}
${\mathcal O}_{kl}$ is the inverse of the overlap matrix,
$({\mathcal O}^{-1})_{kl}=\prodkl$.  They are displayed on the
right hand side of \figref{P-3.2-1} and compared to harmonic
oscillator eigenstates (dashed lines).  One observes that the
occupied single-particle states $\ket{\psi_m}$ consist of an s-,
p- ,d- and an f-state, all very close to harmonic oscillator
states.  The difference between both sets can be made
arbitrarily small by letting $d/\sqrt{a_0}$ approach zero.
The f-state is for example essentially the first gaussian minus
the second plus the third minus the fourth. All others  are
similar combinations. As already stressed in the two-body case
in the previous section, when the wave packets overlap,
${r}_l$ and ${p}_l$ loose their classical meaning of
position and momentum of particle $l$. In the limiting case
$d\rightarrow 0$ all ${r}_l\rightarrow 0$ and all
${p}_l\rightarrow 0$ and the harmonic oscillator shells emerge.
The distributions in coordinate and momentum space are the quantum
mechanically correct ones of four spin polarized fermions in a
harmonic oscillator.

\subsubsection{Fermi-Dirac distribution due to antisymmetrization}
\label{sec-2-2-2}

\begin{figure}[ht]
\begin{center}
\epsfig{file=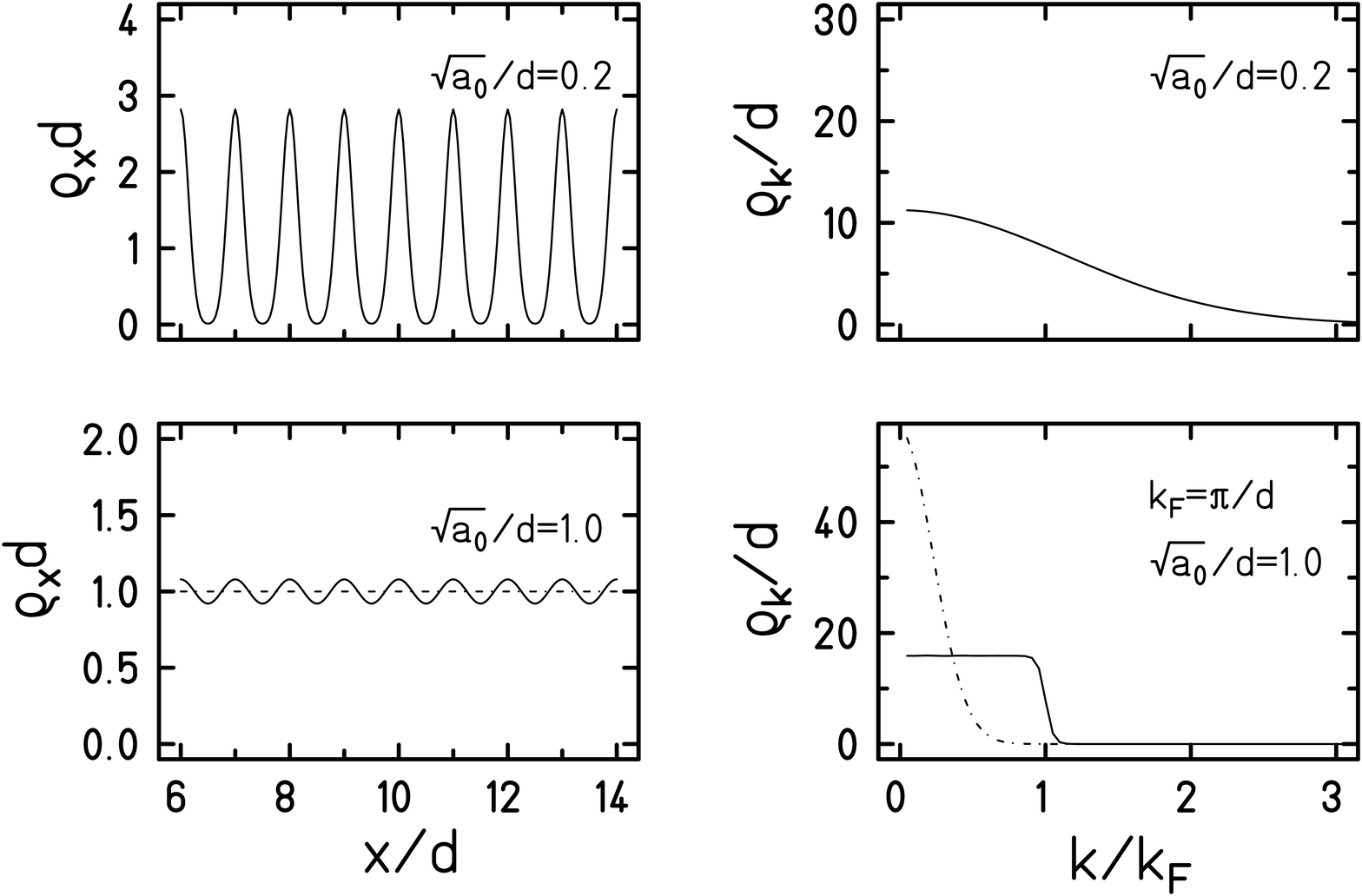,height=90mm}
\mycaption{
Densities in coordinate and momentum space.
Full line with antisymmetrization, dash-dotted line without.
Upper part: section of spatial density of hundred gaussians
(not overlapping in coordinate space) and
corresponding momentum distribution. Distributions with and without
antisymmetrization are identical.
Lower part: same as above but for overlapping gaussians.
For details see text.}{P-3.2-2}
\end{center}
\end{figure}

A second example is illustrated in \figref{P-3.2-2}, where we
consider 100 equally spaced gaussians in one dimension.  Again
all mean momenta are zero and the width $a_0$ is real.  In the
upper part of \figref{P-3.2-2} the width $\sqrt{a_0}$ is 0.2 of
the mean distance $d$ so that the wave packets are well
separated. Therefore the spatial density
$\rho_x=\bra{x}\op{\rho}^{(1)}\ket{x}$ and the momentum density
$\rho_k=\bra{k}\op{\rho}^{(1)}\ket{k}$ are not changed by
antisymmetrization.  In the lower part the width has been
increased to $\sqrt{a_0}=d$. Without antisymmetrization (dash
dotted line) the spatial density is uniform and the momentum
distribution is that of a single packet. After
antisymmetrization (full lines) one obtains the typical shell
model oscillations in coordinate space and a Fermi distribution
in momentum space.  It is amazing to see how in
\eqref{Onebds} the superposition of gaussians by means of the
inverse overlap matrices can create a fully occupied momentum
state, see for example in \figref{P-3.2-2} the lower right
momentum distribution at $k=0.8k_F$, where the individual
gaussians give practically zero probability to measure this
momentum.  We also calculated the eigenstates of the kinetic
energy in the occupied space and got perfect sinusoidal waves.

These two examples illustrate nicely that Slater determinants of
localized single-particle states with zero mean momentum can
describe the harmonic oscillator shell model or even the Fermi
motion of a gas of fermions in which plane waves are occupied up
to the Fermi momentum.
If one wants to simulate this effect by a ``Pauli potential",
disregarding the momentum distribution in each wave packet,
the resulting ground state momentum distribution is 
unsatisfactory \cite{DDR87}. The densities in coordinate and momentum
space are just not given by the
distributions of the ${r}_l$ and ${p}_l$.

At this point we would like to remind the reader of the
different meaning of the three ``momentum quantities" used in the
article:
\begin{enumerate}
\item momentum operator of particle $l$ (not observable):
$\op{\vek{k}}(l)=\EinsOp\otimes\cdots\otimes\op{\vek{k}}\otimes\cdots\EinsOp$,
\item momentum variable $\vek{k}$ as used in the momentum
representation $\braket{\vek{k}}{q}$ and
\item momentum parameter $\vek{p}_l$ which characterizes the
state $\ket{q_l}=\ket{\vek{r}_l,\vek{p}_l,a_l,\chi_l}$.
\end{enumerate}

\subsubsection*{Resum\'{e}}

\begin{enumerate}
\item The coordinate distribution $\rho_x(\vek{x})$ is given by
the observable
\begin{equation}
\rho_x(\vek{x})
=
\frac{\bra{Q}
\sum_{l=1}^N\,\delta\left(\vek{x}-\op{\vek{x}}(l)\right)\ket{Q}}
{\braket{Q}{Q}}
\end{equation}
and not by the (eventually time averaged) distribution of the
$\vek{r}_l$. Analogue the momentum distribution is is given by
\begin{equation}
\rho_k(\vek{k})
=
\frac{\bra{Q}
\sum_{l=1}^N\,\delta\left(\vek{k}-\op{\vek{k}}(l)\right)\ket{Q}}
{\braket{Q}{Q}}
\end{equation}
and not by the distribution of the $\vek{p}_l$. All $\vek{p}_l$
may be zero and there is still Fermi motion.
\item Fermi motion is not random motion in $\vek{r}_l$ and
$\vek{p}_l$.
\item $\vek{r}_l$ and $\vek{p}_l$ resume the classical meaning
only when the system is so dilute in phase space that the Pauli
principle has no consequences any more.
\end{enumerate}

\subsubsection{Dynamical considerations}
\label{sec-2-2-3}

The equations of motion for the most simple antisymmetric trial
state \fmref{E-2-0-1} are, see \eqref{E-2-0-3},
\begin{eqnarray}
i \sum_{\nu} {\mathcal C}_{\mu\nu}\left(Q^*,Q \right)\, \dot{q}_{\nu} 
=
\pp{\mathcal T}{q^*_{\mu}}
+
\pp{\mathcal V}{q^*_{\mu}}
\ .
\end{eqnarray}
Like in the two-body case, applying the inverse of 
${\mathcal C}\left(Q^*,Q \right)$ to the derivatives of the kinetic energy
yields the simple result
\begin{eqnarray}
\label{E-2-2-3}
\left(
\begin{array}{c} 
   \dot{a}_1
   \\ 
   \dot{\vek{b}}_1
   \\ 
   \dot{\chi}_1
   \\ 
   \dot{a}_2
   \\ 
   \dot{\vek{b}}_2
   \\ 
   \dot{\chi}_2
   \\ 
   \vdots
\end{array}
\right)
=
\left(
\begin{array}{c} 
   \frac{i}{m} \\ 0 \\ 0 \\   \frac{i}{m} \\ 0 \\ 0 \\ \vdots
\end{array}
\right)
-
i\;{\mathcal C}^{-1}\;
\left(
\begin{array}{c} 
   \pp{\mathcal V}{a_1^*} 
   \\
   \pp{\mathcal V}{\vek{b}_1^*}
   \\ 
   \pp{\mathcal V}{\chi_1^*}
   \\ 
   \pp{\mathcal V}{a_2^*} 
   \\ 
   \pp{\mathcal V}{\vek{b}_2^*}
   \\ 
   \pp{\mathcal V}{\chi_2^*}
   \\ 
   \vdots
\end{array}
\right)
\ .
\end{eqnarray}
Although in the case of free motion $\vek{r}_l$ and $\vek{p}_l$
follow the classical path, this is once more the exact solution
of the \se\ with $\dot{a}_l=i/m$ and $\vek{b}_l$  and the spin
parameters $\chi_l$ being constant. The reason is of course that
the many-body time evolution operator, which commutes with the
antisymmetrization operator, is a product of single-particle
time evolution operators and the free single-particle motion of
gaussians is exact.

The Pauli principle appears, like in the two-body case, in the
interaction part in a twofold way. The interaction energy
${\mathcal V}\left(Q^*,Q \right)$ has exchange terms and the
metric ${\mathcal C}^{-1}\left(Q^*,Q \right)$ couples all
generalized forces $\pp{\mathcal V}{q^*_{\mu}}$ because it is in
general not diagonal in the particle indices. Therefore the
second term on the r.h.s. of \eqref{E-2-2-3} acts like a
$N$-body force. 

Looking at the structure of \eqref{E-2-2-3} it seems more
natural to substitute the second term on the r.h.s. by a
two-body interaction instead of considering the structure of the
kinetic energy as done in \secref{sec-2-1}. In nuclear physics
Wilets and coworkers \cite{WHK77,WYC78} were the first who
proposed a space and momentum dependent two-body ``Pauli core"
which was applied to atomic physics later \cite{WiC98}.

\newpage
\section{Models in nuclear and atomic physics}
\label{models}

In nuclear physics at non-relativistic
energies the nucleons cannot be treated as classical particles
on trajectories. Their phase space density is too large to ignore
the Pauli principle. Therefore various molecular dynamics
models which incorporate the Pauli principle on different levels
of sophistication are proposed in the literature. They will be 
discussed in more detail in the following subsections.        

Molecular dynamics calculations are also widespread in atomic
physics to describe interacting atoms and molecules. But here
the degrees of freedom are usually the classical c.m. positions
and velocities of the nuclei. The quantal electrons which move
in the electric field of the atoms and provide the attraction
are often treated by mean-field methods. There are also
applications in which each electron is treated as an individual
entity localized in phase space. Like for nucleons, if the
density is high enough, they can become a degenerate Fermi gas
for which a classical molecular dynamics picture cannot be
applied. 

\subsection{Antisymmetrized wave packets in nuclear physics}
\label{secawp}

The time-dependent Hartree Fock (TDHF)  method \cite{DDK85}
was expected to be suited for the description of colliding
nuclei because stationary Hartree Fock calculations represent
successfully ground state properties for all nuclei. The TDHF
equations are obtained from the variational principle
\eqref{extremal} when the single-particle states $\ket{q}$ which
form the single Slater determinant $\ket{Q}$ are varied in an
unrestricted way. However, it turned out that, even at low beam
energies where the relative speed between the two colliding
nuclei is small compared to the Fermi velocity and hence a mean
field is always well established, the amount of fluctuations in
the collective variables, like energy loss, scattering angle or
mass distribution, was much too small compared to measured data
\cite{DDK85,BaV92,ReS92,LCA98}. Also the inclusion of collision
terms, extended TDHF (see proceedings of ``Time-dependend
Hartree-Fock and beyond" \cite{GoR82}), did not improve the
situation. The reason is that the TDHF equations contain a common
mean field which does not allow fluctuations to grow. As discussed
in \secref{quantbranch} quantum branching into
other Slater determinants with different mean fields is
missing. Even the collision integral which induces fractional
occupation of states and hence a mixing of Slater determinants
does not change that because one common mean field is again
calculated from this mixture. 

Calculations which treat the TDHF time-evolution for each member
of a thermal ensemble with its own specific mean field
\cite{KnS84,KnW88} can describe large fluctuations which develop
during the expansion. The initial ensemble of Slater
determinants, which could be envisaged as the result of quantum
branching, was only assumed but not dynamically calculated.

The shortcomings of a common mean field applies also to models
which solve the Vlasov   
equation augmented by a collision integral like VUU (Vlasov
Uehling Uhlenbeck) or BUU (Boltzmann Uehling Uhlenbeck), see for
instance 
\cite{AiB85,StG86,BeD88,WBC90,GFH99}. 
Although no explicit antisymmetrization like in TDHF is present,
Liouville's theorem and Pauli blocking in the collision term
prevent an overoccupation of the single-particle phase space. 
The inherent lack of fluctations in these one-body descriptions
led to molecular dynamics models for fermions which will be
discussed in the following subsections.

\subsubsection{Time-dependent $\alpha$-cluster model}

The first molecular dynamics model which uses antisymmetrized
many-body states of 
localized constituents in nuclear physics is an extension  
of the $\alpha$-cluster model, which successfully characterizes
$\alpha$-particle nuclei, to the time-dependend case. 
In the model nuclei are represented as Slater
determinants of wave packets for $\alpha$-particles. The width
parameter of the gaussian single-particle wave packets is
chosen to be fixed \cite{CGS82,SKF83} as well as time-dependent
\cite{DOP82}. Even superpositions of two Slater determinants of
wave packets for $\alpha$-particles with time-dependent width
are investigated \cite{BCG85}.
Since these models are applicable only for nuclei with
pronounced $\alpha$-substructure new models were developed in
the late eighties, which address one wave packet to each
single nucleon.

\subsubsection{Fermionic Molecular Dynamics -- FMD}
\label{secfmd}

The model of Fermionic Molecular Dynamics (FMD) was suggested in
1990 \cite{Fel90,FBS95,FeS97} in order to describe ground
states of atomic nuclei and heavy-ion reactions in the energy
regime below particle production. The many-body trial state of
FMD is a Slater determinant $\ket{Q(t)}$ of single-particle gaussian wave
packets $\ket{q_l(t)}$ where $q_l(t)$ denotes the set of
single-particle parameters
$q_l(t)=\{\vek{b}_l(t),a_l(t),\chi_l(t),\xi_l\}$, 
$\vek{b}_l(t)=\vek{r}_l(t)+ia_l(t)\vek{p}_l(t)$,
which contains mean position, mean momentum and complex width.
The spin degrees of freedom are represented by a spinor
$\ket{\chi_l(t)}$ (see \eqref{E-2-0-2}). The isospin part
$\ket{\xi_l}$ is taken to be time-independent and identifies
either a proton or a neutron.   

The equations of motion for all parameters are obtained from the
variational principle as described in \secref{TDVP} (see \eqref{eom}). The
hamiltonian is an effective one, because the strong short range
repulsion in the nucleon-nucleon interaction causes 
correlations which cannot be described by the trial state
$\ket{Q(t)}$ \cite{FNR98}. Also the spin correlations caused by the
strong tensor force are only poorly represented in a single
Slater determinant. 

In FMD the ground state is defined by the deepest minimum of the
energy ${\cal H}(Q^*,Q)=\bra{Q}\op{H}\ket{Q}$ . Since all
generalized forces $\partial{\cal H}/\partial q^*_l=0$
vanish in the minimum this state is stationary, all
$\dot{q}_l=0$.  There is no ``Fermi motion" in the parameters
$\vek{r}_l,\ \vek{p}_l, \cdots.$
Ground state properties like binding energies and rms
radii can be reproduced equally well with a variety of effective
nucleon-nucleon interactions which differ mainly in their momentum
dependence. But the intrinsic structure of the nuclei depends on 
the interaction. Superpositions of single-particle states as well as of
Slater determinants can be used in order to obtain a more
refined description of nuclear structure, \eg for halo nuclei
\cite{NFR99}.

FMD is able to model a variety of heavy-ion reactions ranging from fusion
to dissipative reactions and multifragmentation
\cite{FeS97}. Different from TDHF fluctuations occur in these
reactions but the results show also that
initial correlations given by the intrinsic structure of
the ground states play a major r\^{o}le in the simulation of
fragmentation reactions \cite{NFR99}. 

The time-dependent width parameters is an important
non-classical degree of freedom 
\cite{FBS95}, especially to allow for evaporation of nucleons, a
process that otherwise is strongly hindered, because each
escaping wave packet takes away at least its zero-point
energy. Inside a nucleus this zero-point energy is typically
10~MeV but evaporated nucleons have on the mean only 2~MeV
kinetic energy. Therefore the packet has to spread during the
evaporation process.

Themodynamic equilibrium properties can be determined in FMD by
means of time averaging, see \secref{sec-4-3} and Ref. \cite{ScF97}.

As already explained in \secref{quantbranch} the restricted
parameterization leads to barriers which for the exact solution
would not exist. Especially the splitting of wave packets is an
important source for quantum branching. The lack of this
dynamical freedom is a serious 
hindrance of forming clusters \cite{KiD96} which is also
observed in Antisymmetrized Molecular Dynamics and FMD
investigations of spinodal decomposition 
of nuclear matter and quenches possibly important
reaction mechanisms in multifragmentation reactions \cite{CoC98}. 
Further development in this direction is needed.

\subsubsection{Antisymmetrized Molecular Dynamics -- AMD}
\label{secamd}

Antisymmetrized Molecular Dynamics (AMD) is similar to FMD with
respect to the choice of the trial state but includes random
branching between trial states.
For details the reader is referred to
Refs.~\cite{OHM92A,OHM92B,OHM93,OnH96A,OnH96B}. AMD describes
the nuclear many-body system by a Slater determinant
$\ket{Q(t)}$ of gaussian wave packets chracterized by the
parameter set $q_l(t)=\{\vec{Z}_l(t),\chi_l,\xi_l\}$,  
where in AMD notation the complex parameters
$\vec{Z}_l=\frac{1}{\sqrt{2a_0}}\vec{b}_l=
\frac{1}{\sqrt{2a_0}}\vec{r}_l+i\sqrt{\frac{a_0}{2}}\vec{p}_l$
denote the time-dependent centroids of the wave 
packets and $\chi_l,\xi_l$ the time-independent
spin-isospin components which can be either proton or neutron,
spin up or down.
The width parameter $\nu=\frac{1}{2a_0}$ is real and time-independent and
the same for all wave packets. 

The time evolution of the $\vec{Z}_l(t)$
is determined by the time-dependent variational principle
\eqref{extremal} which leads to the equations of motion given in \eqref{eom}.
The Hamilton function ${\cal H}$ used in AMD is the
expectation value of a hamiltonian $\op{H}$
plus an additional term which removes the spurious zero-point
energy of the center of mass wave packets for the different
clusters. This c.m. energy, which is of the order of 10~MeV for
all clusters, is an artifact of all product states.

The smooth variation of the $\vec{Z}_l(t)$ due to the
generalized forces $\partial{\cal H}/\partial\vec{Z}_l^*$ is
supplemented by different stochastic forces. These can be
regarded as a phenomenological ansatz for quantum branching
between different trial states $\ket{Q_j(t)}$ as discussed in
\secref{quantbranch}. One branching procedure takes care of
deviations caused by the short range repulsion between nucleons.
For that a collision term is introduced which randomly changes
the relative canonical momenta of a pair of wave
packets. In order not to enter Pauli-forbidden regions in phase
space approximate canonical variables are used which for two
particles reduce
to those discussed in \secref{sec-2-1-4} , compare
\figref{F-2-1-1}.
 
Another branching simulates the spreading and splitting of wave
packets, which is an essential process for an adequate
description of evaporation and absorption, but cannot be
accomplished by the trial state   
\cite{OhR95,OnH96A,OhR97A,OhR97B}.

AMD is able to reproduce the essential properties of nuclear
ground states (minima in ${\cal H}(Q^*,Q)$) and, when extended to
trial states which use superpositions of single-particle states
or Slater determinants, even the structure of halo nuclei,
see \eg \cite{KHO95}.

Multifragmentation reactions are investigated for beam energies
around the Fermi energy. Before comparing with experimental data
{from} heavy-ion collisions the result of a simulated collision is
fed into a statistical decay program to account for long time
processes \cite{Ono98,Pue77}.

Since the numerical effort of Antisymmetrized Molecular Dynamics
as well as of FMD grows with $N^4$, for the 
calculation of systems with more than $N=80$ nucleons
approximations are needed. Recently the AMD group developed a ``triple
loop approximation", which converts the fourfold sum of
the potential energy (see \eqref{E-2-0-5}) into a threefold one,
so that systems like 
Au~$+$~Au are feasible now \cite{Ono98}.

\newpage
\subsection{Product states of wave packets -- QMD}
\label{secqmd}

Models which parameterize the many-fermion trial state by a
simple product of gaussian wave packets are called Quantum
Molecular Dynamics in nuclear physics. First QMD versions were
invented in the eighties \cite{AiS86,ARP87,Aic91,KOM92}. They all
employ a product state  
\begin{equation} 
\label{E-3-1-1}
\ket{Q(t)} =
\ket{q_{1}(t)}\otimes\ket{q_{2}(t)}\otimes\cdots\otimes\ket{q_{N}(t)}
\end{equation}
of single-particle states
$\ket{q_l(t)}=\ket{\vek{r}_l(t),\vek{p}_l(t)}$ defined in \eqref{gaussian},
where only the mean positions $\vek{r}_l(t)$ and the mean
momenta $\vek{p}_l(t)$ are time-dependent. The width is fixed
and the same for all wave packets. 

The resulting equations of motion are the classical ones, given
in \eqref{Hamiltoneof}. Also 
the interpretation of $\vek{r}_l(t)$ and $\vek{p}_l(t)$ is
purely classical and the particles are considered
distinguishable. This simplifies the collision term, which acts as
a random force, and at higher energies the description of transitions from
nucleons into resonances.
 
All QMD versions use a collision term with Pauli blocking in
addition to the classical dynamics. 
Some versions consider spin and isospin,
others use nucleons with an average electric charge.  Several
QMD versions try to incorporate the Pauli principle by means of
a Pauli potential that prevents nucleons of the same kind to
come too close in phase space, see \secref{sec-2-1}. Due to
these simplifications QMD has the  advantage that the 
numerical effort grows only with $N^2$ and thus allows
the simulation of large systems. In addition all QMD versions
use a statistical decay program for the long time dynamics.

\subsubsection{Versions}

QMD models are widely used in nuclear physics and exist in
many versions. We would like to mention some of them and
apologize for not mentioning all others. A more elaborate overview on
QMD models is provided in Ref. \cite{HPA98}.

\begin{enumerate}
\item Using experience with VUU/BUU models \cite{AiB85} one of
the first QMD versions \cite{AiS86,Aic91} was suggested. It
exploited the trial state \fmref{E-3-1-1} only insofar that the
interaction gets an effective range due to the folding with the
wave packets. In any other respect the model propagates point
particles on classical trajectories. The zero-point energy
originating from localization is omitted. As initial states
random distributions of mean coordinates and momenta are taken
according to the experimental ground-state density profile and
binding energy. This distribution is however not the ground
state of the model hamiltonian but an unstable excited state. The model
ground state is highly over-bound. This QMD version
does not distinguish between protons and neutrons, all nucleons
carry an average charge.
\item The IQMD version treats the isospin explicitly but is the
same in all other respects. The version was designed for the
analysis of collective flow and pions \cite{HZN89}.
\item Another branch of the QMD evolution uses a Pauli potential
and takes the trial state \fmref{E-3-1-1} more seriously
\cite{PKS92,KSG95}. Since nucleons are kept apart in phase space 
by the Pauli potential \cite{PRS91} the minimum of the 
hamiltonian determines the nuclear
ground state. This is not a mere theoretical beauty, but
very important if one wants to investigate the survival of initial
ground-state correlations in the final products. An ensemble of
random initial states is not able to answer such questions.
\item In the same spirit the Japanese QMD
\cite{MOH92,NCM95,CIF96,MNI96} is 
constructed. Supplementary the width parameter which is
time-independent in all other versions is chosen to be
time-dependent here (EQMD). Analogous to AMD the c.m. zero-point
energies are subtracted from the hamiltonian, which is difficult
because of the changing width.
Some observations in EQMD are similar to those in
FMD. Especially evaporation processes are described much better
than with fixed width. Also for fusion reactions the
time-dependent width plays a major r\^{o}le because the fusion cross
sections are too small with fixed width. 
\item Another version of QMD was developed in Copenhagen and
called NMD \cite{BIM95}. 
\item Many attempts have been undertaken to extend the applicability
of QMD towards higher (relativistic) energies. Models are for
instance RQMD \cite{SSG89,LPF95,Sor95} and UrQMD \cite{BBB98A}.
\end{enumerate}

\subsubsection{Decoupling of center of mass and relative motion}

A prominent problem of many-body trial states expressed in terms
of single-particle quantities, irrespective whether they are
antisymmetrized or not, is the center of mass motion which does
not separate from the relative motion.  One attempt to solve the
problem is the construction of a trial state, where the
single-particle width parameters are replaced by a width matrix
$A_{kl}(t)$ \cite{KiD96} 
\begin{eqnarray} \label{E-1-3-2}
\braket{ \vek{x}_1,\dots,\vek{x}_N }{Q(t)}
=
\exp\left\{
-\sum_{k,l}
(\vek{x}_k-\vek{r}_k(t))\,A_{kl}(t)\,(\vek{x}_l-\vek{r}_l(t)) 
+ i \sum_{k}
\vek{p}_k(t) \, \vek{x}_k
\right\}\, \ket{\chi} \, ,
\end{eqnarray}
$\ket{\chi}$ is a normalized spin-isospin state. 
The dynamical freedom of the matrix elements $A_{kl}(t)$ allows this
state to factorize into c.m. and intrinsic degrees 
of freedom for subgroups of particles. For two particles
the advantage of ansatz \fmref{E-1-3-2} can be seen
immediately
\begin{eqnarray} 
\braket{\vek{x}_1, \vek{x}_2}{Q} 
= \exp\left\{
- A_{cm} \left(\vec{X}-\vec{R}\, \right)^2 +i\,\vec{P}\vec{X}\right\}\,
\exp\left\{
-A_{rel} \left(\vek{x}-\vek{r}\right)^2 +i\,\vek{p}\vek{x}
\right\}\, \ket{\chi} \, .
\end{eqnarray}
With the usual definitions of relative
($\vek{x},\vek{r},\vek{p}$)
and c.m. ($\vec{X},\vec{R},\vec{P}$) coordinates
the width matrix-elements are related by
$A_{cm} = 2(A_{11}+A_{12})$, $A_{rel} = (A_{11}-A_{12})/2$,
and $A_{11}=A_{22}$.  Independent of the relative motion, which
can be in a bound state with width $A_{rel}$, the c.m. wave
packet width $A_{cm}(t)$ can spread according to free motion.
For product states the variance in the c.m.
coordinate is always connected to that of the relative
motion. For the product of two identical
gaussian wave packets the relation is $A_{cm}(t)=4 A_{rel}(t)$
\cite{KiD96}.

The separation of internal and c.m. variables should enhance
fragment production which is otherwise suppressed due to the 
localization energy of the c.m. motion. The authors could
demonstrate a significant improvement in the 
description of light fragments. Unfortunately the  proposed
ansatz could not be extended to antisymmetrized states up to now. 
The trial state
(\ref{E-1-3-2}) is  also not flexible enough to describe the
splitting of wave packets neccessary to model particle
capture as discussed in connection with quantum branching in
\secref{quantbranch}.

\subsubsection{Approximate canonical variables -- Pauli potential}

The historically first models that tried to describe
fragmentation reactions on the basis of single-particle motion
were classical models, see for instance
\cite{BoP77,BPM80}. Problems arising from the fact, that 
classical particles neither obey the Heisenberg uncertainty
relation nor the Pauli principle of identical fermions are
attacked by introducing two-body interactions
$\sum_{i<j}\,{\mathcal
V}_{uncertainty}(\vek{r}_{ij},\vek{p}_{ij},a)$
\cite{WHK77,WYC78}
and 
$\sum_{i<j}\,{\mathcal V}_{Pauli}(\vek{r}_{ij},\vek{p}_{ij},a,S_{ij})$
\cite{WHK77,WYC78,DDR87,DoR87,BoG88,PRS91,MOH92,NCM95,EbM97}
which imitate the two quantum effects, see \eqref{E-2-1-21}. 

The method is quite successful in calculating energies and
reasonable single-particle occupations in momentum space for the
free Fermi gas at finite temperatures \cite{DDR87,DoR87}, but as
already mentioned in \secref{sec-2-1}, care should be taken in
using these variables in Hamilton's equation of motion as if
they were canonical. In addition classical models cannot
correctly describe quantum statistical properties in general,
like occupation numbers, mean energy, specific heat etc.

Nevertheless, the idea of simulating Heisenberg uncertainty and
Pauli exclusion principle by means of two-body interactions is
still being used in nowadays applications of classical dynamics
to many-fermion problems, \eg \cite{LBB94,WiC98}. 

\newpage
\subsection{Atomic physics}
\label{secap}

In atomic physics molecular dynamics applications are widespread
because the de Broglie wave length of the atoms (molecules) is
often much shorter than variations in the inter-molecular
potential. Therefore a trial state for the center of mass
coordinates, which is a product of well localized gaussians as
discussed in \secref{classicalmech}, 
is well suited 
\begin{eqnarray} 
\ket{Q_{atom}^\prime}= 
\ket{\vek{B}_1}\otimes\cdots\otimes\ket{\vek{B}_N}
\ \ \mbox{with}\ \
\vek{B}_l=\vek{R}_l+iA_0\vek{P}_l
\ .
\end{eqnarray}
The electrons are much lighter so that quantum effects like
Pauli principle and uncertainty are important. Their trial state
might be thought of as an antisymmetrized many-body state
$\ket{Q_{el}^\prime;Q_{atom}^\prime}$
which depends on electronic degrees of freedom summarized in the
set $Q_{el}^\prime=\{q_0,q_1,q_2,\cdots\}$, e.g. characterizing
different orbits. It also 
depends on the variables of the atoms $\{\vek{R}_l,\vek{P}_l\}$
which mark for example the phase space centers of the
orbits. The dependence on $\vek{P}_l$ is usually  
neglected because the velocities of the electrons are much
larger than those of the atoms. The total trial state is the
product
\begin{eqnarray} 
\ket{Q^\prime}=
\ket{Q_{atom}^\prime}\otimes\ket{Q_{el}^\prime;Q_{atom}^\prime}
\ .
\end{eqnarray}
Different from nuclear physics the hamiltonian is known. It can be
well approximated by (spin and other relativistic effects neglected)
\begin{eqnarray} 
\op{H}&=&\sum_{l=1}^N \left[ \frac{\vek{\op{K}}^2(l)}{2M_l}  
- \sum_{i=1}^{N_{el}} \frac{Z_l e^2}{|\vek{\op{X}}(l)-\vek{\op{x}}(i)|}
\right]\ 
+ \sum_{l < k}^{N} \frac{Z_l Z_k e^2}{|\vek{\op{X}}(l)-\vek{\op{X}}(k)|}
+ \ \op{H}_{el} \\
\op{H}_{el} &=&
\sum_{i=1}^{N_{el}} \frac{\vek{\op{k}^2}(i)}{2m_{el}} +
\sum_{i < j}^{N_{el}}
   \frac{e^2}{|\vek{\op{x}}(i)-\vek{\op{x}}(j)|}
\ ,
\end{eqnarray}
where capital and small letters denote atomic and electronic
variables, respectively. From
the Lagrange function ${\cal L}^\prime$, \eqref{Lagrange}, one
obtains with the appropriate approximations 
the well known Quantum Molecular Dynamics equations
for atomic physics. The atomic variables $\vek{R}_l$ and
$\vek{P}_l$ follow classical equations of motion (width $A_0$
small) under the influence of electron potentials given by 
\begin{eqnarray} \nonumber
{\cal V}_{atom-el}&=&\bra{Q_{el}^\prime;Q_{atom}^\prime}
\sum_{l=1}^N \left[
-i\dot{\vek{B}_l}\frac{\partial}{\vek{B}_l}+
\sum_{i=1}^{N_{el}} \frac{Z_le^2}{|\vek{R}_l-\vek{\op{x}}(i)|} 
   \right] \ket{Q_{el}^\prime;Q_{atom}^\prime} \\
&+&\bra{Q_{el}^\prime;Q_{atom}^\prime}\op{H}_{el}
\ket{Q_{el}^\prime;Q_{atom}^\prime} \ .
\end{eqnarray}
The coupled equations of motion for the electronic degrees of
freedom, $Q_{el}^\prime$, are approximated at various levels of
sophistication. 

These quantum molecular dynamics models are not subject of this
review because they do not treat the electrons (fermions) as being
localized in phase space.

\subsubsection{Product states of wave packets in atomic physics}

For simple Coulomb systems \cite{BNN93} and for hydrogen plasmas
models are proposed which employ trial states of the type 
\begin{eqnarray} 
\ket{Q^\prime}=
\ket{Q_{atom}^\prime}\otimes\ket{Q_{el}^\prime}
\ ,
\end{eqnarray}
where the electronic state $\ket{Q_{el}^\prime}$ consists of
localized gaussian wave packets, \eqref{gaussian}. 
These models are QMD models in the sense of \secref{secqmd},
either with time-independent width and without Pauli potential 
\cite{EbM97} or with Pauli potential \cite{EbS97} or with
time-dependent width and Pauli potential
\cite{KTR94A,KTR94B}. Only the electronic part is represented by
a wave function, the protons of the hydrogen plasma move on
classical trajectories.

An interesting application, which is referred to in \secref{sec-4-3-2}, is
the study of a plasma under 
extreme conditions, high temperature or pressure 
and phase transitions to the liquid and solid phase
\cite{KTR94A,KTR94B}. Other investigations focus on the
degree of ionization of a partially ionised plasma
\cite{EFP96,EbM97}. 

\subsubsection{Quantum branching}

When the Born-Oppenheimer approximation is valid adiabatic
energy surfaces can be calculated as
\begin{eqnarray} \label{energysurface} 
{\cal V}_{ad}(Q_{atom}^{\prime *},Q_{atom}^\prime;\nu_{el})
&=&\bra{\nu_{el};Q_{atom}^\prime}
\sum_{l=1}^N \sum_{i=1}^{N_{el}} 
\frac{Z_l e^2}{|\vek{R}_l-\vek{\op{x}}(i)|} 
+\op{H}_{el}
\ket{\nu_{el};Q_{atom}^\prime} \ ,
\end{eqnarray}
where $\nu_{el}=0$ denotes the lowest energy state of the
electrons under the influence of the charges $Z_l$ of static
ions positioned at $\vek{R}_l$. $\nu_{el}=1,2,\cdots\ $ numerate
the excited eigenstates of the electronic system, for example vibrational
modes or particle-hole excitations. In principle one can set up
an improved trail state as a linear combination
\begin{eqnarray} \label{trialhopping}
\ket{Q^\prime}
&=&\sum_{\nu_{el}} q_{\nu_{el}} \ket{Q_{atom,\nu_{el}}^\prime}\otimes
\ket{\nu_{el};Q_{atom,\nu_{el}}^\prime} 
\end{eqnarray}
and then try to solve the coupled equations which result from
the variational principle \fmref{varpinciple}
for the complex amplitudes $q_{\nu_{el}}$ and the ion variables
$Q_{atom,\nu_{el}}^\prime=\{\vek{R}_l^{\,\nu_{el}},
\vek{P}_l^{\,\nu_{el}};l=1,\cdots,N;\nu_{el}=1,2,\cdots\}$. From
energy, momentum, and angular momentum conservation it is obvious that one
has to have as many different trajectories 
$\{\vek{R}_l^{\,\nu_{el}}, \vek{P}_l^{\,\nu_{el}} \}$ for the ions as
there are excited states which can be populated, because an inelastic
excitation $\nu_{el}\rightarrow\nu_{el}^\prime$ will change the 
ion trajectories accordingly. 

But these equations of motion are
usually too complex to be solved numerically, one therefore
introduces quantum branching (as discussed in
\secref{quantbranch}) called quasiclassical trajectory surface
hopping method. See for example references in \cite{THA97}. 
Different approaches are tested against an
accurate quantum dynamics calculation of a realistic system by
Topaler et al. \cite{THA97}. They find fair agreement between
the quantum branching methods and results of the quantum
equations which result from a trial state of type
\eqref{trialhopping}. This is expected if electronic coherence
is not important on time scales of the ion motion so that the
hopping between 
the energy surfaces (\xref{energysurface}) acts as a
random Langevin force on the ion trajectories.    

\newpage
\section{Statistical properties}
\label{sec-4-0}

Molecular dynamics models are not only used to simulate
nonequilibrium but also equilibrium situations, especially when
correlations require descriptions which go beyond mean-field
approaches and quasi-particles.
In the context of classical mechanics a vast literature exists
\cite{Hoo85}, even for relativistic cases \cite{BBB98B}, in which
equilibrium properties are studied. So called thermostated time 
evolutions \cite{Nos84,Hoo85,KBB90,Nos91}, in which appropriate
coupling to external degrees of freedom adjusts the temperature,
are on firm grounds in classical mechanics
and have been used successfully for equilibrium, 
e.g. to investigate classical spin systems, 
as well as for non-equilibrium, e.g. to study glass transitions. 

The classical procedures fail for quantum systems and especially
for identical fermions when the phase-space density is not small
anymore and the effects of the Pauli principle become
important. No analogue to thermostated time evolutions exists yet
for quantum systems.  There are few attempts to infer
thermodynamic properties from dynamical simulations in quantum
mechanics, some of them are given in
\cite{Kus93,OhR93,KTR94A,BDH94,ScF96,OnH96B,EbM97,OhR97A,OhR97B,ScF97,Sch98}.
Usually one performs time-averages and relies on the ergodic
assumption. The validity of this method depends of course
crucially on the statistical and ergodic properties of the
dynamical model. Following the mentioned articles one realizes
that the matter is still under debate.

Two general aspects are discussed in detail in the
following sections. One concerns the
thermo{\it static} properties of a molecular dynamics model where
the attribute thermostatic refers to the characteristics of the
static canonical statistical operator, which determines
the quantal partition function
$Z(T)=\Tr(\exp\{-\Operator{H}/T\})$. Once the trace is evaluated
within a given model,
its thermostatic properties can be deduced by standard methods
like partial derivatives of $\ln Z(T)$ with respect to
temperature $T$ or other parameters contained in the Hamilton
operator $\Operator{H}$. 

The other and even more important aspect, which is discussed in
\secref{sec-4-2}, is the dynamical behaviour of a
molecular dynamics model. For example a dissipative system which
is initially far from equilibrium is expected to equilibrate
towards the canonical ensemble.  The simulation of such a system
within the model provides a crucial test of its thermo{\it
dynamic} properties. Often one uses an ergodicity
assumption, \ie that time-averages are equivalent to ensemble
averages, which should be verified.

\subsection{Thermostatics}
\label{thermostatics}

The question of thermo{\it static} properties
can be condensed to the question, whether the set of model
states $\ket{Q}$ is complete, i.e. able to span the many-particle Hilbert
space. If that is the case the unit operator in $N$-particle space can be
written as  
\begin{eqnarray}
\EinsOp^{(N)}&=&\int \mbox{d{\Large $\mu$}}(Q)\;
\frac{\ket{Q} \bra{Q}}{\braket{Q}{Q}}\ ,
\end{eqnarray}
where $\mbox{d{\Large $\mu$}}(Q)$ is a measure which  depends on the
parameter set $Q$ because the trial states $\ket{Q}$ are in
general nonorthogonal. 
For fermions and gaussian wave packets the measure is derived in the
following subsection.

Once the completeness is shown the thermo{\it static} relations have to be
correct, provided the trace of the partion function
\begin{eqnarray}
\label{PartFunc}
Z(T)&=&\Tr\left(\exp\left\{-\Operator{H}/T \right\}\right)
\\
&=&\int \mbox{d{\Large $\mu$}}(Q)\;
\frac{\bra{Q} \exp\left\{-\Operator{H}/T \right\} \ket{Q}}
     {\braket{Q}{Q}}\ .
\nonumber
\end{eqnarray}
is calculated with the trial states $\ket{Q}$ in quantum fashion.

In the case of Fermionic Molecular Dynamics (FMD) and 
Antisymmetrized Molecular Dynamics (AMD)
\cite{ScF96,OHM92A,OHM92B}
the antisymmetric many-body states form an overcomplete
set and provide a full representation for the unit operator. 
As the calculation of the trace does not depend on the representation
all thermostatic properties like Fermi-Dirac distribution,
specific heat, mean energy as a function of temperature
etc. ought to be correct and fully quantal using FMD or AMD
trial states. The difficult task is to calculate 
$\bra{Q}\exp\left\{-\Operator{H}/T\right\}\ket{Q}$, or 
$\bra{Q}\Operator{B}\exp\left\{-\Operator{H}/T\right\}\ket{Q}$,
where $\Operator{B}$ is an observable. The trace integral can
be evaluated by Monte-Carlo methods.

\subsubsection{Completeness relation with coherent states}

In the following it is shown that Slater determinants of coherent
states span the whole Hilbert space for fermions. 

Coherent states $\ket{\vek{z}}$ which are defined as the
eigenstates of the harmonic oscillator destruction operator
$\Operator{\vek{a}}=\sqrt{\frac{1}{2a_0}}\Operator{\vek{x}}+
i\sqrt{\frac{a_0}{2}}\Operator{\vek{k}}$,
\begin{eqnarray}
\Operator{\vek{a}}\; \ket{\vek{z}}&=&\vek{z} \;\ket{\vek{z}}
\ ,\qquad \OphHO =
\frac{1}{m a_0} \; 
\left( \Operator{\vek{a}}^+ \, \Operator{\vek{a}} 
+ \frac{3}{2} \right)
\end{eqnarray}
form an overcomplete set of states. They are the gaussian states
defined in \eqref{gaussian} with 
$\vek{z}=\sqrt{\frac{1}{2a_0}}\vek{r}+
i\sqrt{\frac{a_0}{2}}\vek{p}=\vek{b}/\sqrt{2a_0}$  
for a real width parameter $a_0$.
Their completeness relation reads in single-particle space 
\begin{eqnarray}
\label{KohVoll}
\EinsOp^{(1)}
&=& \int \frac{\dint^3\Re{z}\;\dint^3\Im{z}}{\pi^3}\;
\frac{\ket{\vek{z}}\bra{\vek{z}}}{\braket{\vek{z}}{\vek{z}}}
\\
&=& \int \frac{\mbox{d}^3r\;\mbox{d}^3p}{(2\pi)^3}\;
\frac{\ket{\vek{r},\;\vek{p}}\bra{\vek{r},\;\vek{p}}}
{\braket{\vek{r},\;\vek{p}}{\vek{r},\;\vek{p}}}\ ,
\nonumber
\end{eqnarray}
where $\ket{\vek{z}}$ labels the coherent states by their
eigenvalue with respect to $\Operator{\vek{a}}$, 
and the phase space notation
$\ket{\vek{r},\;\vek{p}}$ labels the states by their
expectations values of coordinate and momentum operators. 
Coherent states are extensively discussed in \cite{KlS85}.

Since we are dealing with fermions the spin degree of freedom
has to be considered and consequently the resolution of unity
changes to
\begin{eqnarray}
\EinsOp^{(1)}
&=& \int \frac{\mbox{d}^3r\;\mbox{d}^3p}{(2\pi)^3}\;\sum_{m}\;
\frac{\ket{q}\bra{q}}{\braket{q}{q}}\ ,
\end{eqnarray}
where the sum runs over the two magnetic quantum numbers
$m=\pm\half$ which are included in the set of parameters denoted
by $q$ (see \eqref{E-2-0-2}).

Proceeding one step further the unit operator in the
antisymmetric part of the two-particle Hilbert space is the
antisymmetric product of two single-particle unit operators 
\begin{eqnarray}
\EinsOp^{(2)}&=&\Operator{A}^{(2)}\; 
\left(\EinsOp^{(1)}\otimes\EinsOp^{(1)}\right)\;
\Operator{A}^{(2)} \\
&=& 
\half\Big(1 - \Operator{P}_{12}\Big)\; 
\left(\EinsOp^{(1)}\otimes\EinsOp^{(1)}\right)\;
\half\Big(1 - \Operator{P}_{12}\Big)\ ,
\nonumber
\end{eqnarray}
which may be expressed with antisymmetric two-particle states
$\ket{q_1,\; q_2}_a$ as
\begin{eqnarray}
\EinsOp^{(2)}&=& 
\int \frac{\mbox{d}^3r_1\;\mbox{d}^3p_1}{(2\pi)^3}\;\sum_{m_1}\;
\int \frac{\mbox{d}^3r_2\;\mbox{d}^3p_2}{(2\pi)^3}\;\sum_{m_2}\;
\frac{\ket{q_1,\; q_2}_a \; {}_a\bra{q_1,\; q_2}}
     {\braket{q_1}{q_1}\braket{q_2}{q_2}}
\\[5mm]
\mbox{where}
&&
\ket{q_1,\; q_2}_a:= 
\half\left(
\ket{q_1}\otimes\ket{q_2} -
\ket{q_2}\otimes\ket{q_1}\right)
\ .
\nonumber
\end{eqnarray}
Following this line the resolution of unity in the antisymmetric
part of the $N$-particle Hilbert space 
can be written as the projection of the
$N$-particle unit operator onto the antisymmetric subspace.
Be $\ket{{Q}}$ the unnormalized Slater determinant \fmref{E-2-0-1} of
single-particle states $\ket{q}$
\begin{eqnarray}
\ket{{Q}}&=&\frac{1}{N!}\sum_{P}\sign(P)
\left(\ket{q_{P(1)}}\otimes\cdots\otimes\ket{q_{P(N)}}\right)
\ ,
\end{eqnarray}
then the unit operator is 
\begin{eqnarray}
\label{ASEinsOp}
\EinsOp^{(A)}&=&\int \mbox{d{\Large $\mu$}}(Q)\;
\frac{\ket{Q} \bra{Q}}{\braket{Q}{Q}}\ ,
\end{eqnarray}
with a measure 
\begin{eqnarray}
\mbox{d{\Large $\mu$}}(Q)
&=& 
\braket{{Q}}{{Q}}
\prod_{k=1}^N
\frac{1}{\braket{q_k}{q_k}}
\frac{\mbox{d}^3r_k\;\mbox{d}^3p_k}{(2\pi)^3}\;\sum_{m_k}\;\ ,
\end{eqnarray}
that accounts for antisymmetrization. In
a sampling where the values of $\vek{r}_k$ and $\vek{p}_k$ are
chosen according to Monte-Carlo methods this measure determines
the probability to find the state $\ket{Q}$ belonging to the
configuration
$Q=\{\vek{r}_1,\vek{p}_1,m_1;\vek{r}_2,\vek{p}_2,m_2;\dots\}$ in Hilbert
space. If for example two fermions with the same spin are close
in $\vek{r}$ and $\vek{p}$ then this measure is small because
the norm $\braket{Q}{Q}=\mbox{det}\{\prodkl\}$ will
be small. \eqref{ASEinsOp} is very useful in calculating traces by
means of Monte-Carlo sampling \cite{OhR93}. 

Coherent states are gaussian wave-packets with fixed width, but
the real and imaginary part of the width, $a_R$ 
and $a_I$, may also be integrated over appropriate ranges 
in order to get an improved coverage of the phase space 
when doing the sampling.
\begin{eqnarray}
\label{ASEinsOpBreite}
\EinsOp^{(1)}
&=& 
\;
\int \frac{\mbox{d}^3r\;\mbox{d}^3p}{(2\pi)^3}\;\sum_{m}\;
\int_{\Omega_R}\frac{\mbox{d}a_R}{\Omega_R} \;
\int_{\Omega_I}\frac{\mbox{d}a_I}{\Omega_I}\;
\frac{\ket{q}\bra{q}}{\braket{q}{q}}\ ,
\end{eqnarray}
where $\Omega_R$ and $\Omega_I$ denote the intervals the width
$a=a_R+ia_I$ is integrated over
\begin{eqnarray}
\int_{\Omega_R} \mbox{d}a_R\; = \Omega_R\ ,\quad
\int_{\Omega_I} \mbox{d}a_I\; = \Omega_I\ .
\nonumber
\end{eqnarray}
Since the width $a$ in the completeness relation \eqref{KohVoll}
is arbitrary, the additional integrations in
\eqref{ASEinsOpBreite} do not change the unit operator.

In the case of nuclear physics one has two types of fermions,
so that the
measure for a system with $N$ neutrons and $Z$ protons is
\begin{eqnarray}
\mbox{d{\Large $\mu$}}(Q)
&=& 
\braket{{Q}}{{Q}}
\prod_{k=1}^N
\frac{1}{\braket{q_k}{q_k}}
\frac{\mbox{d}^3r_k\;\mbox{d}^3p_k}{(2\pi)^3}\;\sum_{m_k}
\prod_{l=N+1}^{A=N+Z}
\frac{1}{\braket{q_l}{q_l}}
\frac{\mbox{d}^3r_l\;\mbox{d}^3p_l}{(2\pi)^3}\;\sum_{m_l}
\;\ .
\end{eqnarray}

Once the resolution of unity is given in terms of model states
the partition function can be evaluated.

It is interesting to
note that the norm $\braket{Q}{Q}$ in the measure cancels
with the norm denominator in \eqref{PartFunc}. Provided the
single-particle states are normalized, $\braket{q_k}{q_k}=1$,
and $\Operator{H}$ does not depend on the spins,
the partion function looks almost classical 
\begin{eqnarray}\label{PartFunc1}
Z(T)&=&\frac{2^N}{(2\pi)^{3N}}
\int\mbox{d}^3r_1\,\mbox{d}^3p_1\, 
    \mbox{d}^3r_2\,\mbox{d}^3p_2\cdots
    \mbox{d}^3r_N\,\mbox{d}^3p_N\,
\bra{Q} \exp\left\{-\Operator{H}/T \right\} \ket{Q} \ ,
\end{eqnarray}
except that there is an operator in the
exponent. \eqref{PartFunc1} is still the exact quantum
expression. There is no contradiction between the fact that the
hamiltonian may have a discrete energy spectrum with gaps
between the levels and the fact that the parameters $\vek{r}_k$
and $\vek{p}_k$ are integrated in a continous fashion
\cite{Sch99}. Only if the expectation value is moved up into the
exponent things become wrong. The Pauli exclusion principle is
fully taken care of by the antisymmetric state $\ket{Q}$. If two
particles with equal spin are at the same point in phase space
$\ket{Q}=0$ and hence $\bra{Q} \exp\left\{-\Operator{H}/T
\right\} \ket{Q}=0$, so that forbidden states do not contribute
to the partition function.

\subsubsection{Example for many fermions}

In the following the above considerations are illustrated with
the example of $N$ identical fermions in a common
\red{ single-particle}
potential.
Starting from the Hamilton operator
\red{
\begin{eqnarray}
\Operator{H}&=&\sum_{l=1}^N
\Operator{h}(l)\ ,
\quad
\Operator{h}(l) = \frac{\Operator{\vek{k}}^2(l)}{2 m}
                  + v(\Operator{\vek{x}}(l))
\end{eqnarray}
}
the mean energy of the $N$-fermion system can be derived from
the partition function $Z(T)$, \eqref{PartFunc}, as the
derivative with respect to $T$
\begin{eqnarray}\label{Hmean}
\EnsembleMean{\Operator{H}}&=&T^2 \frac{\partial}{\partial T} \ln(Z(T))
\nonumber \\ 
&=&\frac{
\int \mbox{d{\Large $\mu$}}(Q)\;{\mathcal W}(T)\;
\sum_{m,n}^N {\mathcal O}_{nm}(T) 
\left[T^2 \frac{\partial}{\partial T} 
\bram\exp\Big\{-\Operator{h}/T\Big\}\ketn\right]
}{
\int \mbox{d{\Large $\mu$}}(Q)\;
{\mathcal W}(T)
}\ ,
\end{eqnarray}
where the two abbreviations ${\mathcal W}(T)$ and 
${\mathcal O}^{-1}(T)$  are introduced as
\red{
\begin{eqnarray}
{\mathcal W}(T)\equiv
\frac{\bra{Q} \exp\{-\Operator{H}/T\}\ket{Q}}
     {\braket{Q}{Q}} &=&
\frac{\mbox{det}\Big(\brak \exp\{-\Operator{h}/T\}\ketl\Big)}
     {\mbox{det}\Big(\prodkl\Big)}\ ,
\\[3mm]
\left({\mathcal O}^{-1}(T)\right)_{kl}
&=&
\brak \exp\{-\Operator{h}/T\}\ketl \ . \label{OT}
\end{eqnarray}
For free motion, i.e. $v(\Operator{\vek{x}})= 0$,
the matrix elements (\ref{OT}) are given by:
\begin{eqnarray} \label{OBexpt}
\brak \exp\left\{-\frac{\Operator{\vek{k}}^2}{2 mT}\right\}\ketl
&=&
\left( \frac{2\pi \ a_k^*a_l}{a_k^*+a_l+\frac{1}{mT}}\right)^\frac{3}{2}
\exp\left\{-\frac{(\vek{b}_k^*-\vek{b}_l)^2}{2(a_k^*+a_l+\frac{1}{mT})} 
\right\} \ .
\end{eqnarray}
For the special case of the three-dimensional harmonic oscillator,
i.e. $v(\Operator{\vek{x}})= \half m \omega^2 \Operator{\vek{x}}^2$,
the matrix elements can easily be calculated
if one sets all $a_k=a_0=1/(m\omega)$ \cite{Sch96}:
}
\begin{eqnarray} \label{OBexph}
\brak \exp\left\{-\frac{\OphHO}{T}\right\}\ketl
&=&\left(\frac{\pi}{m \omega}\right)^\frac{3}{2}
\exp\left\{-\frac{m \omega}{4}(\vek{b}_k^*-\vek{b}_l)^2
-\frac{m\omega}{2}\,\vek{b}_k^*\vek{b}_l (1-e^{-\frac{\omega}{T}}) 
-\frac{3\omega}{2T} \right\}  .
\end{eqnarray}
\red{
For other potentials the matrix elements (\ref{OT})
assume a more complicated form.}

\begin{figure}[ht]
\begin{center}
\epsfig{file=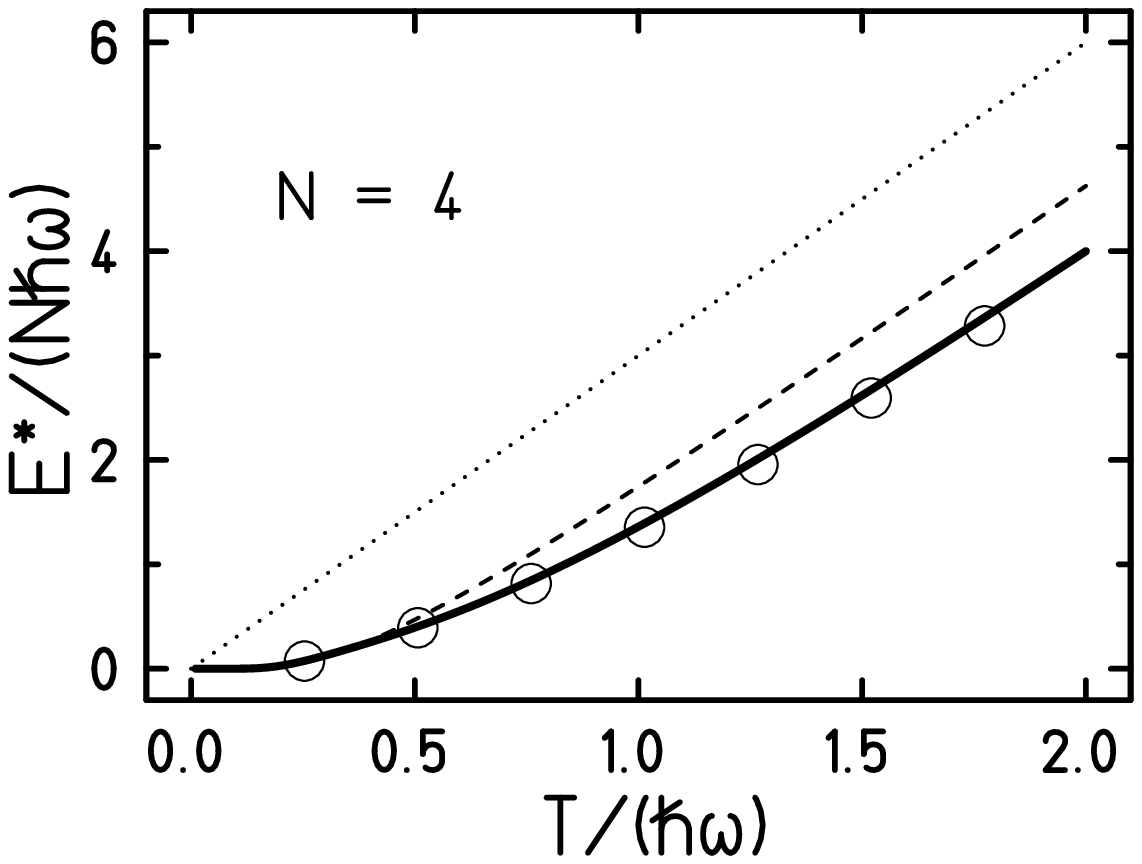,height=65mm}
\mycaption{Excitation energy as a function of temperature,
calculated with antisymmetrized gaussian states (circles) or with
eigenfunctions of the harmonic oscillator (solid line).
The dotted line shows the classical result $E^*/N=3T$ and the
dashed line the result of quantum Boltzmann statistics.
}{P-5.1-1}
\end{center}
\end{figure}

Figure \xref{P-5.1-1} shows the result of a Metropolis
integration \cite{MMR53}
over $\vek{r}_k$ and $\vek{p}_k$ of \eqref{Hmean}
for a system of four identical fermions with equal spins in a harmonic 
oscillator by open circles.
The solid line displays the results when the traces are calculated
by discrete sums over eigenstates of the many-body Hamilton
operator $\OpHHO$ \cite{ScS98}. For comparison the 
classical dependence is shown as a dotted line.
The dashed line represents the result for the quantum Boltzmann
case for distinguishable particles, were the trial state in
\eqref{PartFunc1} is a direct product of single-particle
states. In both quantum cases the specific heat is vanishing at
$T=0$ because of the finite energy gap between ground state and
first excited state.

As expected from the general argument that traces are
independent of the representation, provided one uses a complete
set, the numerical Metropolis integration with a
sampling of $10^6$ points in the 24-dimensional phase space of
the continuous parameters $\vek{r}_k, \vek{p}_k,$ which specify
the Slater determinants $\ket{Q}$, gives the same result as
summing over eigenstates of the many-body hamiltonian. The
continous integration is not in contradiction with Fermi
statistics and the discrete spectrum of the hamiltonian, and the
classically looking \eqref{PartFunc1} is fully quantal.

\subsubsection*{Resum\'{e}}

\begin{enumerate}
\item The thermo{\it static} properties of a model ought to be
correct, if the set of model states $\ket{Q}$ is complete,
i.e. able to span the many-particle Hilbert space.
\item In FMD and AMD the antisymmetric many-body states of
single-particle gaussian wave packets form an overcomplete
set and provide a full representation for the unit operator.
\end{enumerate}

\newpage
\subsection{Thermodynamics}
\label{sec-4-2}

In molecular dynamics the time evolution as given by the \tdvp,
without a collision term or quantum branching, is
deterministic. Given a state
$\ket{Q(t_0)}$ at an initial time $t_0$ the system is described
by the pure state $\ket{Q(t)}$ at all earlier and later
times. Therefore, like with the exact solution of the
Schr\"odinger equation, thermal properties have to be obtained by
coarse graining or time averaging.   

In this section time averaging is compared with the canonical statistical
ensemble for a fermion system.  If the system is ergodic both
are equivalent and equilibrium properties 
can be evaluated by molecular dynamics simulations.
For this the ergodic ensemble is defined by the statistical
operator $\Rerg$ as
\begin{eqnarray}\label{EER}
\Rerg&\; := \;&\lim_{t_2\rightarrow\infty}\;\frac{1}{(t_2 - t_1)}\;
\int_{t_1}^{t_2} \mbox{d}t\;
\frac{\ket{Q(t)}\bra{Q(t)}}{\braket{Q(t)}{Q(t)}}\ .
\end{eqnarray}
Herewith the ergodic mean of an operator $\Operator{B}$ is given by
\begin{eqnarray}\label{EEM}
\hspace{-8mm}\ErgodicMean{\Operator{B}}
:= \;
\Tr\left(\Rerg\;\Operator{B}\right)
=
\lim_{t_2\rightarrow\infty}\;
\frac{1}{(t_2 - t_1)}\;
\int_{t_1}^{t_2} \mbox{d}t\;
\frac{\bra{Q(t)}\Operator{B}\ket{Q(t)}}{\braket{Q(t)}{Q(t)}}\ .
\end{eqnarray}
If the ergodic assumption is fulfilled, the statistical operator
$\Rerg$ should depend only on $\erw{\Operator{H}}$, which is
actually a constant of motion.  Therefore the notation with the
condition ``$\erw{\Operator{H}}$" in \eqref{EEM} is used.

Expectation values are well-defined with \eqref{EEM}
so that one can easily calculate extensive quantities like the
excitation energy. But it is not obvious how an
intensive thermodynamical quantity, such as the temperature,
might be extracted from deterministic molecular dynamics with
wave packets.  

In classical mechanics with momentum-independent
interactions the partition function
\begin{eqnarray}
&&
Z_{\mbox{\scriptsize classical}}(T)=\int\prod_{k=1}^N 
\frac{\mbox{d}^3r_k\;\mbox{d}^3p_k}{(2\pi)^3}\;
\exp\left\{-\frac{1}{T}\,
{\cal H}_{\mbox{\scriptsize classical}}(\vec{r}_1,\vec{p}_1,\cdots)\right\}
\\&=&
\prod_{k=1}^N\int\mbox{d}^3 p_k\;
\exp\left\{-\frac{1}{T}\,\frac{\vec{p}_k^{\,2}}{2 m_k}\right\}
\cdot
\int
\prod_{l=1}^N 
\frac{\mbox{d}^3r_l}{(2\pi)^3}\;
\exp\left\{-\frac{1}{T}\,{\cal V}(\vec{r}_1,\vec{r}_2,\cdots)\right\}
\nonumber
\end{eqnarray}
is a product of a term with the kinetic energy and a term
containing the interactions. Therefore, a fit of the momentum
distribution with a Boltzmann distribution, or the equipartition
theorem, can be used to determine the temperature $T$. For
example, in simple formulations of the Nos\'e-Hoover thermostat
\cite{Hoo85,KBB90,Nos91} the equipartition theorem serves as
basic ingredient.  

In the quantum case \eqref{PartFunc} has to be employed which 
does not show this
factorization. For an interacting finite system one cannot, in
analogy to the Boltzmann case, fit a Fermi function to the
momentum distribution to determine the temperature. An example is the
ground state of a nucleus for which the momentum distribution
has a smeared out Fermi edge due to the finite size and the
two-body interaction and not
because of temperature.

\subsubsection{Ergodic ensemble of fermions in a harmonic
oscillator}

In this section the ideal gas of fermions in a common
one-dimensional harmonic oscillator potential is used for
demonstration. The hamiltonian $\HHO$ is written in second
quantization with the fermion creation operator $\Operator{c}_n^+$
\red{(which creates a fermion in the $n$-th single-particle
eigenstate of $\OphHO$)}
as
\begin{eqnarray}
\HHO&=&\sum_{l=1}^N \OphHO(l)=
\red{
\sum_{l=1}^N \left( \frac{\Operator{\vek{k}}^2(l)}{2 m}
        + \half m \omega^2 \Operator{\vek{x}}^2(l) \right)
}
=\omega \; \sum_{n=0}^\infty
\Big(n + \half\Big)\; \Operator{c}_n^+\Operator{c}_n
\end{eqnarray}
and the trial state $\ket{Q(t)}$ describes
four fermions with equal spins. To test the fermionic
nature of the dynamical evolution the ergodic ensemble averages
of the occupation numbers, 
$\ErgodicMean{\Operator{c}_n^+\Operator{c}_n}$, 
see \eqref{EEM}, are evaluated and compared with 
$\EnsembleMean{\Operator{c}_n^+\Operator{c}_n}$
of the canonical ensemble discussed in the
previous section. 

The occupation numbers, which range from 0 to 1 for fermions,
are chosen on purpose to make clear from the beginning, that the
equations of motion for the parameters  
$Q(t)=\{\vec{r}_1(t),\vec{p}_1(t), \cdots\}$, as given by
\eqref{eom}, might be generalized Hamilton equations, but the
observables always have to be calculated with the quantum state
$\ket{Q(t)}$. If $\vec{r}_k(t)$ and $\vec{p}_k(t)$ were taken as the
particle coordinates in a classical picture, the question for the
mean occupation number of the $n$-th eigenstate of the harmonic
oscillator could even not be posed.

As pointed out already, in Fermionic Molecular Dynamics (FMD) 
the time evolution of gaussian wave packets
in a common oscillator is exact, and thus the occupation
probabilities of the eigenstates of the Hamilton operator do not
change in time.  In order to equilibrate the system a repulsive
short-range two-body interaction is introduced.  The
strength of the interaction is chosen weak enough so that the
ideal gas picture is approximately still valid.

In the initial state $\ket{Q(t_0)}$, 
which is far from equilibrium, three wave packets
with a width of $a(t_0) = 1/m\omega$ are put close to the origin at
$x(t_0) = (-d, 0, d)$ with $d=0.5/\sqrt{m\omega}$, while the
fourth packet is pulled away from the center.

Figure \xref{OccNumTime} gives an impression of
the chaotic time dependence of 
$\bra{Q(t)} \Operator{c}_n^+\Operator{c}_n \ket{Q(t)}$
for $n=0, 3\; \mbox{and}\; 6$.

\begin{figure}[ht]
\begin{center}
\epsfig{file=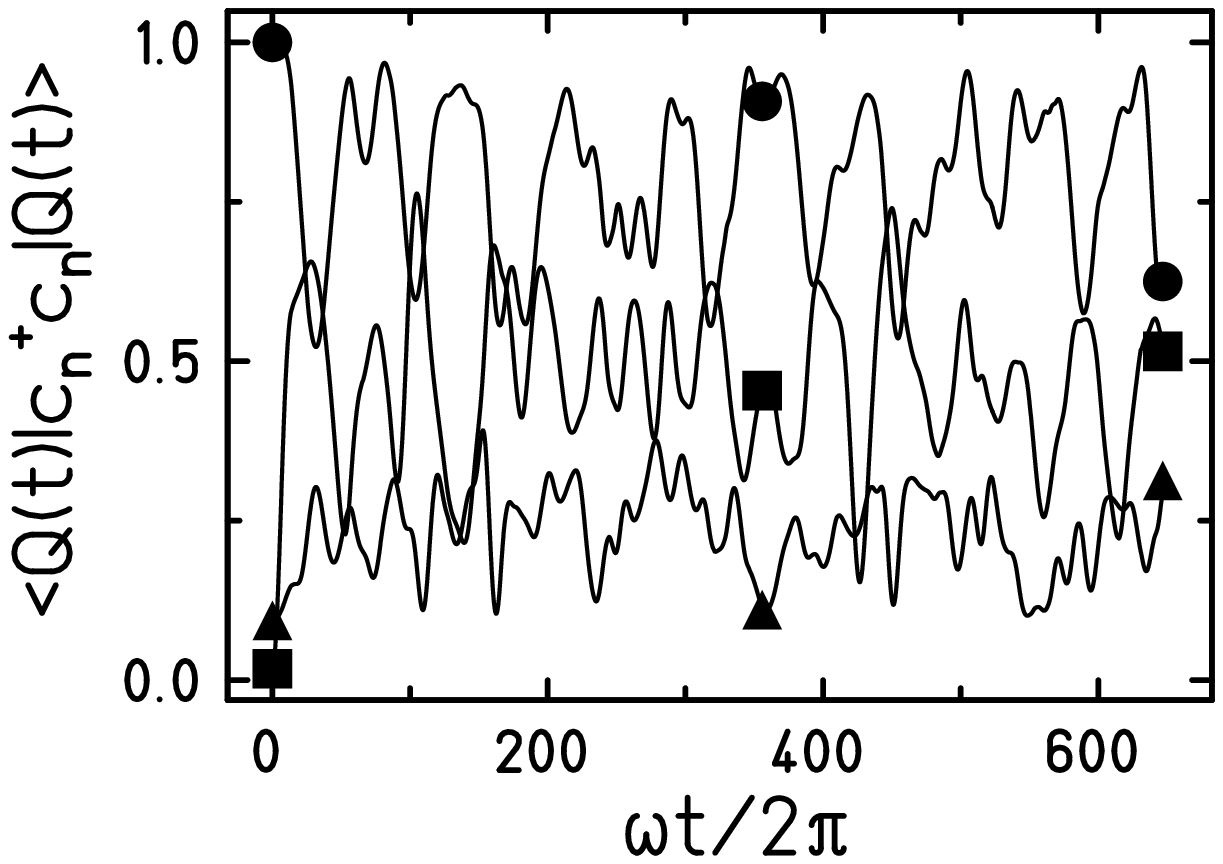,height=50mm}
\mycaption{Occupation probabilities
versus time --
$n=0$: circles,
$n=3$: squares,
$n=6$: triangles.
}{OccNumTime}
\end{center}
\end{figure}
\begin{figure}[ht]
\begin{center}
\epsfig{file=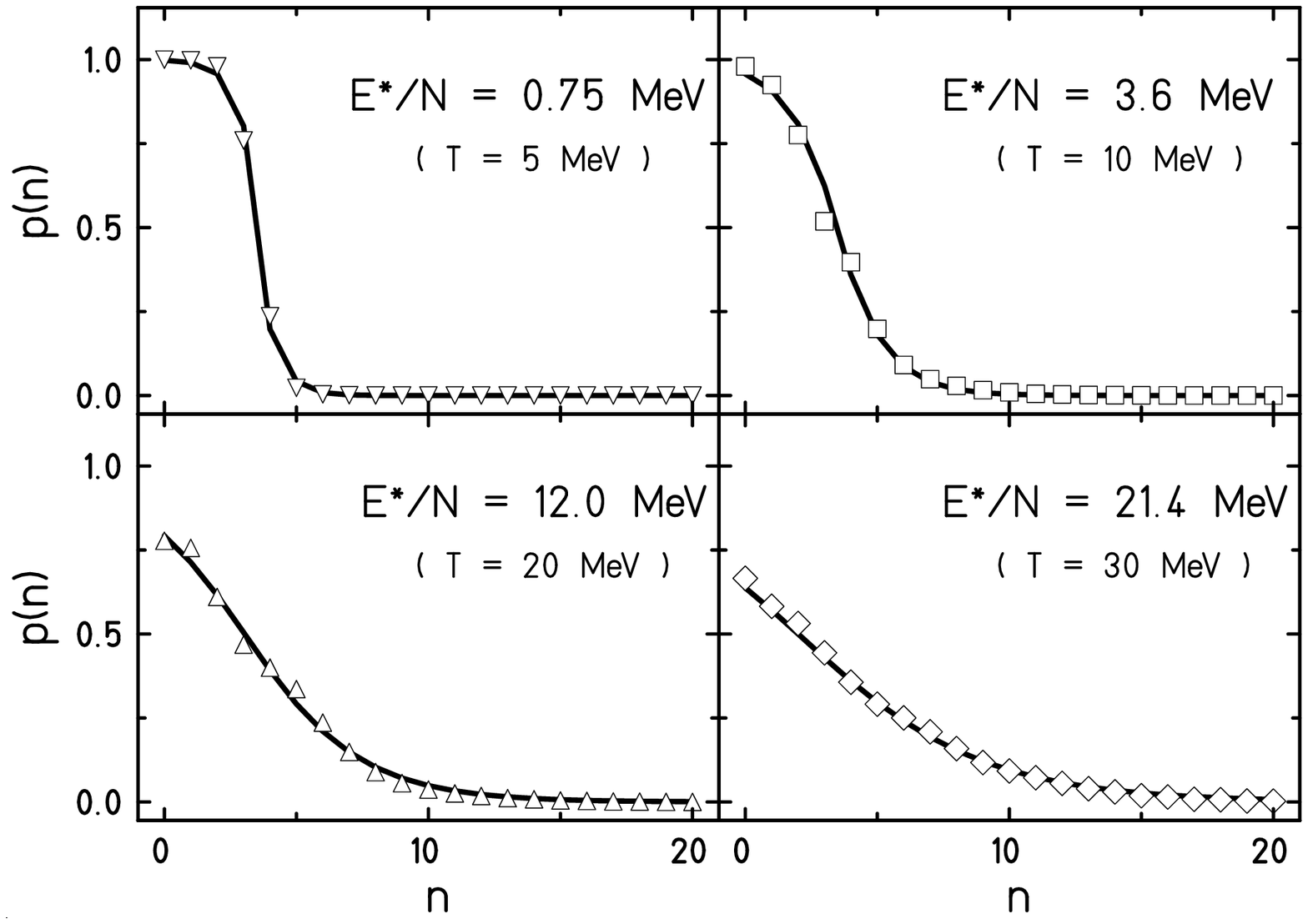,width=120mm}
\mycaption{Occupation probabilities calculated in the ergodic ensemble
$\ErgodicMean{\Operator{c}_n^+\Operator{c}_n}$ (symbols) compared
with the canonical ensemble
$\EnsembleMean{\Operator{c}_n^+\Operator{c}_n}$ (solid line).
}{HOBesetz}
\end{center}
\end{figure}

The result of time averaging is seen in \figref{HOBesetz}
(symbols) for four different initial displacements which
correspond to four different excitation energies of the fermion
system.  To each case we assign a canonical ensemble which has
the same mean energy, i.e.
$E^* = \ErgodicMean{\HHO - E_0}=\;\;\EnsembleMean{\HHO - E_0}$.
The solid lines in \figref{HOBesetz} show
the corresponding distributions of occupation probabilities for
these canonical ensembles.  Their temperatures $T$ are also
quoted in the figure. The one to one
correspondence between the occupation probabilities of the
ergodic ensemble and the ones of the canonical ensemble, which
has the same mean energy $\erw{\Operator{H}}$ as the pure state,
demonstrates that the system is ergodic and
that the fermion molecular dynamics trajectory covers the
many-body phase space
according to Fermi-Dirac statistics.

This result is not trivial because, firstly, the system is very
small, consisting of only four particles, and secondly, the
equations of motion are approximated by FMD.

\subsubsection{Describing the system with fixed-width trial states}
\label{SectionAMD}

As explained
already in \secref{thermostatics}, both type of trial states
with dynamical and fixed widths
span the whole Hilbert space and thus the thermostatic
properties are the same. Reducing the degrees of freedom by
keeping all width parameters at a fixed value $a_0$ leads
however to a different dynamical behavior as discussed in \secref{sec-2-1}.  
The equations of motion are no longer exact solutions for the
case of free motion. In a common oscillator the exact solution
is obtained only if  all $a_l=1/(m\omega)$ because $da_l/dt$
is zero in that special case, see \secref{secamd}. If the width
has a different value, spurious scattering occurs like in the
case of free motion whenever two particles come too close in
phase space.

The left hand part of \figref{FixedWidth} displays for the very
same system as in the previous section the result of the
time evolution but without the randomizing two-body interaction. If
the widths are chosen to be $a_l=1/(m\omega)$, 
the resulting exact  time evolution is
just a unitary transformation in the one-body space and the
occupation probabilities are stationary (circles).  
But if the widths are 
taken as $a_l=1.2 /(m\omega)$ the occupation probabilities
change in time and the spurious scatterings equilibrate the
system even without an interaction.  The right hand part of
\figref{FixedWidth} shows the mean occupation probabilities when
the interaction is switched on (triangles).
Apparently the nature of the randomizing force is not relevant,
it may even be a spurious force which originates from a too
restricted trial state.

\begin{figure}[ht]
\begin{center}
\epsfig{file=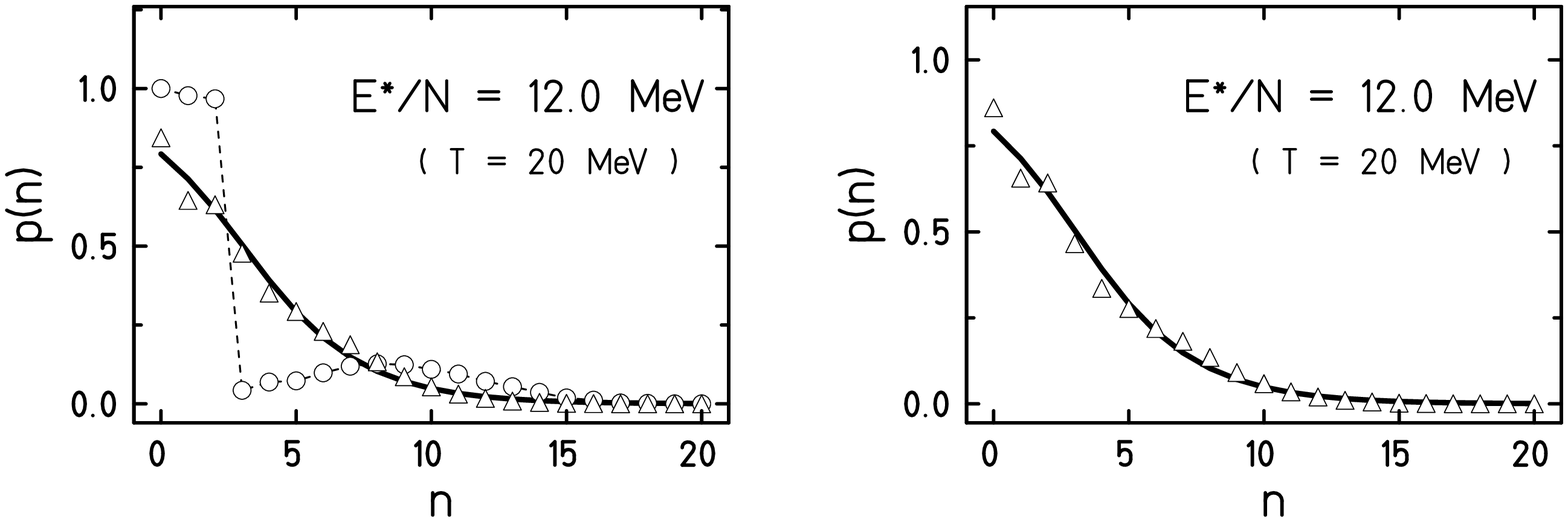,height=50mm}
\mycaption{Occupation probabilities calculated in the ergodic ensemble
using trial states with fixed widths (symbols)
compared to those calculated in the
canonical ensemble (solid line).
L.h.s.: Without interaction, circles: $a_l=1/(m\omega)$,
triangles: $a_l=1.2/(m\omega)$.
R.h.s.: With interaction and $a_l=1.2/(m\omega)$.
}{FixedWidth}
\end{center}
\end{figure}

In Antisymmetrized Molecular Dynamics (AMD) \cite{OHM92A,OHM92B}
trial states with time-independent widths are used, and as
expected from the above simple example the thermodynamic
properties of the model comply with Fermi-Dirac statistics.  It
would be interesting to see how collision term and branching
influence the dynamical statistical properties of AMD.  As the
Pauli-blocking prescription is consistent with the AMD state we
expect a faster equilibration due to the additional
randomization.

\subsubsection{Canonical and ergodic ensemble for distinguishable
particles}

To complete the discussion in this section the fermions are replaced by 
distinguishable particles, i.e. the antisymmetrized many-body state
is replaced by a product state of gaussian wave packets.
\red{
In this case the relation between temperature $T$ and excitation energy
$E^*/N$ is known and given by
$E^*/N=\frac{1}{2}\omega[\mbox{coth}(\frac{\omega}{2T})-1]$.
}

\begin{figure}[hh]
\begin{center}
\epsfig{file=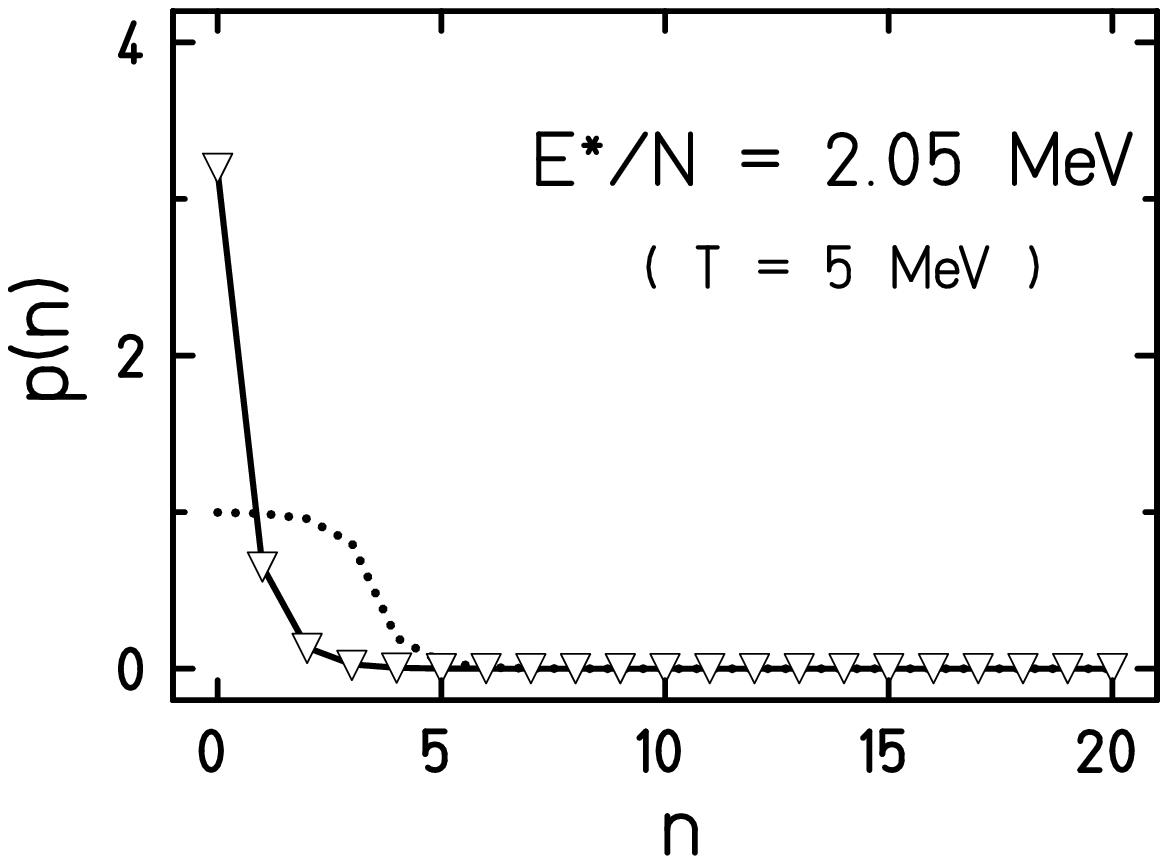,height=50mm}
\mycaption{Mean occupation numbers for a product state
(Boltzmann statistics). Ergodic ensemble (triangles),
canonical ensemble (solid line), both for an excitation energy
of $E^*/N=2.05$~MeV which corresponds to a temperature
$T=5$~MeV in the canonical ensemble.
The dotted line shows the result for Fermi-Dirac statistics
at the same temperature.}{HOBoltzmann}
\end{center}
\end{figure}

The ergodic ensemble is again investigated at
different energies and compared with the corresponding canonical
ensembles with the same mean energies \cite{ScF96,FeS97}. The time
evolution
of the system exhibits also for this case ergodic behaviour at
all excitation energies.  As an example \figref{HOBoltzmann} is 
showing the case of $E^*/N = 2.05$~MeV
\red{(which corresponds to $T = 5$~MeV)}
after a
time averaging of about 2000 periods.  The ergodic ensemble
(triangles) and the Boltzmann canonical ensemble (solid line)
are the same within the size of the symbols.
\red{This means that the ergodic ensemble is again equivalent to the quantum
canonical ensemble and not to the classical one, because for
$E^*/N = 2.05$~MeV and $T = 5$~MeV one is still in the quantal
regime according to the relation given above. The classical relation
$E^*/N=T$ for the one-dimensional oscillator holds only for
$E^*/N\gg\omega$ (here $\omega=8$~MeV).}

However, since distinguishable particles are not affected by 
the Pauli principle, the occupation numbers for the
many-body ground state look quite different from those of the 
Fermi-Dirac distribution at the same temperature
(dotted line
\red{in \figref{HOBoltzmann}}).



\subsubsection*{Resum\'{e}}

When discussing statistical properties of molecular dynamics
with gaussian wave-packets (coherent states) one should always
keep in mind that any observable or statistical weight has to be
calculated with the trial state according to quantum
mechanics. One should not fall into a completely classical
approach, misled by the ``classical" appearance of equations of
motion or phase space integrals, which is due to the
representation of the coherent states in terms of $\vek{r}_k$
and $\vek{p}_k$.

Statistical properties of molecular dynamics for fermions can be
deduced from simulations of equilibrium situations but due to
quantum effects a measure for intensive quantities like
temperature is not readily available.

\newpage
\subsection{Thermal properties of interacting systems by time averaging}
\label{sec-4-3}

\red{
In the previous section only noninteracting or weakly
interacting systems are considered for which partition 
functions can often be calculated because the Hamilton operator 
is a one-body operator. If the interaction between the particles 
is strong enough, for example attractive enough to form
self-bound many-body systems, the ideal gas picture is no longer
valid and solid and liquid phases appear besides the vapor
phase. Here the partition function $Z(T)$ cannot be evaluated
analytically anymore because it would amount to solving the full 
interacting many-body problem
$\op{H} \ket{\Psi_n} = E_n \ket{\Psi_n}$ in
the desired range of energies.

Take for example a fermion system in a large spherical container with
repulsive walls. At zero temperature the lowest eigenstate 
$\ket{\Psi_{n=1}}$ describes a self-bound system in its internal
ground state located at the center of the container. 
With increasing energies not only internal excitations and c.m. motion
of this drop occur but there is also the possibility to have two
or more bound objects which are separated from each other
and surrounded by vapor. 
They can be in different internal excitations
with various c.m. energies and vapor energies,
all adding up to the total eigenenergy $E_n$. 
This means that the quantum number $n$ enumerates not only the
excited eigenstates but also the c.m. degrees of freedom, the partition
into different drops, and the fermion vapor state.
The number of eigenstates in an energy interval increases
rapidly with excitation energy.

In principle one can deduce all thermodynamic properties from the
level density, but it is obvious that for those
complex and highly correlated states $\ket{\Psi_n}$
it is impossible to solve the
eigenvalue problem, neither analytically nor numerically. Therefore one
tries to simulate correlated many-body systems by 
means of molecular dynamics (which describes the time evolution 
in an approximate way) and replaces the ensemble average by 
a time average.      
}
\subsubsection{FMD - the nuclear caloric curve}

As discussed in the section \ref{sec-4-2} 
it can be shown that even
a small system with only a few particles distributed over two
different harmonic oscillator potentials is ergodic.  So one can
use one subsystem, for which the relation between excitation
energy and temperature is known,
as a thermometer to determine
the temperature of the other. This idea has been used in FMD
simulations of phase transitions in nuclei where the ideal Fermi
gas picture does not apply 
\red{
because the nucleons are interacting by strong two-body forces.
The construction of a thermometer is necessary because  
}
the temperature cannot be determined 
\red{
from the momentum distribution
} 
\cite{ScF97}.
\red{
In an interacting small fermion system the Fermi-edge is
broadened not only due to temperature but also due to the finite size
of the system and two-body correlations.

A thermostat (large heat bath coupled to the system) is not advisable 
because phase transitions in small systems are recognized best in a 
micro-canonical situation where the energy distribution is within
a narrow energy range and variations of the level density
$\rho(E)$, which indicate a phase transition,
are not blurred by the Boltzmann factor $e^{-E/T}$ of the
canonical ensemble.
}   

The thermometer consists of a quantum system of distinguishable
particles which move in a common harmonic oscillator potential
and interact with the nucleons.  The nucleus itself is confined
by a wide harmonic oscillator potential which serves as a
containment.  This is an important part of the setup because it
keeps the evaporated nucleons (vapor) in the vicinity of the
remaining liquid drop so that equilibration with the surrounding
vapor can take place.  The coupling between nucleons and
thermometer particles is chosen to be weak, repulsive and of
short range. It has to be as weak as possible in order not to
influence the nuclear system too much. On the other hand it has
to be strong enough to allow for reasonable equilibration times.

The determination of the relation between excitation energy
$E^*$ and temperature $T$ (caloric curve) is done by time-averaging
of the energy of the nucleonic system as well as of the
thermometer over a long period according to \eqref{EEM}. The
time-averaged energy of the thermometer $E_{th}$ determines the
temperature $T$ through the known relation
\red{$E_{th}/N_{th}=
\frac{3}{2}\omega_{th}\mbox{coth}(\omega_{th}/2T)$}
for an ideal gas of
distinguishable particles in a common harmonic oscillator
potential with frequency $\omega_{th}$ (quantum Boltzmann statistics).

\begin{figure}[ht]
\begin{center}
\epsfig{file=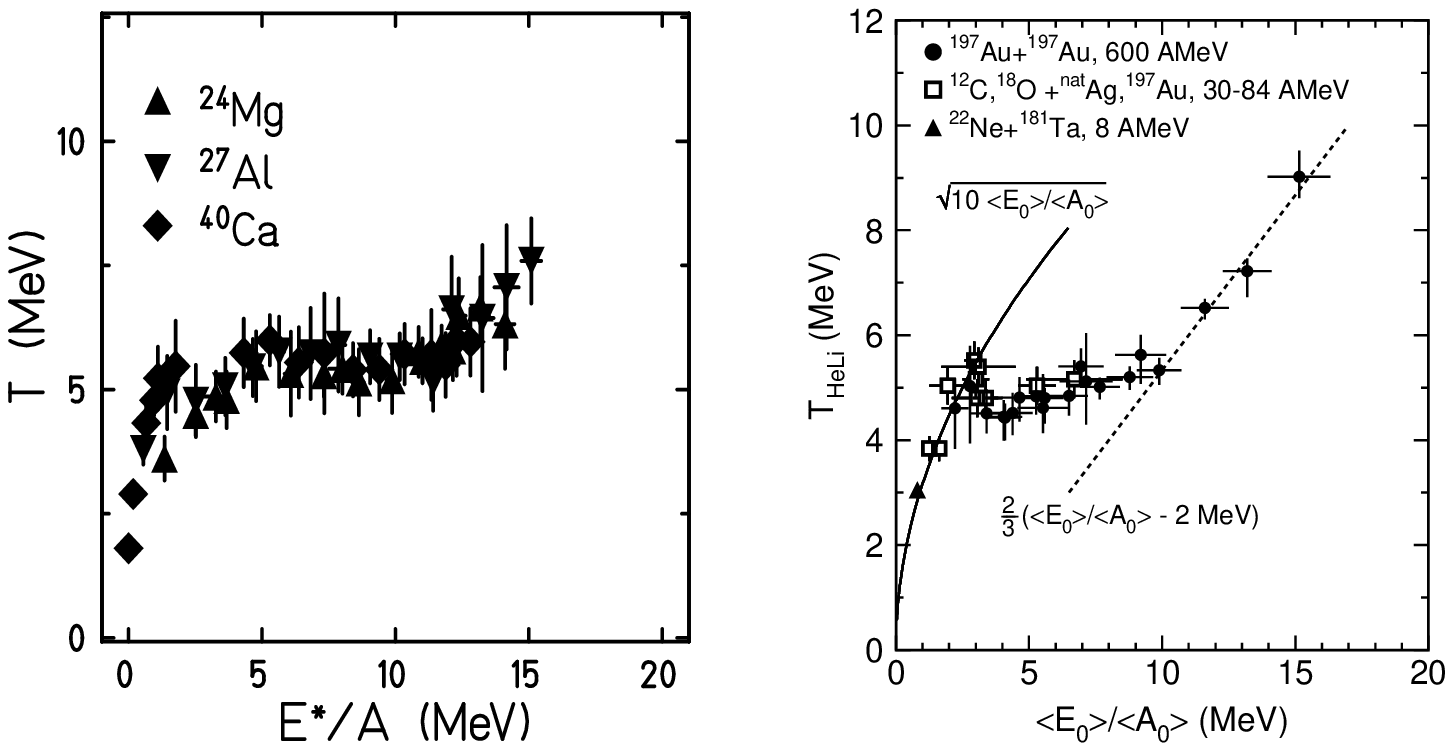,height=75mm}
\vspace*{3mm}
\mycaption{Caloric curve, l.h.s.: of \element{24}{Mg},
\element{27}{Al} and \element{40}{Ca} at $\hbar\omega=1~\MeV$,
from \cite{ScF97},
r.h.s.: as determined by the Aladin group
from the decay of spectator nuclei, from \cite{Poc95}.}{F-4-3-1}
\end{center}
\end{figure}

The resulting caloric curves for the nuclei \element{24}{Mg},
\element{27}{Al} and \element{40}{Ca} are displayed on the left hand side of
\figref{F-4-3-1}. All caloric curves clearly exhibit three different
parts. Beginning at small excitation energies the temperature
rises steeply with increasing energy as expected for a Fermi gas
in the shell model. There the nucleons remain bound in the
excited nucleus which behaves like a drop of liquid.  At an
excitation energy of $3~\MeV$ per nucleon the curve flattens and
stays almost constant up to about $11~\MeV$. This plateau at
$T\approx$ 5 to 6 MeV marks the coexistence phase where at low
excitation energy one big drop is surrounded with low density
vapor. With increasing energy the drop disolves more and more
into vapor until all nucleons are unbound and the system has
reached the vapor phase. The latent heat is hence about 8~MeV at
a pressure which is estimated to be close to zero.

The caloric curve shown on the l.h.s. of \figref{F-4-3-1} has a striking
similarity with the caloric curve determined experimentally from
the fragmentation of colliding nuclei by the ALADIN group
\cite{Poc95,Poc97}. Their results are displayed on the r.h.s. of
the same figure. The position and the extension of the plateau
agree quite well with the FMD calculation. Nevertheless, there
are important differences between the experimental setup and the
one used in the simulations. For further details see
\cite{Poc97,ScF97}.

\subsubsection{Phase transitions of hydrogen plasma}
\label{sec-4-3-2}

Hydrogen plasma under extreme conditions, high temperature or
pressure is of great current interest, since it shows new
structural, dynamical and electronic properties like
orientational ordering, pressure induced metallization or
changes in the vibronic spectra \cite{HeM92,KTR94A,KRT99}.

One model employed in this context, is called
wave-packet molecular dynamics. It is using gaussian packets with 
time-dependent widths and a Pauli potential, \red{which is derived
from the antisymmetrization of pairs of nucleons,} see \secref{secap}
and \cite{KTR94A}.  256 protons and 256 electrons are
distributed in a cubic box with periodic continuation in all
directions. Since the protons are classical their temperature
is simply given by their kinetic energy via the equipartition
theorem.

The equations of motion are followed over $6\cdot 10^{-14}$~s.
One observable which is sampled over this time period is the
proton pair distribution function $g_{pp}(r)$ which is related to the
probability of finding two particles at the distance $r$. The
proton pair distribution reveals details about the binding and
short range correlations in the system.
\begin{figure}[ht]
\begin{center}
\epsfig{file=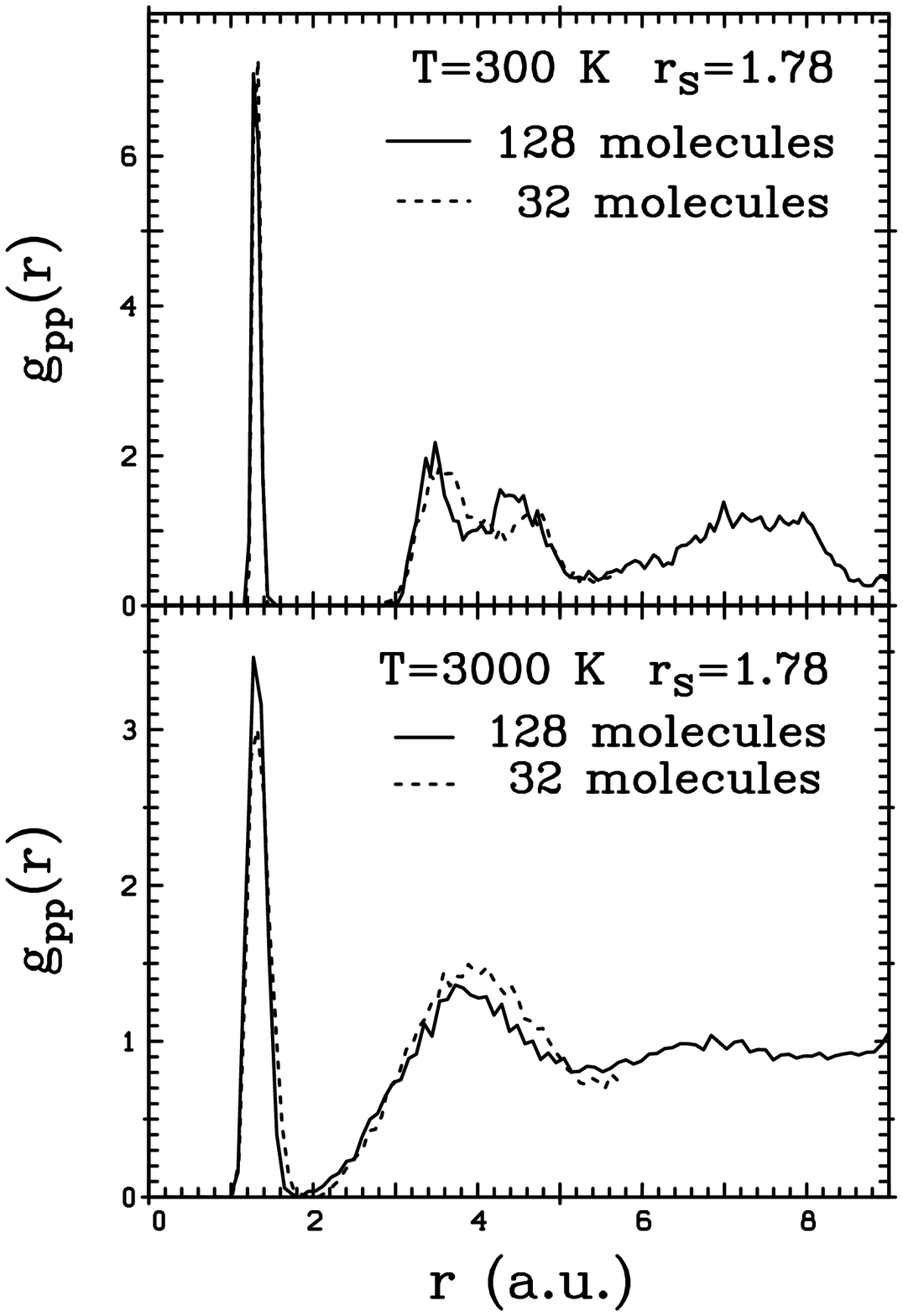,height=100mm}
\mycaption{The proton pair distribution function $g_{pp}(r)$ at
$T=300$~K (top) and $T=3000$~K (bottom). The solid line shows
the simulation with 128 molecules, the dashed one with 32 molecules.
Figure provided by P.-G. Reinhardt.}{F-PGR-1}
\end{center}
\end{figure}
In \figref{F-PGR-1} the pair distribution function $g_{pp}(r)$
is plotted for two temperatures. The peak at
$r=1.3$~a.u. signals that the protons are bound in H$_2$ molecules. The
authors find that due to medium effects this bond length is
shifted to a smaller value compared to the free one which is
$r=1.47$~a.u. in their model. At $T=300$~K pronounced peaks can
be seen around $r=4$~a.u. which indicate a solid structure of the
H$_2$ molecules. That these peaks are smoother at the higher
temperature is interpreted as a signal that the system is now in
the liquid phase. 

The authors also find that wave-packet molecular dynamics is
very efficient for the discussed problem and faster than the
Car-Parrinello method \cite{CaP85} while providing comparable
results \cite{KTR94A}.

\newpage
%
\acknowledgments For supplying us with references or figures we
would like to thank S.~Bass, M.~Belkacem, Ph.~Chomaz,
P.~Danielewicz, W.~Ebeling, O.~Knospe, Tomoyuki Maruyama, A.~Ono, J.~Randrup,
E.~Rass, P.-G.~Reinhard, U.~Saalmann, R.~Schmidt and H.~St\"ocker.
We also thank R.~Roth for carefully reading
the manuscript.

%


\begin{references}
\newcommand{\titel}[1]{{"#1," }}
\addcontentsline{toc}{section}
{{\numberline {REFERENCES\hskip 0pt plus1fill minus1fill\relax }{}}}
\harvarditem{Aichelin and Bertsch}{1985}{AiB85}
	Aichelin~J., and G.~Bertsch, 1985,
	\titel{Numerical simulation of medium energy heavy ion
	reactions} 
	Phys. Rev. C {\bf 53}, 1730.
%
\harvarditem{Aichelin and St\"ocker}{1986}{AiS86}
	Aichelin~J., and H.~St\"ocker, 1986,
	\titel{Quantum Molecular Dynamics - a novel approach to
	$N$-body correlations in heavy ion collisions}
	Phys. Lett. B   {\bf 176}, 14.
%
\harvarditem{Aichelin {\em et al.}}{1987}{ARP87}
	Aichelin~J., A.~Rosenhauer, G.~Peilert,
	H.~St\"ocker, and W.~Greiner, 1987,
	\titel{Importance of momentum-dependent interactions for
	the extraction of the nuclear equation of state from
	high-energy heavy-ion collisions}
	Phys. Rev. Lett. {\bf 58}, 1926.
%
\harvarditem{Aichelin}{1991}{Aic91}
	Aichelin~J., 1991,
	\titel{Quantum Molecular Dynamics -- a dynamical
	microscopic $N$-body approach to investigate fragment
	formation and the nuclear equation of state in heavy ion
	collisions}
	Phys. Rep. {\bf 202}, 233.
%
\harvarditem{Arnol'd}{1989}{Arn89}
	Arnol'd~V.I., 1989,
	{\it Mathematical Methods of Classical Mechanics},
	(Springer-Verlag, Berlin).
%
\harvarditem{Balian and Veneroni}{1988}{BaV88}
        Balian~R., and M.~Veneroni, 1988,
	\titel{Static and dynamic variational principles for
        expectation values of observables}
	Ann. Phys. (N.Y.) {\bf 187}, 29.
%
\harvarditem{Balian and Veneroni}{1992}{BaV92}
        Balian~R., and M.~Veneroni, 1992,
	\titel{Correlations and fluctuations in static and
        dynamic mean-field approaches}
	Ann. Phys. (N.Y.) {\bf 216}, 351.
\harvarditem{Bauhoff {\em et al.}}{1985}{BCG85}
	Bauhoff~W., E.~Caurier, B.~Grammaticos, and
	M.~P\l{}oszajczak, 1985,
	\titel{Description of light-ion collisions in the
	time-dependent cluster model}
	Phys. Rev. C {\bf 32}, 1915.
%
\harvarditem{Bass {\em et al.}}{1998}{BBB98A}
	Bass~S.A., M.~Belkacem, M.~Bleicher, M.~Brandstetter,
	L.~Bravina, C.~Ernst, L.~Gerland, M.~Hofmann,
	S.~Hofmann, J.~Konopka, G.~Mao, L.~Neise, S.~Soff,
	C.~Spieles, H.~Weber, L.A.~Winckelmann, H.~St\"ocker,
	W.~Greiner, Ch.~Hartnack, J.~Aichelin, and N.~Amelin, 1998,
	\titel{Microscopic models for ultrarelativistic heavy ion collisions}
	Prog. Part. Nucl. Phys. {\bf 41}, 225.
%
\harvarditem{Belkacem {\em et al.}}{1998}{BBB98B}
	Belkacem~M., M.~Brandstetter, S.A.~Bass, M.~Bleicher,
	L.~Bravina, M.I.~Gorenstein, J.~Konopka, L.~Neise,
	C.~Spieles, S.~Soff, H.~Weber, H.~St\"ocker, and
	W.~Greiner, 1998,
	\titel{Equation of state, spectra and composition of hot
	and dense infinite hadronic matter in a microscopic
	transport model}
	Phys. Rev. C {\bf 58}, 1727.
%
\harvarditem{Bertsch and Das Gupta}{1988}{BeD88}
	Bertsch~G.F., and S.~Das Gupta, 1988,
	\titel{A guide to microscopic models for intermediate
	energy heavy ion collisions}
	Phys. Rep. {\bf 160}, 189.
%
\harvarditem{Blaise {\em et al.}}{1994}{BDH94}
	Blaise~ P., P.~Durand, and O.~Henri-Rousseau, 1994,
	\titel{Irreversible evolution towards equilibrium of
	coupled quantum harmonic oscillators. A coarse grained
	approach} 
	Physica A {\bf 209}, 51.
%
\harvarditem{Boal and Glosli}{1988}{BoG88}
	Boal~D.H., and J.N.~Glosli, 1988,
	\titel{Quasiparticle model for nuclear dynamics studies:
	Ground state properties}
	Phys. Rev. C {\bf 38}, 1870.
%
\harvarditem{Bodmer and Panos}{1977}{BoP77}
	Bodmer~A.R., and C.N.~Panos, 1977,
	\titel{Classical microscopic calculations of high-energy
	collisions of heavy ions}
	Phys. Rev. C {\bf 15}, 1342.
%
\harvarditem{Bodmer {\em et al.}}{1980}{BPM80}
	Bodmer~A.R., C.N.~Panos, and A.D.~MacKellar, 1980,
	\titel{Classical-equations-of-motion calculations of
	high-energy heavy-ion collisions}
	Phys. Rev. C {\bf 22}, 1025.
%
\harvarditem{Bondorf {\em et al.}}{1995}{BIM95}
	Bondorf~J.P., D.~Idier, and I.N.~Mishustin, 1995,
	\titel{Self-organization in expanding nuclear matter}
	Phys. Lett. B {\bf 359}, 261.
%
\harvarditem{Broeckhove {\em et al.}}{1988}{BLK88}
	Broeckhove~J., L.~Lathouwers, E.~Kesteloot, and P.~van
	Leuven, 1988,
	\titel{On the equivalence of the time-dependent variational
	principles} 
	Chem. Phys. Lett.   {\bf 149}, 547.
%
\harvarditem{Broeckhove {\em et al.}}{1989}{BLL89}
	Broeckhove~J., L.~Lathouwers, and P.~van Leuven, 1989,
	\titel{Time-dependent variational principles and conservation
	laws in wavepacket dynamics}
	J.~Phys.~A: Math. Gen.  {\bf 22}, 4395.
%
%
\harvarditem{Car and Parrinello}{1985}{CaP85}
	Car~R., and M.~Parrinello, 1985,
	\titel{Unified approach for molecular dynamics and density
	functional theory}
	Phys. Rev. Lett. {\bf 55}, 2471.
%
%
\harvarditem{Caurier {\em et al.}}{1982}{CGS82}
	Caurier~E., B.~Grammaticos, and T.~Sami, 1982,
	\titel{The time-dependent cluster model}
	Phys. Lett. B {\bf 109}, 150.
%
\harvarditem{Chiba {\em et al.}}{1996}{CIF96}
	Chiba~S., O.~Iwamoto, T.~Fukahori, K.~Niita,
	T.~Maruyama, T.~Maruyama, and A.~Iwamoto, 1996, 
	\titel{Analysis of proton-induced fragment production
	cross-sections by the quantum molecular dynamis plus
	statistical decay model}
	Phys. Rev. C {\bf 54}, 285.
%
\harvarditem{Chomaz {\em et al.}}{1996}{CCG96}
        Chomaz, Ph., M.Colonna, and A.Guarnera, 1996,
        \titel{Spinodal deomposition of atomic nuclei}
        in {\it Advances in Nuclear Dynamics 2},
        edited by B. Arruñada and G. Westfall,
        (Kluwer Academic/Plenum).
%
\harvarditem{Colonna and Chomaz}{1998}{CoC98}
	Colonna~M., and Ph.~Chomaz, 1998,
	\titel{Spinodal decomposition in nuclear molecular dynamics}
	Phys. Lett. B {\bf 436}, 1.
%
\harvarditem{Davies {\em et al.}}{1985}{DDK85}
	Davies~K.T.R., K.R.S.~Devi, S.E.~Koonin, and
	M.R.~Strayer, 1985,
	\titel{TDHF calculations of heavy-ion collisions}
	in {\it Treatise of Heavy-Ion Science}, Vol 3,
	edited by D.A.~Bromley, (Plenum, New York), p. 3.
%
\harvarditem{Dorso {\em et al.}}{1987}{DDR87}
	Dorso~C., S.~Duarte, and J.~Randrup, 1987,
	\titel{Classical simulation of the Fermi gas}
	Phys. Lett. B   {\bf 188}, 287.
%
\harvarditem{Dorso and Randrup}{1987}{DoR87}
	Dorso~C., and J.~Randrup, 1987,
	\titel{Classical simulation of nuclear systems}
	Phys. Lett. B   {\bf 215}, 611.
%
\harvarditem{Dorso and Randrup}{1989}{DoR89}
	Dorso~C., and J.~Randrup, 1989,
	\titel{Quasi-classical simulation of nuclear
	dynamics. Phase evolutuion of disassembling nuclei}
	Phys. Lett. B   {\bf 232},  29.
%
\harvarditem{Dro\.{z}d\.{z} {\em et al.}}{1982}{DOP82}
	Dro\.{z}d\.{z}~S., J.~Oko\l{}owicz, and
	M.~P\l{}oszajczak, 1982,
	\titel{The time-dependent cluster theory -- application
	to the $\alpha$--$\alpha$ collision}
	Phys. Lett. B {\bf 109}, 145.
%
\harvarditem{Dro\.{z}d\.{z} {\em et al.}}{1986}{DPC86}
	Dro\.{z}d\.{z}~S., M.~P\l{}oszajczak, and E.~Caurier, 1986,
	\titel{Variational approach to the Schr\"odinger dynamics in
	the Klauder's continuous representations}
	Ann. Phys. {\bf 171}, 108.
%
\harvarditem{Ebeling {\em et al.}}{1996}{EFP96}
	Ebeling~W., A.~F\"orster, and V.Yu.~Podlipchuk, 1996,
	\titel{Quantum wave-packets simulation of ionization
	processes in dense plasmas}
	Phys. Lett. A {\bf 218}, 297.
%
\harvarditem{Ebeling and Militzer}{1997}{EbM97}
	Ebeling W., and B.~Militzer, 1997,
	\titel{Quantum molecular dynamics of partially ionized plasmas}
	Phys. Lett. A {\bf 226}, 298.
%
\harvarditem{Ebeling and Schautz}{1997}{EbS97}
	Ebeling W., and F.~Schautz, 1997,
	\titel{Many particle simulations of the quantum electron
	gas using momentum-dependent potentials}
	Phys. Rev. E {\bf 56}, 3498.
%
\harvarditem{Feldmeier}{1990}{Fel90}
	Feldmeier H., 1990,
	\titel{Fermionic molecular dynamics}
	Nucl. Phys. A {\bf 515}, 147.
%
\harvarditem{Feldmeier  {\em et al.}}{1995}{FBS95}
	Feldmeier H., K.~Bieler, and J.~Schnack, 1995,
	\titel{Fermionic molecular dynamics for ground states
	and collisions of nuclei} 
	Nucl. Phys. A  {\bf 586}, 493.
\harvarditem{Feldmeier and Schnack}{1997}{FeS97}
	Feldmeier H., and J.~Schnack, 1997, 
	\titel{Fermionic molecular dynamics}
	Prog. Part. Nucl. Phys. {\bf 39}, 393.
%
\harvarditem{Feldmeier  {\em et al.}}{1998}{FNR98}
	Feldmeier H., T.~Neff, R.~Roth, and J.~Schnack, 1998,
	\titel{A unitary correlation operator method} 
	Nucl. Phys. A {\bf 632}, 61.
%
\harvarditem{Frenkel}{1934}{Fre34}
	J.~Frenkel J., 1934,
	{\it Wave mechanics}, Advanced theory,
	(Claredon Press, Oxford), p 235.
%
\harvarditem{Gaitanos {\em et al.}}{1999}{GFH99}
        Gaitanos T., C.~Fuchs, and H.H.~Wolter, 1999, 
        \titel{Heavy ion collisions with nonequilibrium
        Dirac-Brueckner mean fields} 
        Nucl. Phys. A {\bf 650}, 97.
%
\harvarditem{Goeke and Reinhard}{1982}{GoR82}
	Goeke K., and  P.G.~Reinhard, 1982, Eds.,
	{\it Time-dependend Hartree-Fock and beyond},
	Lecture Notes in Physics {\bf 171}.
%
\harvarditem{Griffin {\em et al.}}{1980}{GLD80}
	Griffin J.J., P.C.~Lichtner, and M.~Dworzecka, 1980,
	\titel{Time-dependent $S$-matrix Hartree-Fock theory of
	complex reactions}
	Phys. Rev. C {\bf 21}, 1351.
%
\harvarditem{Gerjuoy, Rau, and Spruch}{1983}{GRS83}
	Gerjuoy E., A.R.P.~Rau, and Larry~Spruch, 1983,
	\titel{A unified formulation of the construction of
        variational principles}
	Rev. Mod. Phys. {\bf 55}, 725.
\harvarditem{Hartnack {\em et al.}}{1989}{HZN89}
	Hartnack C., L.~Zhuxia, L.~Neise, G.~Peilert,
	A.~Rosenhauer, H.~Sorge, J.~Aichelin, H.~St\"ocker,
	and W.~Greiner, 1989, 
	\titel{Quantum molecular dynamics: a microscopic model
	from UNILAC to CERN energies}
	Nucl.~Phys.~A {\bf 495}, 303.
%
\harvarditem{Hartnack {\em et al.}}{1998}{HPA98}
	Hartnack C., R.K.~Puri, J.~Aichelin, J.~Konopka,
	S.A.~Bass, H.~St\"ocker, and W.~Greiner, 1998, 
	\titel{Modelling the many-body dynamics of heavy ion
	collisions: Present status and future perspective}
	Eur. J. Phys. A  {\bf 1}, 151.
%
\harvarditem{Heller}{1975}{Hel75}
	Heller E.J., 1975,
	\titel{Time-dependent approach to semiclassical dynamics}
	J.~Chem. Phys. {\bf 62}, 1544.
%
\harvarditem{Hemley and Mao}{1992}{HeM92}
	Hemley R.J., and H.K.~Mao, 1992,
	\titel{Anomalous low-frequency excitations in
	diamond-cell studies of hydrogen at megabar pressures}
	Phys. Lett. A {\bf 163}, 429.
%
\harvarditem{Hoover}{1985}{Hoo85}
	W.G.~Hoover, 1985,
	\titel{Canonical dynamics: Equilibrium phase-space
	distributions}
	Phys. Rev. A {\bf 31}, 1695.
%
\harvarditem{Hoover}{1986}{Hoo86}
	Hoover W.G., 1986,
	\titel{Molecular dynamics}
	Lecture Notes in Physics  {\bf 258},
	(Springer, Berlin)
%
%
\harvarditem{Kanada-En'yo {\em et al.}}{1995}{KHO95}
	Kanada-En'yo Y., H.~Horiuchi, and A.~Ono, 1995,
	\titel{Structure of Li and Be isotopes studied with
	antisymmetrized molecular dynamics}
	Phys. Rev. C {\bf 52}, 628.
%
\harvarditem{Katz}{1965}{Kat65}
	Katz A., 1965,
	{\it Classical Mechanics, Quantum Mechanics, Field Theory},
	(Academic Press, New York).
%
\harvarditem{Kerman and Koonin}{1976}{KeK76}
   Kerman A.K., and S.E.~Koonin, 1976,
   \titel{Hamiltonian formulation of time-dependent variational
   principles for the many-body system}
   Ann. Phys. {\bf 100}, 332.
%
\harvarditem{Khoa {\em et al.}}{1992}{KOM92}
	Khoa D.T., N.~Ohtsuka, M.A.~Matin, A.~Faessler,
	S.W.~Huang, E.~Lehmann, and R.K.~Puri, 1992,
	\titel{In-medium effects in the description of heavy-ion
	collisions with realistic NN interactions}
	Nucl. Phys. A {\bf 548}, 102.
%
\harvarditem{Kiderlen and Danielewicz}{1996}{KiD96}
	Kiderlen D., and P.~Danielewicz, 1996,
	\titel{Fragments in gaussian wave-packet dynamics with
	and without correlations} 
	Nucl. Phys. A {\bf 620}, 346.
%
\harvarditem{Klakow {\em et al.}}{1994}{KTR94A}
	Klakow D., C.~Toepffer, and P.-G.~Reinhard, 1994,
	\titel{Hydrogen under extreme conditions}
	Phys. Lett. A {\bf 192}, 55.
%
\harvarditem{Klakow {\em et al.}}{1994}{KTR94B}
	Klakow D., C.~Toepffer, and P.-G.~Reinhard, 1994,
	\titel{Semiclassical molecular dynamics for strongly 
        coupled Coulomb systems}
	J. Chem. Phys. {\bf 101}, 1.
%
\harvarditem{Klauder and Skagerstam}{1985}{KlS85}
   Klauder J.R., and B.-S.~Skagerstam, 1985,
   {\it Coherent states},
   (World Scientific Publishing, Singapore).
%
\harvarditem{Knaup {\em et al.}}{1999}{KRT99}
   Knaup, M., P.-G. Reinhard, and Ch. Toepffer, 1999,
   \titel{Wave packet molecular dynamics simulations of hydrogen
    near the transition to a metallic fluid}
    Contrib. Plasma Phys. {\bf 39}, 57.
%
\harvarditem{Knoll and Strack}{1984}{KnS84}
   Knoll J., and B. Strack, 1984,
   \titel{The dynamics of the nuclear disassembly in a 
      field-theoretical model at finite entropies}
   Phys. Lett. B {\bf 149}, 45.
%
\harvarditem{Knoll and Wu}{1988}{KnW88}
   Knoll J., and J. Wu, 1988,
   \titel{Expansion dynamics and multifragmentation of a
   saturating system of fermions} 
   Nucl. Phys. A {\bf 481}, 173.
%
\harvarditem{Konopka {\em et al.}}{1995}{KSG95}
	Konopka J., H.~St\"ocker, and W.~Greiner, 1995,
	\titel{On the impossibility of temperature extraction 
        from heavy ion induced particle spectra}
	Nucl. Phys. A {\bf 583}, 357c.
%
\harvarditem{Kramer and Saraceno}{1981}{KrS81}
   Kramer P., and M.~Saraceno, 1981,
   \titel{Geometry of the time-dependent variational principle
   in quantum mechanics}
   Lecture Notes in Physics   {\bf 140},
   (Springer, Berlin).
%
%
\harvarditem{Kusnezov {\em et al.}}{1990}{KBB90}
	Kusnezov D., A.~Bulgac, and W.~Bauer, 1990,
	\titel{Canonical ensembles from chaos}
	Ann. of Phys. {\bf 204}, 155.
%
\harvarditem{Kusnezov}{1993}{Kus93}
	Kusnezov D., 1993,
	\titel{Quantum ergodic wave functions from a thermal
	non-linear Schr\"odinger equation}
	Phys. Lett. A {\bf 184}, 50.
%
\harvarditem{Lacroix, Chomaz, and Ayik}{1998, 1999}{LCA98}
	Lacroix D., Ph.~Chomaz, and S.~Ayik, 1998, 1999,
	\titel{Quantal extension of mean-field dynamics}
	Proceedings XXXVI. Int. Winter Meeting on Nuclear
	Physics, Bormio/Italy p. 485, and
        \titel{On the simulation of extended TDHF theory}
        Nucl. Phys. A {\bf 651}, 369.
%
\harvarditem{Latora {\em et al.}}{1994}{LBB94}
	Latora V., M.~Belkacem, and A.~Bonasera, 1994,
	\titel{Dynamics of instabilities and intermittency}
	Phys. Rev. Lett. {\bf 73}, 1765.
%
\harvarditem{Lehmann {\em et al.}}{1995}{LPF95}
	Lehmann E., R.K.~Puri, A.~Faessler, G.~Batko, and
	S.W.~Huang, 1995,
	\titel{Consequences of a covariant description of heavy
	ion reactions at intermediate energies}
	Phys. Rev. C {\bf 51}, 2113.
%
\harvarditem{Maruyama {\em et al.}}{1992}{MOH92}
	Maruyama Toshiki, A.~Ohnishi, and H.~Horiuchi, 1992,
	\titel{Evolution of reaction mechanisms in the light
	heavy-ion system}
	Phys. Rev.  C {\bf 45}, 2355.
%
\harvarditem{Maruyama, Ono {\em et al.}}{1992}{MOO92}
	Maruyama Toshiki, A.~Ono, A.~Ohnishi, and H.~Horiuchi, 1992,
	\titel{Fragment mass distribution in intermediate energy
	heavy ion collisions and the reaction time scale}
	Prog. Theor. Phys. {\bf 87}, 1367.
%
\harvarditem{Maruyama {\em et al.}}{1996}{MNI96}
	Maruyama Toshiki, K.~Niita, and A.~Iwamoto, 1996,
	\titel{Extension of quantum molecular dynamics and its
	application to heavy-ion collisions}  
	Phys. Rev. C {\bf 53}, 297.
%
\harvarditem{Metropolis {\em et al.}}{1953}{MMR53}
	Metropolis M., A.W.~Metropolis, and M.N.~Rosenbluth,
	A.H.~Teller, and E.~Teller, 1953,
	\titel{Equation of state calculations by fast computing machines}
	J.~Chem. Phys. {\bf 21}, 1087.
%
\harvarditem{Neff {\em et al.}}{1999}{NFR99}
	Neff T., H. Feldmeier, R. Roth, and J. Schnack, 1999,
	\titel{Realistic interactions and configuration mixing
	in fermionic molecular dynamics}
	Proc. Int. Workshop XXXVII in Hirschegg, ISSN 0720-8715,
	p. 283.
%
\harvarditem{Niita {\em et al.}}{1995}{NCM95}
	Niita K., S.~Chiba, T.~Maruyama, T.~Maruyama, H.~Takada,
	T.~Fukahori, Y.~Nakahara, and A.~Iwamoto, 1995,
	\titel{Analysis of the (N,xN') reactions by quantum
	molecular dynamics plus statistical model}
	Phys. Rev. C {\bf 52}, 2620.
%
\harvarditem{Neumann and Fai}{1994}{NeF94}
	Neumann J.J.  and G. Fai, 1994,
	\titel{Classical lagrangian model of the Pauli principle}
	Phys. Lett. B  {\bf 329}, 419.
%
\harvarditem{Nos\'e}{1984}{Nos84}
	Nos\'e S., 1984,
	\titel{A unified formulation of the constant temperature
	molecular dynamics methods}
	J.~Chem. Phys. {\bf 81}, 511.
%
\harvarditem{Nos\'e}{1991}{Nos91}
	Nos\'e S., 1991,
	\titel{Constant temperature molecular dynamics methods}
	Prog. of Theor. Phys. Suppl. {\bf 103}, 1.
%
\harvarditem{Ohnishi and Randrup}{1993}{OhR93}
	Ohnishi A., and J.~Randrup, 1993,
	\titel{Statistical properties of antisymmetrized
	molecular dynamics}
	Nucl. Phys. A {\bf 565},  474.
\harvarditem{Ohnishi and Randrup}{1995}{OhR95}
	Ohnishi A., and J.~Randrup, 1995,
	\titel{Incorporation of quantum statistical features in
	molecular dynamics}
	Phys. Rev. Lett. {\bf 75}, 596.
%
\harvarditem{Ohnishi and Randrup}{1997}{OhR97A}
	Ohnishi A., and J.~Randrup, 1997,
	\titel{Inclusion of quantum fluctuations in wave packet
	dynamics}
	Ann. Phys. 253, 279.
%
\harvarditem{Ohnishi and Randrup}{1997}{OhR97B}
	Ohnishi A., and J.~Randrup, 1997,
	\titel{Quantum fluctuations affect the critical
	properties of noble gases} 
	Phys. Rev. A {\bf 55}, R3315.
%
\harvarditem{Ono {\em et al.}}{1992a}{OHM92A}
	Ono A., H.~Horiuchi, Toshiki Maruyama, and A.~Onishi, 1992,
	\titel{Fragment formation studied with antisymmetrized
	version of molecular dynamics with two nucleon collisions}
	Phys. Rev. Lett. {\bf 68}, 2898.
%
\harvarditem{Ono {\em et al.}}{1992b}{OHM92B}
	Ono A., H.~Horiuchi, Toshiki Maruyama, and A.~Onishi, 1992,
	\titel{Antisymmetrized version of molecular dynamics
	with two-nucleon collisions and its application to heavy
	ion reactions}
	Prog. Theor. Phys. {\bf 87}, 1185.
%
\harvarditem{Ono {\em et al.}}{1993}{OHM93}
	Ono A., H.~Horiuchi, Toshiki Maruyama, and A.~Onishi, 1993,
	\titel{Fragment formation studied with antisymmetrized
	version of molecular dynamics with two-nucleon collisions}
	Phys. Rev. C {\bf 47}, 2652.
%
\harvarditem{Ono and Horiuchi}{1996a}{OnH96A}
	Ono A., and H.~Horiuchi, 1996,
	\titel{Antisymmetrized molecular dynamics of wave
	packets with stochastic incorporation of Vlasov
	equation}
	Phys. Rev. C {\bf 53}, 2958.
\harvarditem{Ono and Horiuchi}{1996b}{OnH96B}
	Ono A., and H.~Horiuchi, 1996,
	\titel{Statistical properties of AMD for non-nucleon-emission
	and nucleon-emission process}
	Phys. Rev. C {\bf 53}, 2341.
%
\harvarditem{Ono}{1998}{Ono98}
	Ono A., 1998,
	\titel{Antisymmetrized molecular dynamics with quantum
	branching processes for	collisions of heavy nuclei} 
        Phys. Rev. {\bf C59}, 853.
%
\harvarditem{Peilert {\em et al.}}{1991}{PRS91}
	Peilert G., J.~Randrup, H.~St\"ocker, and W.~ Greiner, 1991,
	\titel{Clustering in nuclear matter at subsaturation densities}
	Phys. Lett. B {\bf 260}, 271.
%
\harvarditem{Peilert {\em et al.}}{1992}{PKS92}
	Peilert G., J.~Konopka, M.~Blann, M.G.~Mustafa,
	H.~St\"ocker, and W.~Greiner, 1992, 
	\titel{Dynamical treatment of Fermi motion in a
	microscopic description of heavy ion collisions}
	Phys. Rev. C {\bf 46}, 1457.
%
\harvarditem{Pochodzalla {\em et al.}}{1995}{Poc95}
	Pochodzalla J., T.~M\"ohlenkamp, T.~Rubehn, A.~Sch\"uttauf,
	A.~W\"orner, E.~Zude, M.~Begemann-Blaich, T.~Blaich,
	C.~Gross, H.~Emling, A.~Ferrero, G.~Imme, I.~Iori,
	G.J.~Kunde, W.D.~Kunze, V.~Lindenstruth, U.~Lynen,
	A.~Moroni, W.F.J.~M\"uller, B.~Ocker, G.~Raciti, H.~Sann,
	C.~Schwarz, W.~Seidel, V.~Serfling, J.~Stroth,
	A.~Trzcinski, W.~Trautmann, A.~Tucholski, G.~Verde,
	and B.~Zwieglinski, 1995, 
	\titel{Probing the nuclear liquid-gas phase transition}
	Phys. Rev. Lett. {\bf 75}, 1040.
%
\harvarditem{Pochodzalla}{1997}{Poc97}
	Pochodzalla J., 1997,
	\titel{The search for the liquid-gas phase transition in nuclei}
	Prog. Part. Nucl. Phys. {\bf 39}, 443.
%
\harvarditem{P\"uhlhofer}{1977}{Pue77}
	P\"uhlhofer F., 1977,
	\titel{On the interpretation of evaporation residue mass
	distributions in heavy-ion induced fusion reactions} 
	Nucl. Phys. A {\bf 280}, 267.
%
\harvarditem{Reinhard and Suraud}{1992}{ReS92}
        Reinhard P.-G., and E.~Suraud,
	\titel{Stochastic TDHF and the Boltzmann-Langevin equation}
	Ann. Phys. (NY) {\bf 216}, 98.
%
\harvarditem{Saraceno {\em et al.}}{1983}{SKF83}
   Saraceno M., P.~Kramer, and F.~Fernandez, 1983,
   \titel{Time-dependent variational description of
   $\alpha\alpha$ scattering} 
   Nucl. Phys.  A {\bf 405}, 88.
%
\harvarditem{Schmidt and Schnack}{1998}{ScS98}
	Schmidt H.-J., and J.~Schnack,
	\titel{Investigations on finite ideal quantum gases} 
	Physica A {\bf 260}, 479.
%
\harvarditem{Schnack and Feldmeier}{1996}{ScF96}
	Schnack J., and H.~Feldmeier, 1996,
	\titel{Statistical properties of Fermionic Molecular Dynamics}
	Nucl. Phys. A {\bf 601}, 181.
%
\harvarditem{Schnack and Feldmeier}{1997}{ScF97}
	Schnack J., and H.~Feldmeier, 1997,
	\titel{The nuclear liquid-gas phase transition within Fermionic
	Molecular Dynamics}
	Phys. Lett. B {\bf 409}, 6.
%
%
\harvarditem{Schnack}{1996}{Sch96}
	Schnack J., 1996,
	\titel{Kurzreichweitige Korrelationen in der
	Fermionischen Molekulardynamik}
	dissertation, Technical University Darmstadt.
%
\harvarditem{Schnack}{1998}{Sch98}
	Schnack J., 1998,
	\titel{Molecular dynamics investigations on a quantum
	system in a thermostat}
	Physica A {\bf 259}, 49.
%
\harvarditem{Schnack}{1999}{Sch99}
	Schnack J., 1999,
	\titel{Thermodynamics of the harmonic oscillator using
	coherent states} 
	Europhys. Lett. {\bf 45}, 647.
%
\harvarditem{Sorge {\em et al.}}{1989}{SSG89}
	Sorge H., H.~St\"ocker, and W.~Greiner, 1989,
	\titel{Poincar\'e invariant  hamiltonian  dynamics:
	Modelling multi-hadronic interactions in a phase space
	approach}
	Ann. Phys. (NY) {\bf 192}, 266.
%
\harvarditem{Sorge}{1995}{Sor95}
	Sorge H., 1995,
	\titel{Flavor production in Pb (160AGeV) on Pb
	collisions: Effect of color ropes and hadronic
	rescattering} 
	Phys. Rev. C {\bf C52}, 3291.
%
\harvarditem{St\"ocker and Greiner}{1986}{StG86}
	St\"ocker H., and W.~Greiner, 1986,
	\titel{High-energy heavy ion collisions: probing the
	equation of state of highly excited hadronic matter}
	Phys. Rep. {\bf 137}, 277.
%
\harvarditem{Suarez Barnes {\em et al.}}{1993}{BNN93}
	Suarez Barnes I.M., M.~Nauenberg, M.~Nockleby, and
	S.~Tomsovic, 1993,
	\titel{Semiclassical theory of quantum propagation: The
	Coulomb potential}
	Phys. Rev. Lett. {\bf 71}, 1961.
%
\harvarditem{Topaler {\em et al.}}{1997}{THA97}
	Topaler M.S., M.D.~Hack, T.C.~Allison, Yi-Ping Liu,
	S.L.~Mielke, D.W.~Schwenke, and D.G.~Truhlar, 1997,
	\titel{Validation of trajectory surface hopping methods
	against accurate quantum mechanical dynamics and
	semiclassical analysis of electronic-to-vibrational
	energy transfer}
	J.~Chem. Phys. {\bf 106}, 8699.
%
\harvarditem{Tsue and Fujiwara}{1991}{TsF91}
	Tsue Y., and Y.~Fujiwara, 1991,
	\titel{Time-dependent variational approach in terms of
	squeezed coherent states}
	Prog. Theor. Phys. {\bf 86}, 443.
%
\harvarditem{Tully}{1990}{Tul90}
	Tully J.C., 1990,
	\titel{Molecular dynamics with electronic transitions}
	J. Chem. Phys. {\bf 93}, 1061.
%
\harvarditem{Wilets {\em et al.}}{1977}{WHK77}
	Wilets L., E.M.~Henley, M.~Kraft, and A.D.~MacKellar,
	1977, 
	\titel{Classical many-body model for heavy-ion
	collisions incorporating the Pauli principle}
	Nucl. Phys. A {\bf 282}, 341.
%
\harvarditem{Wilets {\em et al.}}{1978}{WYC78}
	Wilets L., Y.~Yariv, and R.~Chestnut, 1978, 
	\titel{Classical many-body model for heavy-ion
	collisions (II)}
	Nucl. Phys. A {\bf 301}, 359.
%
\harvarditem{Wilets and Cohen}{1998}{WiC98}
	Wilets L., and J.S.~Cohen, 1998,
	\titel{Fermion molecular dynamics in atomic, molecular
	and optical physics}
	Contemporary Physics {\bf 39}, 163.
%
\harvarditem{Wolf {\em et al.}}{1990}{WBC90}
        Wolf Gy., G.~Batko, W.~Cassing, U.~Mosel, K.~Niita, and
        M.~Sch\"afer, 1990,
	\titel{Dilepton production in heavy-ion collisions}
	Nucl. Phys. A {\bf 517}, 615.
%
\end{references}
\end{document}